\documentclass[11pt,a4paper]{article}
\usepackage{epsfig}
\usepackage[T1]{fontenc}    
\usepackage{graphics}
\usepackage{graphicx}
\usepackage{subfigure}
\usepackage{psfrag}
\usepackage{pstricks,pst-coil,pst-fill,pst-plot}
\usepackage{pgf}  
\usepackage[fleqn]{amsmath}    
\usepackage{amssymb}    
\usepackage{amsfonts}   
\usepackage{verbatim}   
\usepackage{mathrsfs}   
\usepackage{dsfont}
\usepackage{euscript}
\usepackage{yfonts}
\usepackage{txfonts}
\usepackage{marvosym}
\usepackage{vmargin}        

\setmarginsrb{1.8cm}{2cm}{1.8cm}{2cm}{1cm}{1cm}{1cm}{1.6cm}
 \makeatletter
 \@addtoreset{equation}{section}
 \makeatother

\let\ua=\uparrow
\let\da=\downarrow
\let\tend=\rightarrow

\newtheorem{prop}{Proposition}[section]

\newtheorem{lemme}{Lemma}[section]

\def\Proof{\medskip\noindent {\it Proof --- \ }}

\def\qed{\hfill\rule{2mm}{2mm}}

\providecommand{\bysame}{\leavevmode\hbox to3em{\hrulefill}\thinspace}

\newcommand\beq{\begin{equation}}
\newcommand\enq{\end{equation}}
\newcommand\bem{\begin{multline}}
\newcommand\enm{\end{multline}}

\def\beqa{\begin{eqnarray}}
\def\eeqa{\end{eqnarray}}
\def\ba{\begin{array}}
\def\ea{\end{array}}

\def\det{\operatorname{det}}

\newcommand{\f}[2]{{\ensuremath{%
    \mathchoice%
    {\dfrac{#1}{#2}}
    {\dfrac{#1}{#2}}
    {\frac{#1}{#2}}
    {\frac{#1}{#2}}
}}}
\newcommand{\tf}[2]{\ensuremath{#1/#2}}
\newcommand{\pa}[1]{\ensuremath{\left(#1\right)}}
\newcommand{\paa}[1]{\ensuremath{\left\{#1\right\}}}
\newcommand{\pac}[1]{\ensuremath{\left[#1\right]}}
\newcommand{\paf}[2]{\ensuremath{\left(\f{#1}{#2}\right)}}

\def\a{\alpha}

\def\be{\beta}

\def\Ga{\Gamma}
\def\de{\delta}

\def\De{\Delta}
\def\eps{\epsilon}
\def\veps{\varepsilon}
\def\la{\lambda}
\def\La{\Lambda}

\def\sg{\sigma}

\def\om{\omega}

\newcommand{\mc}[1]{\ensuremath{\mathcal{#1}}}
\newcommand{\mf}[1]{\ensuremath{\mathfrak{#1}}}
\newcommand{\msc}[1]{\ensuremath{\mathscr{#1}}}

\newcommand{\bs}[1]{\ensuremath{\boldsymbol{#1}}}

\newcommand{\ov}[1]{\ensuremath{\overline{#1}}}
\newcommand{\wt}[1]{\ensuremath{\widetilde{#1}}}
\newcommand{\wh}[1]{\ensuremath{\widehat{#1}}}

\newcommand{\Int}[2]{\ensuremath{\int\limits_{#1}^{#2}}}
\newcommand{\Oint}[2]{\ensuremath{\oint\limits_{#1}^{#2}}}

\newcommand{\sul}[2]{\ensuremath{\sum\limits_{#1}^{#2}}}
\newcommand{\pl}[2]{\ensuremath{\prod\limits_{#1}^{#2}}}

\newcommand{\R}{\ensuremath{\mathbb{R}}}
\newcommand{\Cx}{\ensuremath{\mathbb{C}}}

\newcommand{\Dp}[1]{\ensuremath{\partial_{#1}}}

\newcommand{\limit}[2]{\ensuremath{\underset{#1 \tend #2}{\longrightarrow} }}

\newcommand{\s}[1]{\ensuremath{\sinh(#1)}}
\newcommand{\sd}[1]{\ensuremath{\mathfrak{s}(#1)}}

\newcommand{\ex}[1]{\ensuremath{\e{e}^{#1}}}

\newcommand{\bra}[1]{ \langle \,#1\,|}
\newcommand{\ket}[1]{ |\,#1\, \rangle}

\newcommand{\sbra}[1]{( #1 \! \mid }
\newcommand{\sket}[1]{\mid \! #1  ) }

\newcommand{\braket}[2]{\ensuremath{ \langle #1 \big| \,  #2 \rangle  }}

\def\tr{\operatorname{tr}}

\newcommand{\ddet}[2]{\ensuremath{\det_{#1}\pac{#2}}}

\newcommand{\moy}[1]{\ensuremath{\langle #1 \rangle}}

\newcommand{\dd}{\mathrm{d}}
\newcommand{\e}[1]{\ensuremath{\mathrm{#1}}}

\newcommand{\intff}[2]{\ensuremath{\left [ \, #1 \,; #2 \, \right ] }}

\newcommand{\intof}[2]{\ensuremath{\left ] \, #1 \,; #2 \, \right ] }}
\newcommand{\intoo}[2]{\ensuremath{\left ] \, #1 \,; #2 \, \right [ }}

\newcommand{\intn}[2]{\ensuremath{[\![ \, #1 \,;\, #2 \,]\!]}}

\newcommand{\fa}{\mf{a}}
\newcommand{\vev}[1]{\left\langle #1 \right\rangle}

\def\qp{\wh{q}^{\,(+)}}
\def\qm{\wh{q}^{\,(-)}}

\usepackage{ifpdf}

\ifpdf
\usepackage{epstopdf}
\usepackage[pdftex,ps2pdf,dvips,colorlinks,urlcolor=blue,citecolor=blue,linkcolor=blue]{hyperref}
\else
\usepackage[hypertex,ps2pdf,dvips,colorlinks,urlcolor=blue,citecolor=blue,linkcolor=blue]{hyperref}
\fi
\begin{document}

\begin{flushright}

\end{flushright}
\par \vskip .1in \noindent

\vspace{14pt}

\begin{center}
\begin{LARGE}
{\bf Surface free energy of the open XXZ spin-$\tf{1}{2}$ chain.}
\end{LARGE}

\vspace{30pt}

\begin{large}

{\bf K.~K.~Kozlowski}\footnote[1]{Universit\'{e} de Bourgogne, Institut de Math\'{e}matiques de Bourgogne, UMR 5584 du CNRS, France,
karol.kozlowski@u-bourgogne.fr},~~ {\bf
B.~Pozsgay}\footnote[2]{University of Amsterdam, Institute
for Theoretical Physics, Amsterdam, The Netherlands, b.s.pozsgai@uva.nl}
\par

\end{large}

\vspace{40pt}

\centerline{\bf Abstract} \vspace{1cm}
\parbox{12cm}{\small

We study the boundary free energy of the XXZ spin-$\tf{1}{2}$ chain
subject to diagonal boundary fields.  
We first show that the representation for its finite Trotter number
approximant obtained by G\"{o}hmann, Bortz and Frahm  is 
related to the partition function of the six-vertex model with reflecting ends. Building on the Tsuchiya 
determinant representation for the latter quantity we are able to take the infinite
Trotter number limit. This yields a representation for the surface free energy which 
involves the solution of the non-linear integral equation
that governs the thermodynamics of the XXZ spin-$\tf{1}{2}$ 
chain subject to periodic boundary conditions. We show that this integral representation allows one 
to extract the low-$T$ asymptotic behavior of the boundary magnetization at finite external magnetic field
on the one hand and numerically plot this function on the other hand. }

\end{center}

\vspace{40pt}

\section*{Introduction\label{INT}}

Among various models arizing in physics, one-dimensional models seem to play a very specific role. 
On the one hand, the reduced dimensionality of the system allows one to use various approximation techniques
(correspondence with a Luttinger liquid \cite{HaldaneLuttingerLiquidCaracterofBASolvableModels} or a conformal field theory \cite{CardyConformalExponents}) to be able to provide predictions
for various quantities describing a given gapless quantum
Hamiltonian in certain limiting cases:
structure of the low-lying excitation, low-temperature behavior,
long-distance asymptotic behavior of the correlation functions, etc. 
On the other hand, as it was originally observed by H. Bethe \cite{BetheSolutionToXXX} on the example of the so-called XXX
chain, there exists a rather large class of one-dimensional models that have the property of quantum integrability. 
In other words, one is able to characterize the eigenvectors and eigenvalues of the associated Hamiltonians with the help 
of solutions to certain algebraic equations, the so-called Bethe equations. The latter provide a 
very effective description of the spectrum of the model when the large volume limit $L\tend +\infty$ is considered
(\textit{ie} when one deals with a model having a very large amount of pseudo-particles in its ground state). 
Such results allow one to identify the universality classes of various integrable 
models and, as such, check the predictions of the aforementioned approximation techniques, at least in some cases. 
For instance, using the Bethe Ansatz description of the spectrum, it was shown to be possible to access to the $\tf{1}{L}$
corrections to the ground and excited states just above it, hence allowing one to identify the central charge
and the various scaling dimensions \cite{KlumperBatchelorPearceCentralChargesfor6And19VertexModelsNLIE,
KlumperWehnerZittartzConformalSpectrumofXXZCritExp6Vertex}. Such types of results gave explicit predictions for the critical exponents 
governing the long-distance asymptotic behavior of two-point functions. The long-distance asymptotic behavior was then confirmed 
by extracting it directly from the exact representations for the correlators first at the free fermion points
\cite{McCoySomeAsymptoticsForXYCorrelators,TracyVaidyaRedDensityMatrixSpaceAsymptImpBosonsT=0,WuAsymptoticsSpinSpinIsingModel} 
and then for general interacting models
\cite{KozKitMailSlaTerXXZsgZsgZAsymptotics,KozMailletSlaLowTLimitNLSE}. 


The question of the thermodynamics of Bethe Ansatz solvable models was first addressed by Yang \& Yang 
\cite{Yang-YangNLSEThermodynamics}. These authors derived a non-linear integral equation 
whose solution allows one to compute the free energy of the so-called non-linear Schr\"{o}dinger
model \cite{LiebLinigerCBAForDeltaBoseGas}
at finite temperature $T$ and in the presence of an external chemical potential. 
The reasoning of Yang \& Yang was raised to the level of a theorem by Dorlas, Lewis and Pul\'{e} 
\cite{DorlasLewisPuleRigorousProofYangYangThermoEqnNLSE} with the help of large deviation techniques. 
The approach of Yang \& Yang has been generalized to the study of the thermodynamics 
of the XXZ spin-$\tf{1}{2}$ chain simultaneously and independently by Gaudin \cite{GaudinTBAXXZMassiveInfiniteSetNLIE} and Takahashi
\cite{TakahashiTBAforXXZFiniteTinfiniteNbrNLIE}. Their approach built on the so-called string conjecture
\cite{BetheSolutionToXXX} and allowed them to characterize the thermodynamics of the model in terms of a solution to an infinite hierarchy 
of non-linear integral equations. Since then, it has been applied with success to many other models and bears the name of the 
thermodynamic Bethe Ansatz.

Another path to the study of thermodynamics of spin chains has been proposed by Koma 
\cite{KomaIntroductionQTM6VertexForThermodynamicsOfXXX,KomaIntroductionQTM6VertexForThermodynamicsOfXXZ}. Building on the 
method for mapping quantum Hamiltonians in D-dimensions into models 
of classical statistical physics in D+1 dimensions proposed by Suzuki \cite{SuzukiCorrespondence(D+1)StatPhysDQuantumHamiltonians},
he argued that the computation of the partition function of the XXX and XXZ models in a magnetic field 
is equivalent to obtaining the largest eigenvalue of the transfer matrix associated with a specific inhomogeneous six-vertex model, the so-called quantum transfer matrix (QTM).
Although Koma could not provide at the time a proper analytic framework for taking the so-called infinite Trotter number limit,
he was able to carry out a numerical analysis along with an extrapolation to infinite Trotter numbers.
Then Takahashi refined Koma's approach and was able to take the infinite Trotter number limit analytically. 
This led to a description of the thermodynamics of the XYZ and XXZ spin-$\tf{1}{2}$ chains
in terms of an infinite sequence of numbers that ought to be fixed numerically 
\cite{TakahashiThermoXXZInfiniteNbrRootsFromQTM,TakahashiThermoXYZInfiniteNbrRootsFromQTM}. 
Soon after, Kl\"{u}mper \cite{KlumperNLIEfromQTMDescrThermoXYZOneUnknownFcton} 
proposed an important simplification to the QTM-based approach. 
Namely, building on the method of non-linear integral equations \cite{BatchelorKlumperFirstIntoNLIEForFiniteSizeCorrectionSpin1XXZAlternativeToRootDensityMethod}, he proposed a 
way for sending the Trotter number $N$ to infinity in the Bethe equations describing the 
largest eigenvalue of the QTM. He obtained the full description of the thermodynamics of the XYZ 
(and XXZ) spin-$\tf{1}{2}$ chains in terms of a \textit{single} unknown function 
that satisfies a \textit{single} non-linear integral equation. 
This function allows one to compute the free energy at \textit{any} finite temperature $T$.
Kl\"{u}mper also proposed non-linear integral equations 
describing the sub-leading eigenvalues of the QTM what gave access to the correlation lengths. 
His results confirmed the conformal field theory-based predictions \cite{AffleckCFTPreForLargeSizeCorrPartitionFctonAndLowTBehavior} 
for the low-T behavior of the free energy.

All of the above results were obtained in the case of models subject to periodic boundary conditions. 
It so happens that the situation is definitely much less understood in the case of integrable models 
subject to other types of boundary conditions. 
Models such as the XXZ spin-$\tf{1}{2}$ chain subject to diagonal boundary fields
\beq
J^{-1} \cdot \bs{\mc{H}}= \sum_{m=1}^{M-1} \Big\{ \sigma^x_m \,\sigma^x_{m+1} +
  \sigma^y_m\,\sigma^y_{m+1} + \cosh \eta \,(\sigma^z_m\,\sigma^z_{m+1}+1)\Big\}
  + \sinh \eta \coth\xi_- \, \sg^z_1 + \sinh \eta \coth\xi_+ \,\sg^z_M \; + \f{\cosh\pa{2\eta}}{\cosh\pa{\eta}}  
\label{ecriture Hamiltonien XXZ ac bord}
\enq
 have been solved through the 
coordinated Bethe Ansatz \cite{AlcarazBatchelorBaxterQuispelCBAopenXXZ} and later by the algebraic Bethe Ansatz
\cite{SklyaninABAopenmodels}. Above, $\sg_p^{x}$, $\sg_p^{y}$ and $\sg_p^{z}$ are Pauli matrices acting on the
Hilbert space $V_p \simeq \Cx^2$  attached to the $p^{\e{th}}$
site of the chain.  $\eta$ and $\xi_{\pm}$ are parameters characterizing the anisotropy 
and the boundary fields and $J$ is an overall coupling constant fixing the energy scale. 
The analysis of the Bethe equations arizing in the diagonalization of the Hamiltonian \eqref{ecriture Hamiltonien XXZ ac bord} 
 showed that the boundaries produce additional $\e{O}(1)$ contributions (\textit{ie} ones that do not scale with the volume $M$) to the 
 energies of the ground and excited states in respect to the values obtained for periodic boundary conditions 
 \cite{AlcarazBatchelorBaxterQuispelCBAopenXXZ}. 

When turning the temperature on, one expects that changing the boundary conditions of the XXZ spin-$\tf{1}{2}$ chain will not alter 
the leading part of the  model's free energy, \textit{ie} the one scaling with the volume $M$. 
However, there will arize additional, volume-independent, contributions 
to the free energy. These constitute the so-called surface free energy. 
Various approximation techniques have been developed to 
characterize the surface free energy of the open XXZ spin-$\tf{1}{2}$
chain (and also of more general, not necessarily exactly solvable models). 
 Field theory based method led to predictions for the leading behavior 
at low-temperatures of the surface free energy in the 
vanishing external magnetic field limit
\cite{Affleck-XXZ-boundary-QFT,Bortz:JPhysAMathGen38:2005,Sirker:JStatMech0601P01007:2006,Fujimoto:PhysRevLett92:2004,FurusakiHikihara}.

The estimation of the surface free energy has also been considered in the framework of the
thermodynamic Bethe Ansatz, what allowed one to make predictions in the full range of temperatures. 
Such calculations were first performed in the papers
\cite{TsvelikOpenXXZandKondoimpurityeffect,Frahm-Zvyagin,Zvyagin-Makarova}.
However, their results did not agree with the field-theory based results in the low-temperature limit.
The origin of this discrepancy was elucidated in the papers
\cite{sajat-g,woynarovich,woynarovich-uj} which re-considered 
the derivation of free energy of Bethe Ansatz solvable models, both in the
periodic and the open boundary case.
It was shown in \cite{sajat-g,woynarovich-uj} that a careful
derivation of the partition function leads 
to additional contributions, which are expressed in terms of 
Fredholm-determinants. These results
were in agreement with the series of multiple integrals found previously in
\cite{Dorey:2004xk} in the case of integrable relativistic quantum
field theories. The methods of \cite{sajat-g,woynarovich-uj} have not
yet been applied to the XXZ spin chain, but it is expected that
analogous additional contributions are present in the case of
the open spin chain as well, on top of the
results given in \cite{TsvelikOpenXXZandKondoimpurityeffect,Frahm-Zvyagin,Zvyagin-Makarova}.

However, all the aforementioned approaches to the calculation of the surface free energy 
where neither direct nor exact. 
A approach to obtain exact results for the surface free energy
was initiated by G\"{o}hmann, Bortz and Frahm in \cite{BortzFrahmGohmannSurfaceFreeEnergy}.  
The main result of that paper was a rigorous\footnote{modulo the strongly supported conjecture 
on the non-degeneracy of the quantum transfer matrix's largest eigenvalue and interchangeability 
of the Trotter and infinite volume limits} representation for a finite Trotter number 
approximant of the boundary free energy. The latter was expressed as
an expectation value of a large number of local 
operators forming the so-called finite temperature boundary
operator. This expectation value was to be  
computed  in respect to the eigenvector associated with the dominant eigenvalue of the
quantum transfer matrix arizing in the description of the thermodynamics of the periodic $XXZ$ spin-$\tf{1}{2}$
chain. 
However, the representation given in
\cite{BortzFrahmGohmannSurfaceFreeEnergy} is rather implicit and it
was not clear how to evaluate it at finite Trotter numbers, or how to
take the infinite Trotter number limit.

The present paper is devoted to overcoming these
difficulties. Starting from G\"{o}hmann et \textit{al}'s result  
we show that one can, in fact, interpret the correlation function involving the finite temperature boundary operator 
as a specific case of the partition function of the six-vertex model with reflecting ends. 
Building on the techniques developed by Izergin \cite{IzerginPartitionfunction6vertexDomainWall}
for the partition function of the six-vertex model with domain wall boundary conditions,
Tsuchiya \cite{TsuchiyaPartitionFunctWithReflecEnd} showed that the 
former also admits a finite-size determinant representation. One cannot take 
the Trotter limit immediately on the level of Tsuchiya's determinant. We thus 
recast it in a form where the Trotter limit can be preformed easily. 
This yields the main result of this paper, namely
an exact representation for the surface free energy of the
spin-$\tf{1}{2}$ XXZ chain subject to diagonal boundary fields.  
Using our exact representation, we obtain a simple integral representation for the boundary
magnetization. This allows us to obtain the first terms of the low-temperature asymptotic behavior 
of the boundary magnetization. In particular, we show that we recover the results of 
\cite{JimboKedemKonnoMiwaXXZChainWithaBoundaryElemBlcks,KozKitMailNicSlaTerElemntaryblocksopenXXZ}
for the $T=0$ case. We also use our representation to carry out numerical plots of this quantity in the massive regime
of the model. 


This paper is organized as follows. 
In section \ref{Section Preliminary definitions}, we briefly review the setting of the 
algebraic Bethe Ansatz framework for integrable models subject to the so-called diagonal boundary conditions. 
This allows us to set the quantum transfer matrix-based approach to the thermodynamics of the model
and then review the representation for the finite Trotter number approximant of the 
surface free energy obtained by  G\"{o}hmann, Bortz and Frahm. 
In section \ref{Section New representation boundary free energy}, building on 
a factorization of G\"{o}hmann et \textit{al}'s formula, we recast the aforementioned 
 quantity in terms of Tsuchiya determinants. 
In section \ref{Section Taking infinite Trotter number}, we carry out several transformations 
on the formula obtained in section  \ref{Section New representation boundary free energy}, 
what ultimately allows us to take the infinite Trotter number limit. 
 Finally, in section \ref{Section Boundary magnetization}, 
we use our exact representation for the surface free energy so as to obtain an integral representation 
for the boundary magnetization at finite temperature. 
This integral representation allows us to extract the low-temperature behavior of the boundary magnetization in the gapless phase 
as well as to plot it, on the basis of numerical calculations, versus various parameters of the model 
such as the boundary magnetic field or the anisotropy.

\section{Preliminary definitions and the algebraic Bethe Ansatz framework}
\label{Section Preliminary definitions}
\subsection{Definition of the surface free energy}

One can diagonalize the Hamiltonian $\bs{\mc{H}}$  \eqref{ecriture Hamiltonien XXZ ac bord}
within the algebraic Bethe Ansatz approach. For this purpose, 
one builds a one-parameter commutative family of operators, the so-called boundary transfer matrix \cite{SklyaninABAopenmodels}.
The latter is expressed as a trace over a two-dimensional auxiliary space $V_a$ labelled by a roman index $a$
of a product of bulk monodromy matrices $T_a(\la)$, $\wh{T}_a(\la)$ and scalar solutions $K^{\pm}_a(\la)$
to the so-called reflection equations \cite{CherednikReflectionEquationFactorisabilityOfScattering}:
\beq
\bs{\tau}(\la)  \; = \;  \e{tr}_a\Big[  K^+_a(\la) T_a(\la) K^-_a(\la) \wh{T}_a(\la) \Big]   \;. 
\label{ecriture boundary transfer matrix}
\enq
The bulk monodromy matrices are $2\times 2$ matrices in the auxiliary space whose entries are operators acting on the quantm 
space $\mf{h}=\otimes_{p=1}^{M}V_p$ of the chain. Thus, $\bs{\tau}(\la)$ is also an operator on $\mf{h}=\otimes_{p=1}^{M}V_p$.
The bulk monodrmoy matrices associated with the XXZ chain are built as ordered products of the 
$6$-vertex type $R$-matrix\footnote{which, for further convenience, we chose to write in its polynomial normalization}
\beq
R(\la) = \pa{ \ba{cccc}  \s{\la+\eta} &  0 &0 &0 \\
							0 & \s{\la}& \s{\eta} & 0 \\
								0 & \s{\eta}& \s{\la} & 0 \\
							0 & 0 & 0 & \s{\la+\eta} \ea  } \;.
\enq
This $R$-matrix can be seen as the matrix representation of an operator acting on a tensor product of two dimensional vector 
spaces $\Cx^2\otimes \Cx^2$. In the following, the subscripts $a,b$ in $R_{ab}(\la)$ mean that it acts non-trivially 
(\textit{ie} not as the identity) solely on the tensor product $V_a \otimes V_b$, where $V_a \simeq V_b \simeq \Cx^2$. 
Having introduced enough notations, we are now in position to write the explicit realisation for $T_a(\la)$ and $\wh{T}_a(\la)$:
\beq
T_a\pa{\la} = R_{aM}\pa{\la-\xi_M}\dots R_{a1}\pa{\la-\xi_1} \qquad \e{and} \qquad 
\wh{T}_a\pa{\la} = R_{1a}\pa{\la+\xi_1} \dots R_{Ma}\pa{\la+\xi_M} \;. 
\enq
There $\xi_k$ represent inhomogeneity parameters. We do stress that the roman index $a$ refers 
to an auxiliary two-dimensional space whereas the indices $1,\dots,M$ refer
to the various quantum spaces $V_1,\dots, V_M$ associated with the sites of the chain. 
As we have already specified, when these indices occur in a matrix, they label the spaces (auxiliary and quantum) where the
latter acts non-trivially.

Lastly, the expression for $\bs{\tau}(\la)$ also involves the diagonal solutions of the reflection equations that have been first found by Cherednik \cite{CherednikReflectionEquationFactorisabilityOfScattering}
\beq
K^{\pm}_a\pa{\la} = K_a\pa{\la +\tf{\eta}{2} \pm \tf{\eta}{2} ; \xi_{\pm}  } \qquad \e{with} \quad 
K_a\pa{\la;\xi} = \pa{  \ba{cc} \s{\la+\xi} & 0 \\ 0 & \s{\xi-\la} \ea
}_{\pac{a}} \;.
\label{Kdef}
\enq
 These $K$-matrices can be checked to satisfy 
\beq
\e{tr}_a\big[ K^+_a(0) \big] = 2 \s{\xi_+} \cosh(\eta) \qquad \e{and}  \qquad 
\e{tr}_a\big[ K^-_a(0) \big] = 2 \s{\xi_-}  \;.
\label{ecriture traces matrices K}
\enq
%
%
The $R$-matrix given above satisfies a certain amount of properties. For instance, when the spectral parameter $\la$
is set to zero, they reduce to permutation operators 
\beq
R_{ab}(0) \; = \; \s{\eta} \, \mc{P}_{ab} \;,  
\label{ecriture reduction R vers permutation}
\enq
where $\mc{P}_{ab}$ is the permutation operator on $V_a\otimes V_b$. They also fullfill a crossing relation 
\beq
\sg_1^y \, R_{12}^{t_1}(\la-\eta) \, \sg_1^y = - R_{21}(-\la) \;. 
\enq
Above, the superscript $t_1$ in the $R$-matrix refers to the transposition on the first space. 
The repeated use of the crossing relation results in the alternative representation for the boundary transfer matrix,
\beq
\bs{\tau}\pa{\la} = (-1)^M\e{tr}_a\pac{  K^+_a\pa{\la} T_a\pa{\la}
  K^-_a\pa{\la} \sigma_a^y \, T_a^{t_a}\pa{-\la-\eta} \sigma_a^y }   \;. 
\label{crossed transfer matrix}
\enq
Further, bulding on the identities \eqref{ecriture reduction R vers permutation}-\eqref{ecriture traces matrices K},
one can show \cite{CherednikReflectionEquationFactorisabilityOfScattering} that, 
in the homogeneous limit ($\xi_k=0$, $k=1,\dots,M$), $\bs{\tau}\pa{\la}$ enjoys  the properties 
\beq
\bs{\tau}\pa{0}  = \f{ \e{tr}_a\pac{ K^+_a\pa{0} } \e{tr}_a\pac{ K^-_a\pa{0} } }{2} \pac{ \sinh(\eta) }^{2M}  \bs{id}
\qquad \e{and} \qquad 
\bs{\mc{H}} = \f{J \sinh \eta }{ \bs{\tau}\pa{ 0 } } \f{\dd }{\dd \la} \bs{\tau} \pa{\la} _{\mid_{\la=0}} \;. 
\enq
There $\bs{id}$ stands for the identity operator on $\mf{h}$ and $\bs{\mc{H}} $ 
is given by \eqref{ecriture Hamiltonien XXZ ac bord}. 
As a consequence, 
\beq
\pac{ \bs{\tau}\pa{-\tf{\be}{N} } \cdot \bs{\tau}^{-1}\!\pa{0}  }^N = \ex{ - \f{ \bs{\mc{H}} }{T} } \cdot 
\pa{1 + \e{O}\big( N^{-1} \big) } \qquad \e{with} \qquad \be = \f{ J\sinh(\eta) }{ T }.
\enq
The above estimates allow one to write a Trotter limit-based representation for the partition function 
of the open XXZ spin-$\tf{1}{2}$ chain in a uniform external magnetic field:
\beq
Z_M \; = \; \e{tr}_{1,\dots,M} \Big[ \ex{-\f{ \bs{\mc{H}} }{T} } \Big] \qquad \e{with} \qquad 
							\bs{\mc{H}}_h \; = \; \bs{\mc{H}} - \f{h}{2T} \sul{k=1}{N}\sg_k^z \;. 
\enq
This representation takes the form 
\beq
Z_M =  \lim_{N\tend +\infty}  \e{tr}_{1,\dots,M} \bigg\{ 
	\Big[ \bs{\tau}\pa{-\tf{\be}{N} } \cdot \bs{\tau}^{-1} \!\pa{0}  \Big]^N   \pl{a=1}{M} \ex{ \f{h}{2T}\sg_a^z } \bigg\} \; . 
\label{ecriture fction partition comme trace}
\enq

\begin{figure}[t]
\centering
\begin{pgfpicture}{0cm}{-0.2cm}{14cm}{2.2cm}
\pgfsetendarrow{\pgfarrowto}

\pgfline{\pgfxy(0.5,1)}{\pgfxy(1.5,1)}  
\pgfline{\pgfxy(1,0.5)}{\pgfxy(1,1.5)}

\pgfputat{\pgfxy(1.8,1)}{\pgfbox[center,center]{$+$}}
\pgfputat{\pgfxy(1,0.2)}{\pgfbox[center,center]{$+$}}
\pgfputat{\pgfxy(1,1.8)}{\pgfbox[center,center]{$+$}}
\pgfputat{\pgfxy(0.2,1)}{\pgfbox[center,center]{$+$}}

\pgfputat{\pgfxy(3.3,1)}{\pgfbox[center,center]{$\equiv \f{ \sinh(u-v+\eta) }{ \sinh(\eta) }$}}

\pgfline{\pgfxy(5.6,1)}{\pgfxy(6.6,1)}  
\pgfline{\pgfxy(6.1,0.5)}{\pgfxy(6.1,1.5)}

\pgfputat{\pgfxy(6.9,1)}{\pgfbox[center,center]{$+$}}
\pgfputat{\pgfxy(6.1,0.2)}{\pgfbox[center,center]{$-$}}
\pgfputat{\pgfxy(6.1,1.8)}{\pgfbox[center,center]{$-$}}
\pgfputat{\pgfxy(5.3,1)}{\pgfbox[center,center]{$+$}}

\pgfputat{\pgfxy(7.4,1)}{\pgfbox[left,center]{$\equiv \f{ \sinh(u-v) }{ \sinh(\eta)} $}}

\pgfline{\pgfxy(10.5,1)}{\pgfxy(11.5,1)}  
\pgfline{\pgfxy(11,0.5)}{\pgfxy(11,1.5)}

\pgfputat{\pgfxy(11.8,1)}{\pgfbox[center,center]{$-$}}
\pgfputat{\pgfxy(11,0.2)}{\pgfbox[center,center]{$-$}}
\pgfputat{\pgfxy(11,1.8)}{\pgfbox[center,center]{$+$}}
\pgfputat{\pgfxy(10.2,1)}{\pgfbox[center,center]{$+$}}

\pgfputat{\pgfxy(12.3,1)}{\pgfbox[left,center]{$ \equiv 1$}}
\end{pgfpicture}
\caption{The weights of the 6-vertex model. Here $u$ is attached to
  the horizontal line and $v$ to the vertical one.}
\label{fig:6}
\end{figure}

\begin{figure}[t]
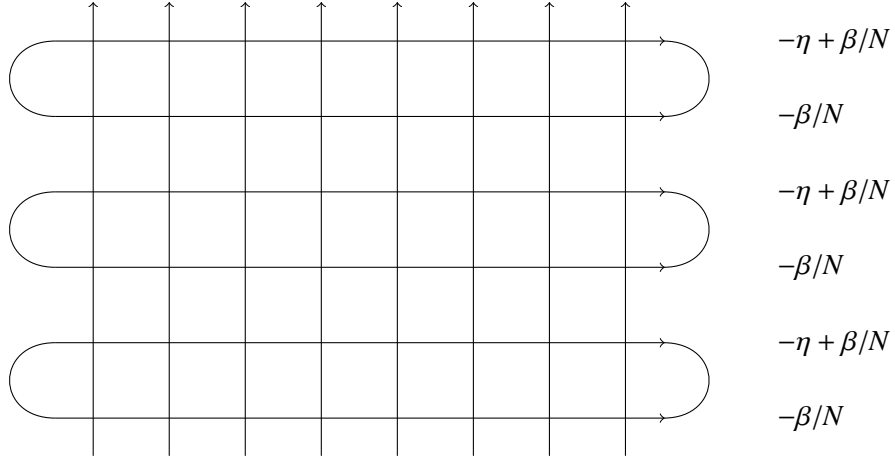

\centering
\begin{pgfpicture}{0cm}{0cm}{12cm}{7cm}

\pgfsetendarrow{\pgfarrowto}

\pgfline{\pgfxy(1.5,1)}{\pgfxy(9.5,1)}
\pgfline{\pgfxy(1.5,2)}{\pgfxy(9.5,2)}
\pgfline{\pgfxy(1.5,3)}{\pgfxy(9.5,3)}
\pgfline{\pgfxy(1.5,4)}{\pgfxy(9.5,4)}
\pgfline{\pgfxy(1.5,5)}{\pgfxy(9.5,5)}
\pgfline{\pgfxy(1.5,6)}{\pgfxy(9.5,6)}

\pgfputat{\pgfxy(11,1)}{\pgfbox[left,center]{$-\beta/N$}}
\pgfputat{\pgfxy(11,3)}{\pgfbox[left,center]{$-\beta/N$}}
\pgfputat{\pgfxy(11,5)}{\pgfbox[left,center]{$-\beta/N$}}
\pgfputat{\pgfxy(11,2)}{\pgfbox[left,center]{$-\eta+\beta/N$}}
\pgfputat{\pgfxy(11,4)}{\pgfbox[left,center]{$-\eta+\beta/N$}}
\pgfputat{\pgfxy(11,6)}{\pgfbox[left,center]{$-\eta+\beta/N$}}

\pgfline{\pgfxy(2,0.5)}{\pgfxy(2,6.5)}
\pgfline{\pgfxy(3,0.5)}{\pgfxy(3,6.5)}
\pgfline{\pgfxy(4,0.5)}{\pgfxy(4,6.5)}
\pgfline{\pgfxy(5,0.5)}{\pgfxy(5,6.5)}
\pgfline{\pgfxy(6,0.5)}{\pgfxy(6,6.5)}
\pgfline{\pgfxy(7,0.5)}{\pgfxy(7,6.5)}
\pgfline{\pgfxy(8,0.5)}{\pgfxy(8,6.5)}
\pgfline{\pgfxy(9,0.5)}{\pgfxy(9,6.5)}

\pgfclearendarrow

\pgfmoveto{\pgfxy(9.5,5)}
\pgfcurveto{\pgfxy(10.3,5)}{\pgfxy(10.3,6)}{\pgfxy(9.5,6)}
\pgfstroke
\pgfmoveto{\pgfxy(9.5,3)}
\pgfcurveto{\pgfxy(10.3,3)}{\pgfxy(10.3,4)}{\pgfxy(9.5,4)}
\pgfstroke
\pgfmoveto{\pgfxy(9.5,1)}
\pgfcurveto{\pgfxy(10.3,1)}{\pgfxy(10.3,2)}{\pgfxy(9.5,2)}
\pgfstroke

\pgfmoveto{\pgfxy(1.5,5)}
\pgfcurveto{\pgfxy(0.7,5)}{\pgfxy(0.7,6)}{\pgfxy(1.5,6)}
\pgfstroke
\pgfmoveto{\pgfxy(1.5,3)}
\pgfcurveto{\pgfxy(0.7,3)}{\pgfxy(0.7,4)}{\pgfxy(1.5,4)}
\pgfstroke
\pgfmoveto{\pgfxy(1.5,1)}
\pgfcurveto{\pgfxy(0.7,1)}{\pgfxy(0.7,2)}{\pgfxy(1.5,2)}
\pgfstroke

\end{pgfpicture}
\caption{Graphical representation of the partition function
  \eqref{ecriture fction partition comme trace} at zero overall
  magnetic field. The $M$ vertical lines
  correspond to the quantum spaces of the spins, whereas the
$2N$  horizontal lines joined by the curly end-curves correspond to the auxiliary space
  occurring in the definition \eqref{ecriture boundary transfer matrix}
  for the boundary transfer matrix.  
  The bulk weights are represented on Fig.~\ref{fig:6} whereas the boundary ones are represented on Fig.~\ref{Kfig}. 
  Periodic boundary conditions are imposed in the  vertical direction. }
\label{startingpoint}
\end{figure}
\begin{figure}[t]
\centering
  \begin{pgfpicture}{-1cm}{0cm}{17cm}{2cm}
    \pgfmoveto{\pgfxy(1.5,0)}
\pgfcurveto{\pgfxy(0.7,0)}{\pgfxy(0.7,1)}{\pgfxy(1.5,1)}
\pgfstroke

\pgfputat{\pgfxy(1.8,1)}{\pgfbox[left,center]{$+$}}
\pgfputat{\pgfxy(1.8,0)}{\pgfbox[left,center]{$-$}}

\pgfputat{\pgfxy(2.5,0.5)}{\pgfbox[left,center]
{$\displaystyle = \frac{\sinh(-\beta/N+\xi_-)}{\sqrt{2}\sinh \xi_-}$}}

   \pgfmoveto{\pgfxy(9.5,0)}
\pgfcurveto{\pgfxy(8.7,0)}{\pgfxy(8.7,1)}{\pgfxy(9.5,1)}
\pgfstroke

\pgfputat{\pgfxy(9.8,1)}{\pgfbox[left,center]{$-$}}
\pgfputat{\pgfxy(9.8,0)}{\pgfbox[left,center]{$+$}}

\pgfputat{\pgfxy(10.5,0.5)}{\pgfbox[left,center]
{$\displaystyle= \frac{\sinh(\beta/N+\xi_-)}{\sqrt{2}\sinh \xi_-}  $}}
  \end{pgfpicture}

  \begin{pgfpicture}{-1cm}{0cm}{17cm}{2cm}
    \pgfmoveto{\pgfxy(1.5,0)}
\pgfcurveto{\pgfxy(2.3,0)}{\pgfxy(2.3,1)}{\pgfxy(1.5,1)}
\pgfstroke

\pgfputat{\pgfxy(0.8,1)}{\pgfbox[left,center]{$+$}}
\pgfputat{\pgfxy(0.8,0)}{\pgfbox[left,center]{$-$}}

\pgfputat{\pgfxy(2.5,0.5)}{\pgfbox[left,center]
{$\displaystyle= \frac{\sinh(-\beta/N+\eta+\xi_+)}{\sqrt{2}\sinh\xi_+\cosh(\eta)}  $}}

    \pgfmoveto{\pgfxy(9.5,0)}
\pgfcurveto{\pgfxy(10.3,0)}{\pgfxy(10.3,1)}{\pgfxy(9.5,1)}
\pgfstroke

\pgfputat{\pgfxy(8.8,1)}{\pgfbox[left,center]{$-$}}
\pgfputat{\pgfxy(8.8,0)}{\pgfbox[left,center]{$+$}}

\pgfputat{\pgfxy(10.5,0.5)}{\pgfbox[left,center]
{$\displaystyle=\frac{\sinh(\beta/N-\eta+\xi_+)}{\sqrt{2}\sinh\xi_+\cosh(\eta)} $}}
  \end{pgfpicture}
\caption{The boundary weights of the partition function depicted in
  Fig. \ref{startingpoint}. 
}
\label{Kfig}
\end{figure}

We stress that the subscript $1,\dots, M$ is there to indicate that the traces in \eqref{ecriture fction partition comme trace}
are taken over all the quantum spaces $V_p$, \textit{ie} over the whole quantum space $\mf{h}$ of the chain. 
Equation \eqref{ecriture fction partition comme trace} admits a graphical representation given in 
Fig. \ref{startingpoint}. There, the summation over all possible $\pm$-values of the intermediate lines
in undercurrent and periodic boundary conditions are imposed in the vertical direction. 
The expression for the bulk (resp. boundary) weights is depicted
in Fig. \ref{fig:6}  (resp. Fig. \ref{Kfig}).

After some manipulations, as shown in \cite{BortzFrahmGohmannSurfaceFreeEnergy}, one  is able to recast 
\eqref{ecriture fction partition comme trace} as 
\beq
Z_M =  \lim_{N\tend +\infty} \Bigg\{  
\bigg( \f{ 2 \pac{  \s{\eta} }^{-2M} }{ \e{tr}_a\pac{ K^+_a\pa{0} } \e{tr}_a\pac{ K_a^-\pa{0} }  } \bigg)^N
\e{tr}_{a_1,\dots,a_{2N}} \bigg[  P_{a_1 a_2}\pa{ \tf{-\be}{N} } \dots P_{a_{2N-1} a_{2N} } \pa{ \tf{-\be}{N} } 
 \pl{p=1}{M} \bs{t}^{\mf{q}}(\xi_p)   \bigg]_{\mid \xi_k=0}   \Bigg\}\;.
\label{equation fonction partition volume fini}
\enq
 Note that $\bs{t}^{\bs{q}}(\la) = \e{tr}\big[ T_k^{\mf{q}}(\la)\big]$ 
stands for the quantum transfer matrix associated with 
the quantum monodromy matrix arizing in the study of the XXZ spin-$\tf{1}{2}$  periodic chain: 
\beq
T^{\mf{q}}_k\!\pa{\xi} =  R^{t_{a_{2N}}}_{a_{2N} k }\!\pa{ - \xi -  \tf{\be}{N} } R_{k a_{2N-1}}\!\pa{\xi - \tf{\be}{N} } \dots 
R_{a_{2} k}^{t_{a_2} }\!\pa{-\xi - \tf{\be}{N} } R_{k a_{1}}\!\pa{\xi - \tf{\be}{N} } \ex{\f{h}{2T}  \sg^z_k } 
 \; = \; \pa{\ba{cc} A(\la) & B(\la) \\ C(\la) & D(\la) \ea  }_{ [k] }\;. 
\enq
The quantum transfer matrix $\bs{t}^{\bs{q}}(\la)$ acts on the tensor product of $2N$ two-dimensional 
spaces $V_{a_1}\otimes \dots \otimes V_{a_{2N}}$. The alternative expresssion \eqref{equation fonction partition volume fini}
for the partition functions also involves the one-dimensional projector
\beq
P_{\! ab}\pa{\la} = K_a^+\!\pa{\la} \mc{P}_{\!\! ab}^{t_a}  \,  K_a^{-}\!\pa{\la} \;.
\label{definition projecteur}
\enq
We remind that $\mc{P}_{\! ab}$ is the permutation operator in $V_a \otimes V_b$. 

It has been argued on the basis of thorough numerical computations, small $\be$ analysis 
and free fermion point calculations \cite{EsslerFrahmGohmanKlumperKorepinOneDimensionalHubbardModel,GohmannPrivateCommJustOfQTMMethod}
that the eigenspace associated with its highest eigenvalue of the QTM for finite $N$ is one 
dimensional. We shall build on this fact and denote by $\ket{\Psi_0}$ an associated eigenvector.
We denote by $\La_0$ the so-called dominant eigenvalue associated to $\ket{\Psi_0}$. 
Then, 
\beq
Z_M = \lim_{N \tend +\infty} \Bigg\{  \paf{ \La_0 }{ \sinh^{2N}(\eta) }^M  
\cdot  \paf{ 2 }{ \e{tr}_a\pac{ K^+_a\pa{0} } \e{tr}_0\pac{ K^-_a\pa{0} }  }^N  
\cdot \f{ \bra{\Psi_0} P_{a_1 a_2}\pa{ \tf{-\be}{N} } \dots P_{a_{2N-1} a_{2N} } \pa{ \tf{-\be}{N} } \ket{\Psi_{0}} }{  \braket{\Psi_0}{\Psi_0} } 
+ \dots \; \Bigg\} .
\label{equation calcul trace dans ZM}
\enq
Building on the rigorous approach to thermodynamic limits of partition function 
(see e.g. the monograph of Ruelle \cite{RuelleRigorousResultsForStatisticalMechanics}),
it has been argued in  \cite{KomaIntroductionQTM6VertexForThermodynamicsOfXXZ,SuzukiArgumentsForInterchangeabilityTrotterAndVolumeLimitInPartFcton} 
that, when computing the thermodynamic ($M\tend +\infty$) limit of $Z_M$, one can exchange the
order of the  $N \tend + \infty$ limit with the infinite volume $M\tend +\infty$ one. 
A consequence of such a fact would be that, when computing 
$ \lim_{M\tend +\infty} \big\{\tf{ (\ln Z_M) }{M} \big\}$, one can simply drop out all the terms in 
\eqref{equation calcul trace dans ZM} that have been included in the $\dots$ symbol. Indeed, 
for a finite Trotter number, all these terms will only produce corrections that are, for fixed $N$, 
exponentially small in $M$ and thus vanish when $M\tend +\infty$ and $N$ is fixed. 
We shall not investigate this question further and simply assume that the exchangeability of limits
does hold.

The surface free energy is then defined as the limit  of the difference
between the free energy of the model subject to periodic and the one subject to open boundary conditions
\beq
 \ex{- \f{ f_{\e{surf}} }{ T } }  \equiv \lim_{M\tend +\infty }  \bigg\{  Z_M \cdot \bigg( \f{ \sinh^{2N}(\eta) }{ \La_0 } \bigg)^M  \bigg\} \;.
\enq
In virtue of the previous arguing, the surface free energy for the open spin-$\tf{1}{2}$ XXZ chain 
is given by the below Trotter limit: 
\beq
 \ex{- \f{ f_{\e{surf}} }{ T } }  \; =  \;  \lim_{N \tend +\infty} \Big\{  \ex{- \f{ f_{\e{surf}}^{(N)} }{ T } }  \Big\} 
\qquad \e{where} \qquad 
\ex{- \f{ f_{\e{surf}}^{(N)} }{ T } }   = 
\f{ \bra{\Psi_0} P_{a_1 a_2}\pa{ \tf{-\be}{N} } \dots P_{a_{2N-1} a_{2N} } \pa{ \tf{-\be}{N} } \ket{\Psi_{0}} }{  \braket{\Psi_0}{\Psi_0} 
\cdot \Big\{ \e{tr}_a\pac{ K^+_a\pa{0} } \e{tr}_a\pac{ K^-_a\pa{0} } /2 \Big\}^N   }   
\label{ecriture energie surface free limit trotter}
\enq
The representation for $\ex{- \f{ f_{\e{surf}}^{(N)} }{ T } } $ as the expectation value
\eqref{ecriture energie surface free limit trotter} constitutes the main result obtained in reference
\cite{BortzFrahmGohmannSurfaceFreeEnergy}. 
There, it was also observed that \eqref{definition projecteur} is indeed a one-dimensional projector. 
This can be easily seen as soon as one observes that the transpose of the permutation matrix 
is as a one-dimensional projector 
\beq
\mc{P}_{\! ab}^{t_a}  =  \pac{ \ket{+}_a\ket{+}_b +  \ket{-}_a\ket{-}_b }  \cdot  \pac{  \bra{+}_a\bra{+}_b   +    \bra{-}_a\bra{-}_b    }
\label{ecriture representation projecteur transpose}
\enq
where $\ket{\pm}_a$ is the canonical spin up/down basis in $V_a$. 
This structure allows one to factorize the  representation for the surface free energy at finite Trotter number. 
Indeed, setting 
\beqa
\ket{ v } &=& \pa{ \ket{+}_{a_1}\ket{+}_{a_2} +  \ket{-}_{a_1}\ket{-}_{a_2} } \otimes \dots \otimes 
\pa{ \ket{+}_{a_{2N-1}}\ket{+}_{a_{2N}} +  \ket{-}_{a_{2N-1}}\ket{-}_{a_{2N} } }   \\
\bra{ v } &=& \pa{ \bra{+}_{a_1}\bra{+}_{a_2} +  \bra{-}_{a_1}\bra{-}_{a_2} } \otimes \dots \otimes 
\pa{ \bra{+}_{a_{2N-1}}\bra{+}_{a_{2N}} +  \bra{-}_{a_{2N-1}}\bra{-}_{a_{2N} } }
\eeqa
one recasts $\ex{- \f{ f_{\e{surf}}^{(N)} }{ T } }$ as 
\bem
\ex{- \f{ f_{\e{surf}}^{(N)} }{ T } }  = 
 \paf{ 2 }{ \e{tr}_0\pac{ K_+\pa{0} } \e{tr}_0\pac{ K_-\pa{0} }  }^N   
\f{ \bra{\Psi_0} K_{a_1}^+\pa{-\tf{\be}{N}} \dots K_{a_{2N-1}}^+\pa{ -\tf{\be}{N} } \ket{v}   }{ \braket{\Psi_0}{\Psi_0}  }    \\ 
\times \bra{v} K_{a_1}^-\pa{-\tf{\be}{N}} \dots K_{a_{2N-1}}^-\pa{ -\tf{\be}{N} } \ket{\Psi_0}
\label{ecriture rep trotter finie SFE}
\end{multline}
The representation \eqref{ecriture rep trotter finie SFE} will constitute the starting point of our analysis. 
Indeed, we will establish that each of the two expectation values occurring in the numerator in \eqref{ecriture rep trotter finie SFE}
are related to the partition function of the six-vertex model with reflecting ends. 
However, we first need to discuss in more details the construction of the eigenvectors of the quantum transfer matrix. 

\subsection{Eigenvectors of the QTM}

The QTM can be diagonalized by means of the algebraic Bethe Ansatz 
\cite{KomaIntroductionQTM6VertexForThermodynamicsOfXXX,KomaIntroductionQTM6VertexForThermodynamicsOfXXZ}. Indeed, 
the eigenstates of $\bs{t}^{ \mf{q} }\!\pa{\xi}= \tr_{k}\big[ T_k^{ \mf{q} }\!\pa{\xi} \big]$ are build
as a repetitive action of $B$ operators on the pseudo-vacuum 
\beq
\ket{0 }=\ket{+}_{a_1}\otimes\ket{-}_{a_2}\otimes\,  \dots \,  \otimes \ket{+}_{a_{2N-1}} \otimes \ket{-}_{a_{2N}}  \;. 
\enq
More precisely, one has that 
\beq
\bs{t}^{ \mf{q} }(\xi) \cdot   B\pa{\la_L}\dots B\pa{\la_1} \ket{ 0 }  \; = \;  
 [ \sinh(\eta) ]^{2N} \tau\pa{\xi \mid \{\la_k\}_1^L} \cdot B\pa{\la_L}\dots B\pa{\la_1} \ket{0 }   \;, 
\enq
where 
\bem
\tau\pa{\xi \mid \{\la_k\}_1^L} = \pa{-1}^N \ex{\f{h}{2T} }\pl{k=1}{L} \f{ \s{\xi-\la_k - \eta} }{ \s{\xi-\la_k } } \cdot 
\paf{  \s{\xi + \tf{\be}{N} } \s{\xi - \tf{\be}{N} + \eta} }{ \sinh^2(\eta) }^N   \\ 
+ \pa{-1}^N  \ex{-\f{h}{2T} } \pl{k=1}{L} \f{ \s{\xi-\la_k + \eta} }{ \s{\xi-\la_k } } \cdot 
\paf{  \s{\xi + \tf{\be}{N} -\eta} \s{\xi - \tf{\be}{N}} }{ \sinh^2(\eta) }^N 
\end{multline}
is the eigenvalue of $\bs{t}^{ \mf{q} }\!\pa{\xi} [ \sinh(\eta) ]^{-2N}$ associated with this choice of $\la_k$'s. 
The parameters $\la_k$ are subject to the Bethe Ansatz equations
\beq
-1 =  \ex{-\f{h}{T} } \pl{k=1}{L} \paa{ \f{ \s{\la_p-\la_k + \eta}  }{  \s{\la_p-\la_k - \eta} }  } \cdot 
 \pac{  \f{ \s{\la_p+\tf{\be}{N}-\eta}  \s{\la_p- \tf{\be}{N}} }{  \s{\la_p-\tf{\be}{N}+\eta}  \s{\la_p + \tf{\be}{N}} } }^N
 \; . 
\label{ecriture Bethe Equations QTM}
\enq
Numerical investigations, analysis at the free fermion point and for $\be$ small indicate that the 
 dominant eigenvalue is given by the choice $L=N$, this for any value of $h$ 
 \cite{GohmannKlumperSeelFinieTemperatureCorrelationFunctionsXXZ}. What changes, however, is the actual
distribution of the roots. These are located on the purely imaginary axis when $h=0$ and occupy 
regions with a more complicated shape as soon as $h\not=0$. 

From now on, we focus on the solution $\{ \la_k \}_1^N$ describing the dominant eigenvalue.
We shall also agree that $\ket{\Psi_0}$ refers to the  Bethe Ansatz-issued eigenvector
 $\ket{\Psi_0} = \prod_{a=1}^{N}B\pa{\la_a}\ket{ 0 }$.  
As proposed in \cite{KlumperNLIEfromQTMDescrThermoXYZOneUnknownFcton}, it is useful 
to introduce the function closely related to the exponent of the counting function 
\beq
\wh{\mf{a}}\pa{\om}= \ex{-\f{h}{T} } \pl{k=1}{N} \f{ \s{\om-\la_k + \eta}  }{  \s{\om-\la_k - \eta} }  
 \pac{  \f{ \s{\om + \tf{\be}{N}-\eta}  \s{\om - \tf{\be}{N}} }{  \s{ \om - \tf{\be}{N} + \eta}  \s{ \om + \tf{\be}{N}} } }^N \; .   
\label{definition counting function}
\enq
%
%
The $\wh{\mf{a}}$ function is $i\pi$-periodic, bounded when $\Re(\om) \tend \pm \infty$ and such that it has, in the case of generic 
parameters, 
\begin{itemize}
\item an $N^{\e{th}}$-order pole at $\om = - \tf{\be}{N}$ \, , 
\item an $N^{\e{th}}$-order pole at $\om = \tf{\be}{N} - \eta$ \, ,
\item N simple poles at $\om = \la_k+\eta$, \quad $k=1,\dots,N$.  
\end{itemize}

It thus follows that $1+\wh{\mf{a}}\pa{\om}$ has $3N$ zeroes. $N$ of these are, by construction, the Bethe roots, \textit{ie}
\beq
 1+\wh{\mf{a}}\pa{\la_k}=0\;\; , \qquad  k=1,\dots,N \; . 
\enq
Numerical analysis and calculations at the free fermion point 
indicate that the roots for the ground  state of the QTM can all be encircled by a unique loop $\msc{C}$ that is moreover $N$ independent 
and such that any additional root to $ 1+\wh{\mf{a}}\pa{ \om } =0$ is located outside of this loop 
\cite{GohmannKlumperSeelFinieTemperatureCorrelationFunctionsXXZ}. 
It appears, on a numerical analysis basis, that all these other roots accumulate around $\pm \eta$. 
In the following, we shall build our analysis on the same assumption. 
We do stress that all of the above considerations should be accommodated so as to be consistent with the natural $i\pi$ periodicity 
of the functions involved. Putting all these information together allows one to conclude that 
the function $\wh{\mf{a}}$ solves the non-linear integral equation
\beq
\ln \wh{\mf{a}}\pa{\om} = -\f{h}{T} \; + \;  N \ln  \pac{  \f{ \s{\om + \tf{\be}{N}+\eta}  \s{\om - \tf{\be}{N}} }{  \s{ \om - \tf{\be}{N} + \eta}  \s{ \om + \tf{\be}{N}} } }
\; + \; \Oint{ \msc{C} }{}   \theta^{\prime}\!\pa{\om - \mu} \ln \pac{1 +  \wh{\mf{a}}\pa{\mu} } \cdot \f{ \dd \mu }{2\pi} \;. 
\label{ecriture NLIE fonction a hat regime general}
\enq
where we have set $\theta\pa{\la} = i \ln \pac{ \tf{ \s{\eta-\la}}{ \s{\eta+\la}}  }$. 
The contour $\msc{C}$ encircles the Bethe roots $\la_1,\dots,\la_N$, the pole at $-\tf{\be}{N}$ but not any other 
singularity of the integrand. We also remind that, in the derivation of the  non-linear integral equation, 
one has to make use of the fact that 
$\ln[1+\wh{\mf{a}}\pa{\om}]$ has a zero monodromy around $\msc{C}$ (the contour encloses the $N^{\e{th}}$ order pole at 
$-\tf{\be}{N}$ and the $N$ simple zeroes at $\la_1,\dots, \la_N$ of $1+\mf{a}(\om)$).

\subsection{The norm of Bethe states}

Under the normalization of the $R$-matrix that we have chosen, the "norm" of the eigenstates
of the QTM admits the determinant representation \cite{KorepinNormBetheStates6-Vertex}:
\beq
 \bra{0}\pl{j=1}{N}C(\la_j) \cdot \pl{j=1}{N} B(\la_j) \ket{0} = 
\pl{j=1}{N} \Bigg\{  \; \f{  \wh{\mf{a}}^{\, \prime}\!(\la_j) }{ \wh{\mf{a}}(\la_j) } 
\cdot \pac{ \sd{\la_j,\tf{\be}{N}} \sd{\la_j,\tf{\be}{N}-\eta}}^N  \Bigg\}  \cdot
 \f{ \pl{a,b=1}{N}\s{\la_a-\la_b+\eta} }{  \pl{a\not= b }{ N } \s{\la_a-\la_b}}
\cdot \det_{ \msc{C} } \big[ I+\ov{K} \big] \;. 
\nonumber
\enq
Here, $\det_{ \msc{C} } \big[ I+\ov{K} \big]$ is the Fredholm determinant of the trace class integral operator $I+\ov{K}$
acting on $L^{2}(\msc{C})$ with the integral kernel 
\beq
\ov{K}\pa{\om,\om^{\prime}} = \f{ K(\om-\om^{\prime})  }{ \big( 1+\wh{\mf{a}}(\om^{\prime}) \big) 2\pi}
\qquad  \e{where} \qquad K(\la) = \theta^{\prime}(\la)=  \f{  i \sinh(2\eta)  }{ \s{\la-\eta} \s{\la+\eta} } \;. 
\label{definition noyau K bar cas discret}
\enq
Finally, the "norm" formula also involves the shorthand notation 
\beq
\mf{s}(\la,\mu) \; = \; \s{\la-\mu}  \s{\la+\mu} \;. 
\label{notation double sinus}
\enq
%
%
%




\section{Rewriting the expectation values}
\label{Section New representation boundary free energy}

In this section we first recast the expectation values occurring in \eqref{ecriture rep trotter finie SFE} 
in terms of a semi-homogeneous limit of the partition function of the six-vertex model with reflecting ends. 
As observed by Tsuchiya \cite{TsuchiyaPartitionFunctWithReflecEnd}, the latter admits a determinant representation. 
A direct calculation of the semi-homogeneous limit through the L'H\^{o}pital rule  is however unadapted 
for our purpose. Hence, in the second part of this section, we apply the Cauchy determinant factorization 
\cite{IzerginKitMailTerSpontaneousMagnetizationMassiveXXZ,KozKitMailSlaTerXXZsgZsgZAsymptotics}
so as to recast the Tsuchiya determinant into a form allowing us to take the semi-homogeneous limit in an elegant manner. 
This ultimately leads us to obtain a relatively simple explicit representation for the finite Trotter number approximant of the 
surface free energy.

\subsection{Relation with the partition function}

\begin{prop}
The expectation values 
\beq
\mc{F}^- \equiv \bra{v} K_{a_1}^-\!\pa{-\tf{\be}{N}} \dots K_{a_{2N-1}}^-\!\pa{ -\tf{\be}{N} } B\pa{\la_1}\dots B\pa{\la_N}\ket{0} 
\label{rewrite-eq-1}
\enq
and 
\beq
\mc{F}^+ \equiv \bra{0} C\pa{\la_1} \dots C\pa{\la_N} K_{a_1}^+\!\pa{-\tf{\be}{N}} \dots K_{a_{2N-1}}^+\!\pa{ -\tf{\be}{N} } \ket{v}  
\enq
occurring in \eqref{ecriture rep trotter finie SFE} can be recast as 

\beq
\label{rewrite-eq-2}
\mc{F}^{-} = 
\ex{-\f{Nh}{2T}} (0\mid \underbrace{  \mc{C}^-\!\pa{-\tf{\be}{N}} \dots \mc{C}^-\!\pa{-\tf{\be}{N}}  }_{ N \; \e{terms} } \mid \ov{0} )
\qquad \e{and} \qquad 
\mc{F}^{+} =  \ex{\f{Nh}{2T}}  
( \ov{0} \mid \underbrace{  \mc{B}^+\!\pa{-\tf{\be}{N}} \dots \mc{B}^+\!\pa{-\tf{\be}{N}}  }_{ N \; \e{terms} } \mid 0 ) \;. 
\enq
There $\mc{C}^-$, resp. $\mc{B}^+$, refer to entries of the  boundary monodromy matrices of $-$, resp. $+$, type:
\beqa
\mc{U}^-_a\!\pa{\la} &=  \quad \mc{T}_a\!\pa{\la} K^-_a\!\pa{\la} \wh{\mc{T}}_a\!\pa{\la} \quad  = &\pa{ \ba{cc}  \mc{A}^-\!\pa{\la} & \mc{B}^-\!\pa{\la} \\ 
										\mc{C}^-\!\pa{\la} & \mc{D}^-\!\pa{\la} \ea }_{\pac{a}}  
\label{definition matrice U moins}\\
\pac{ \mc{U}^+_a\!\pa{\la} }^{t_a} &= \mc{T}^{t_a}_a\!\pa{\la} \pac{K^+_a\!\pa{\la}}^{t_a} \wh{\mc{T}}_a^{t_a}\!\pa{\la}  
=  &\pa{ \ba{cc}  \mc{A}^+\!\pa{\la} & \mc{C}^+\!\pa{\la} \\ \mc{B}^+\!\pa{\la} & \mc{D}^+\!\pa{\la} \ea }_{\pac{a}}  \;,
\eeqa
where 
\beq
\mc{T}_a\pa{u } = R_{aN}\pa{u-\la_N} \dots R_{a1}\pa{u-\la_1} \qquad \quad \e{and} \qquad \quad 
\wh{\mc{T}}_a\pa{u } = R_{1a}\pa{u+\la_1} \dots R_{Na}\pa{u+\la_N}  \;. 
\label{definition Ta et Thata pour Tsuchiya determinant reconstruction}
\enq
Lastly, we agree upon 
\beq
\sket{ \ov{0} }  =   \sket{-}_1\otimes \dots \otimes \sket{-}_N
 \qquad \e{and} \qquad 
\sket{ 0 }  =   \sket{+}_1\otimes \dots \otimes \sket{+}_N \; ,
\enq
with $\sket{\pm}_k$ being the canonical spin up/down basis in the space $V_i \simeq \Cx^2$. 
\label{rewriteTHM1}
\end{prop}

\Proof

Using that $B\pa{\la} = \, _k\sbra{+} T^{ \mf{q} }_k\!\pa{\la} \sket{-}_k$, we get 
\bem
\mc{F}^- = \bra{v}\otimes \sbra{0} K_{a_1}^-\pa{-\tf{\be}{N}} \dots K_{a_{2N-1}}^-\pa{ -\tf{\be}{N} } 
R^{t_{a_{2N}}}_{a_{2N} 1 }\!\pa{ - \la_1 -  \tf{\be}{N} } R_{1 a_{2N-1}}\!\pa{\la_1 - \tf{\be}{N} } \dots \\
\hspace{2cm} \dots R_{a_{2} 1}^{t_{a_2} }\!\pa{-\la_1 - \tf{\be}{N} } R_{1 a_{1}}\!\pa{\la_1 - \tf{\be}{N} } 
\ex{\f{h}{2T}\sg_1^z} \dots 
R_{a_{2} N}^{t_{a_2} }\!\pa{-\la_N - \tf{\be}{N} } R_{N a_{1}}\!\pa{\la_N - \tf{\be}{N} } \ex{\f{h}{2T}\sg_N^z}  
\ket{0}\, \otimes \sket{\ov{0}}  \\
=  \bra{v}\otimes \sbra{0} R^{t_{a_{2N}}}_{a_{2N} 1 }\!\pa{ - \la_1 -  \tf{\be}{N} }\dots R^{t_{a_{2N}}}_{a_{2N} N }\!\pa{ - \la_N -  \tf{\be}{N} }
K_{a_{2N-1}}^-\pa{ -\tf{\be}{N} }  R_{1 a_{2N-1}}\!\pa{\la_1 - \tf{\be}{N} } \dots   \\
R_{N a_{2N-1}}\!\pa{\la_N - \tf{\be}{N} }   \dots K_{a_1}^-\pa{-\tf{\be}{N}}  
R_{1 a_{1}}\!\pa{\la_1 - \tf{\be}{N} } \dots R_{N a_{1}}\!\pa{\la_N - \tf{\be}{N} } 
\pl{k=1}{N} \Big\{ \ex{\f{h}{2T}\sg_k^z} \Big\} \cdot  \ket{0}\, \otimes \sket{\ov{0}}
\end{multline}
By using \eqref{definition Ta et Thata pour Tsuchiya determinant reconstruction}, we reconstruct monodromy matrices in which 
the Bethe roots $\{\la_k\}_1^N$ for the ground state of the quantum transfer matrix play the role of inhomogeneities and $-\tf{\be}{N}$ is 
interpreted as the spectral parameter. This leads to  
\beq
\mc{F}^- =   \ex{-\f{Nh}{2T}} \bra{v}\otimes \sbra{0}  \mc{T}_{a_{2N}}^{t_{a_{2N}}}(-\tf{\be}{N}) K_{a_{2N-1}}^-(-\tf{\be}{N} ) \wh{\mc{T}}_{a_{2N-1}}(-\tf{\be}{N})
\cdots \mc{T}_{a_{2}}^{t_{a_{2}}}(-\tf{\be}{N}) K_{a_{1}}^-(-\tf{\be}{N} ) \wh{ \mc{T} }_{a_{1}}(-\tf{\be}{N}) \cdot 
\ket{0}\otimes \sket{\ov{0}}
\nonumber 
\enq
 It now remains to compute the partial scalar products involving the even spaces $a_{2k}$, $k=1, \dots, N$. 
This can be done thanks to the identity
\beq
\Big\{  \;  _{a_{2k-1}}\!\bra{-} \, _{a_{2k}}\!\bra{-}  \; + \; _{a_{2k-1}}\!\bra{+} \, _{a_{2k}}\!\bra{+}  \Big\} \cdot  \mc{T}_{a_{2k}}^{t_{a_{2k}}}(\la) \ket{-}_{a_{2k}}
=\;   _{a_{2k-1}}\!\bra{-} \mc{T}_{a_{2k-1}}(\la)
\enq
Therefore, 
\bem
\mc{F}^- =   \ex{-\f{Nh}{2T}} \,  _{a_{2N-1}}\!\bra{-} \dots  _{a_{1}}\!\!\bra{-} \otimes \sbra{0} \mc{U}^{-}_{a_{2N-1}}\!\pa{-\tf{\be}{N}}
\dots  \mc{U}^{-}_{a_{1}}\!\pa{-\tf{\be}{N}}  \ket{+}_{a_{2N-1}}\dots \ket{+}_{a_1}  \otimes \sket{\ov{0}}  \\
=  \ex{-\f{Nh}{2T}}  \sbra{0} \mc{C}^{-}\!\pa{-\tf{\be}{N}}
\dots  \mc{C}^{-}\!\pa{-\tf{\be}{N}}  \sket{\ov{0}}  \;.
\end{multline}

Very similar steps can be applied so as to re-write $\mc{F}^+$. Indeed, one can express
$C\pa{\la} = \, _k\sbra{-} T^{\mf{q}}_k\!\pa{\la} \sket{+}_k$ what leads to 
\bem
\mc{F}^+ =  \bra{0}\otimes \sbra{\ov{0}} 
R^{t_{a_{2N}}}_{a_{2N} 1 }\!\pa{ - \la_1 -  \tf{\be}{N} } R_{1 a_{2N-1}}\!\pa{\la_1 - \tf{\be}{N} } \dots 
R_{a_{2} 1}^{t_{a_2} }\!\pa{-\la_1 - \tf{\be}{N} } R_{1 a_{1}}\!\pa{\la_1 - \tf{\be}{N} } \ex{\f{h}{2T} \sg_1^z}\dots  \\
\hspace{2cm} \dots 
R_{a_{2} N}^{t_{a_2} }\!\pa{-\la_N - \tf{\be}{N} } R_{N a_{1}}\!\pa{\la_N - \tf{\be}{N} }  \ex{\f{h}{2T} \sg_N^z}
 K_{a_1}^+\!\pa{-\tf{\be}{N}} \dots K_{a_{2N-1}}^+\!\pa{ -\tf{\be}{N} } \, \ket{v} \, \otimes \sket{0}  \\
=  \ex{\f{N h}{2T}}  \bra{0}\otimes \sbra{\ov{0}} R^{t_{a_{2N}}}_{a_{2N} 1 }\!\pa{ - \la_1 -  \tf{\be}{N} }\dots R^{t_{a_{2N}}}_{a_{2N} N }\!\pa{ - \la_N -  \tf{\be}{N} }
 R_{1 a_{2N-1}}\!\pa{\la_1 - \tf{\be}{N} } \dots R_{N a_{2N-1}}\!\pa{\la_N - \tf{\be}{N} }   \\
 K_{a_{2N-1}}^+\!\pa{ -\tf{\be}{N} }  \dots  R_{1 a_{1}}\!\pa{\la_1 - \tf{\be}{N} } \dots R_{N a_{1}}\!\pa{\la_N - \tf{\be}{N} }  
 K_{a_1}^+\!\pa{-\tf{\be}{N}} \, \ket{v} \, \otimes \sket{0}  \\
 =   \ex{ \f{ N h}{2T}}  \bra{0}\otimes \sbra{\ov{0}}  \mc{T}_{a_{2N}}^{t_{a_{2N}}}\pa{-\tf{\be}{N}}  
\wh{ \mc{T} }_{a_{2N-1}}\pa{-\tf{\be}{N}} K_{a_{2N-1}}^+\pa{ -\tf{\be}{N} } \mc{T}_{a_{2N-2}}^{t_{a_{2N-2}}}\pa{-\tf{\be}{N}}  \\
\dots \mc{T}_{a_{2}}^{t_{a_{2}}}\pa{-\tf{\be}{N}}  \wh{ \mc{T} }_{a_{1}}\pa{-\tf{\be}{N}} K_{a_{1}}^+\pa{ -\tf{\be}{N} } 
			 \, \ket{v} \, \otimes \sket{0}
\end{multline}
In the case of $\mc{F}^+$, one takes the partial scalar products in respect to the odd spaces $a_{2k-1}$, $k=1, \dots, N$. 
This is done by means of the identity 
\beq
\; _{a_{2k-1}}\!\bra{+}  \wh{ \mc{T} }_{a_{2k-1}}(\la) K^{+}_{a_{2k-1}}\!\pa{\la} 
\Big( \;  \ket{-}_{a_{2k-1}} \ket{-}_{a_{2k}}  \; + \;  \ket{+}_{a_{2k-1}} \ket{+}_{a_{2k}}  \Big) =
\pac{ K^{+}_{a_{2k}}\!\pa{\la} }^{t_{a_{2k}}} \wh{ \mc{T} }_{a_{2k}}^{t_{a_{2k}}}\!\pa{\la}   \ket{+}_{a_{2k}} \;.
\enq
Therefore, 
\bem
\mc{F}^+ =   \ex{\f{Nh}{2T} } \,  _{a_{2N}}\!\!\bra{-} \dots \,  _{a_{2}}\!\!\bra{-} \otimes \sbra{\ov{0} } 
\pac{\mc{U}^{-}_{a_{2N}}\!\pa{-\tf{\be}{N}} }^{t_{a_{2N}}} 
\hspace{-3mm} \dots \; \pac{ \mc{U}^{-}_{a_{2}}\!\pa{-\tf{\be}{N}} }^{t_{a_{2}}} \ket{+}_{a_{2N}}\dots \ket{+}_{a_2}  \sket{0}  \\
=    \sbra{ \ov{0}} \mc{B}^{+}\!\pa{-\tf{\be}{N}}
\dots  \mc{B}^{+}\!\pa{-\tf{\be}{N}}  \sket{0}  \;.
\end{multline}
Thus proving the second identity. \qed

\begin{figure}[t]
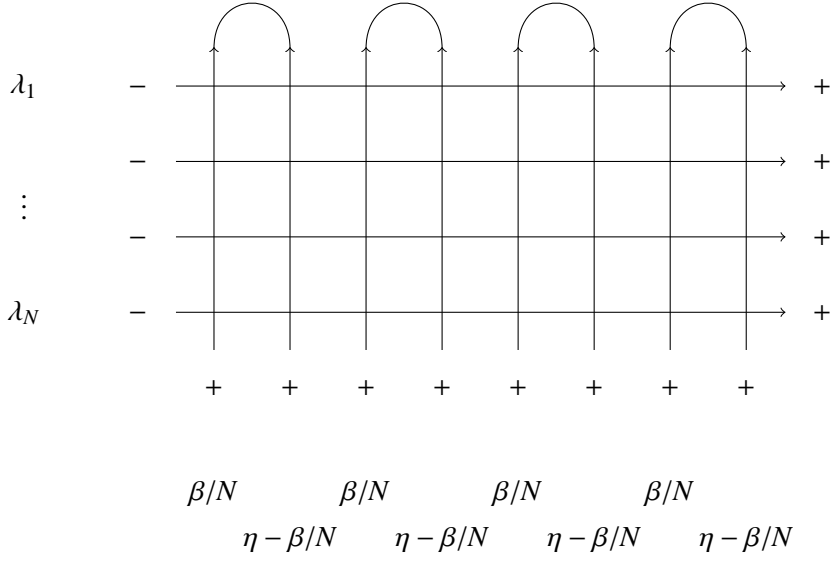

\centering
\begin{pgfpicture}{0cm}{-0.5cm}{14cm}{7cm}

\pgfsetendarrow{\pgfarrowto}

\pgfline{\pgfxy(3.5,3)}{\pgfxy(11.5,3)}
\pgfline{\pgfxy(3.5,4)}{\pgfxy(11.5,4)}
\pgfline{\pgfxy(3.5,5)}{\pgfxy(11.5,5)}
\pgfline{\pgfxy(3.5,6)}{\pgfxy(11.5,6)}

\pgfputat{\pgfxy(4,0.6)}{\pgfbox[center,center]{$\beta/N$}}
\pgfputat{\pgfxy(6,0.6)}{\pgfbox[center,center]{$\beta/N$}}
\pgfputat{\pgfxy(8,0.6)}{\pgfbox[center,center]{$\beta/N$}}
\pgfputat{\pgfxy(10,0.6)}{\pgfbox[center,center]{$\beta/N$}}

\pgfputat{\pgfxy(5,0)}{\pgfbox[center,center]{$\eta-\beta/N$}}
\pgfputat{\pgfxy(7,0)}{\pgfbox[center,center]{$\eta-\beta/N$}}
\pgfputat{\pgfxy(9,0)}{\pgfbox[center,center]{$\eta-\beta/N$}}
\pgfputat{\pgfxy(11,0)}{\pgfbox[center,center]{$\eta-\beta/N$}}

\pgfline{\pgfxy(4,2.5)}{\pgfxy(4,6.5)}
\pgfline{\pgfxy(5,2.5)}{\pgfxy(5,6.5)}
\pgfline{\pgfxy(6,2.5)}{\pgfxy(6,6.5)}
\pgfline{\pgfxy(7,2.5)}{\pgfxy(7,6.5)}
\pgfline{\pgfxy(8,2.5)}{\pgfxy(8,6.5)}
\pgfline{\pgfxy(9,2.5)}{\pgfxy(9,6.5)}
\pgfline{\pgfxy(10,2.5)}{\pgfxy(10,6.5)}
\pgfline{\pgfxy(11,2.5)}{\pgfxy(11,6.5)}

\pgfputat{\pgfxy(3,3)}{\pgfbox[center,center]{$-$}}
\pgfputat{\pgfxy(3,4)}{\pgfbox[center,center]{$-$}}
\pgfputat{\pgfxy(3,5)}{\pgfbox[center,center]{$-$}}
\pgfputat{\pgfxy(3,6)}{\pgfbox[center,center]{$-$}}

\pgfputat{\pgfxy(1.5,3)}{\pgfbox[center,center]{$\lambda_N$}}
\pgfputat{\pgfxy(1.5,6)}{\pgfbox[center,center]{$\lambda_1$}}
\pgfputat{\pgfxy(1.5,4.5)}{\pgfbox[center,center]{$\vdots$}}

\pgfputat{\pgfxy(12,3)}{\pgfbox[center,center]{$+$}}
\pgfputat{\pgfxy(12,4)}{\pgfbox[center,center]{$+$}}
\pgfputat{\pgfxy(12,5)}{\pgfbox[center,center]{$+$}}
\pgfputat{\pgfxy(12,6)}{\pgfbox[center,center]{$+$}}

\pgfputat{\pgfxy(4,2)}{\pgfbox[center,center]{$+$}}
\pgfputat{\pgfxy(5,2)}{\pgfbox[center,center]{$+$}}
\pgfputat{\pgfxy(6,2)}{\pgfbox[center,center]{$+$}}
\pgfputat{\pgfxy(7,2)}{\pgfbox[center,center]{$+$}}
\pgfputat{\pgfxy(8,2)}{\pgfbox[center,center]{$+$}}
\pgfputat{\pgfxy(9,2)}{\pgfbox[center,center]{$+$}}
\pgfputat{\pgfxy(10,2)}{\pgfbox[center,center]{$+$}}
\pgfputat{\pgfxy(11,2)}{\pgfbox[center,center]{$+$}}

\pgfclearendarrow
\pgfmoveto{\pgfxy(10,6.5)}
\pgfcurveto{\pgfxy(10,7.3)}{\pgfxy(11,7.3)}{\pgfxy(11,6.5)}
\pgfstroke
\pgfmoveto{\pgfxy(8,6.5)}
\pgfcurveto{\pgfxy(8,7.3)}{\pgfxy(9,7.3)}{\pgfxy(9,6.5)}
\pgfstroke
\pgfmoveto{\pgfxy(6,6.5)}
\pgfcurveto{\pgfxy(6,7.3)}{\pgfxy(7,7.3)}{\pgfxy(7,6.5)}
\pgfstroke
\pgfmoveto{\pgfxy(4,6.5)}
\pgfcurveto{\pgfxy(4,7.3)}{\pgfxy(5,7.3)}{\pgfxy(5,6.5)}
\pgfstroke


\end{pgfpicture}
\caption{
Partition function of the 6-vertex model with a reflecting
  end. This graph corresponds to the expression
  \eqref{rewrite-eq-1}: the horizontal lines represent the action of
  the $B$-operators on the reference state, whereas the open boundary
  conditions at the top correspond to the action of the local
  $K$-matrices.}
\label{szamoljuk}
\end{figure}

\begin{figure}[t]
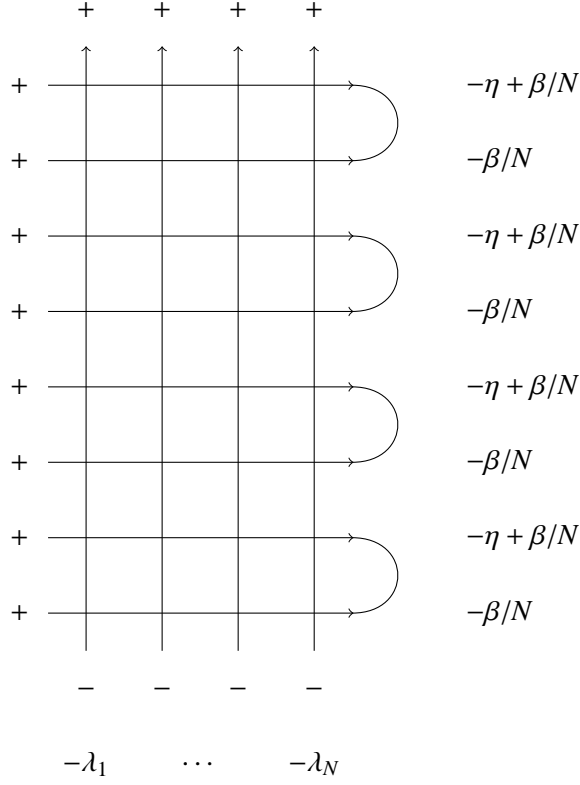

\centering
\begin{pgfpicture}{4cm}{-1.5cm}{12cm}{9cm}

\pgfsetendarrow{\pgfarrowto}

\pgfline{\pgfxy(5.5,1)}{\pgfxy(9.5,1)}
\pgfline{\pgfxy(5.5,2)}{\pgfxy(9.5,2)}
\pgfline{\pgfxy(5.5,3)}{\pgfxy(9.5,3)}
\pgfline{\pgfxy(5.5,4)}{\pgfxy(9.5,4)}
\pgfline{\pgfxy(5.5,5)}{\pgfxy(9.5,5)}
\pgfline{\pgfxy(5.5,6)}{\pgfxy(9.5,6)}
\pgfline{\pgfxy(5.5,7)}{\pgfxy(9.5,7)}
\pgfline{\pgfxy(5.5,8)}{\pgfxy(9.5,8)}

\pgfputat{\pgfxy(5,1)}{\pgfbox[left,center]{$+$}}
\pgfputat{\pgfxy(5,2)}{\pgfbox[left,center]{$+$}}
\pgfputat{\pgfxy(5,3)}{\pgfbox[left,center]{$+$}}
\pgfputat{\pgfxy(5,4)}{\pgfbox[left,center]{$+$}}
\pgfputat{\pgfxy(5,5)}{\pgfbox[left,center]{$+$}}
\pgfputat{\pgfxy(5,6)}{\pgfbox[left,center]{$+$}}
\pgfputat{\pgfxy(5,7)}{\pgfbox[left,center]{$+$}}
\pgfputat{\pgfxy(5,8)}{\pgfbox[left,center]{$+$}}

\pgfputat{\pgfxy(6,0)}{\pgfbox[center,center]{$-$}}
\pgfputat{\pgfxy(7,0)}{\pgfbox[center,center]{$-$}}
\pgfputat{\pgfxy(8,0)}{\pgfbox[center,center]{$-$}}
\pgfputat{\pgfxy(9,0)}{\pgfbox[center,center]{$-$}}

\pgfputat{\pgfxy(6,-1)}{\pgfbox[center,center]{$-\lambda_1$}}
\pgfputat{\pgfxy(7.5,-1)}{\pgfbox[center,center]{$\dots$}}
\pgfputat{\pgfxy(9,-1)}{\pgfbox[center,center]{$-\lambda_N$}}

\pgfputat{\pgfxy(6,9)}{\pgfbox[center,center]{$+$}}
\pgfputat{\pgfxy(7,9)}{\pgfbox[center,center]{$+$}}
\pgfputat{\pgfxy(8,9)}{\pgfbox[center,center]{$+$}}
\pgfputat{\pgfxy(9,9)}{\pgfbox[center,center]{$+$}}

\pgfputat{\pgfxy(11,1)}{\pgfbox[left,center]{$-\beta/N$}}
\pgfputat{\pgfxy(11,3)}{\pgfbox[left,center]{$-\beta/N$}}
\pgfputat{\pgfxy(11,5)}{\pgfbox[left,center]{$-\beta/N$}}
\pgfputat{\pgfxy(11,7)}{\pgfbox[left,center]{$-\beta/N$}}
\pgfputat{\pgfxy(11,2)}{\pgfbox[left,center]{$-\eta+\beta/N$}}
\pgfputat{\pgfxy(11,4)}{\pgfbox[left,center]{$-\eta+\beta/N$}}
\pgfputat{\pgfxy(11,6)}{\pgfbox[left,center]{$-\eta+\beta/N$}}
\pgfputat{\pgfxy(11,8)}{\pgfbox[left,center]{$-\eta+\beta/N$}}

\pgfline{\pgfxy(6,0.5)}{\pgfxy(6,8.5)}
\pgfline{\pgfxy(7,0.5)}{\pgfxy(7,8.5)}
\pgfline{\pgfxy(8,0.5)}{\pgfxy(8,8.5)}
\pgfline{\pgfxy(9,0.5)}{\pgfxy(9,8.5)}

\pgfclearendarrow
\pgfmoveto{\pgfxy(9.5,7)}
\pgfcurveto{\pgfxy(10.3,7)}{\pgfxy(10.3,8)}{\pgfxy(9.5,8)}
\pgfstroke
\pgfmoveto{\pgfxy(9.5,5)}
\pgfcurveto{\pgfxy(10.3,5)}{\pgfxy(10.3,6)}{\pgfxy(9.5,6)}
\pgfstroke
\pgfmoveto{\pgfxy(9.5,3)}
\pgfcurveto{\pgfxy(10.3,3)}{\pgfxy(10.3,4)}{\pgfxy(9.5,4)}
\pgfstroke
\pgfmoveto{\pgfxy(9.5,1)}
\pgfcurveto{\pgfxy(10.3,1)}{\pgfxy(10.3,2)}{\pgfxy(9.5,2)}
\pgfstroke

\end{pgfpicture}
\caption{Partition function of the 6-vertex model with a reflecting
  end. This graph corresponds to the first expression in
  \eqref{rewrite-eq-2}: The horizontal lines together with the
  boundary weight correspond to the action of the $C^-$ operators, and
the fixed boundary conditions on the top and the bottom describe the
two reference states $(0|$ and $|\bar 0)$.
}
\label{szamoljukmar}
\end{figure}

The scalar product in expression \eqref{rewrite-eq-1} is
 depicted on Fig. \ref{szamoljuk}, whereas the
 first expression in
 \eqref{rewrite-eq-2} is
 shown in Fig. \ref{szamoljukmar}. Note that we have used the crossing relation 
 so as to recast part of the weights into a canonical form. 
Note also  that these two figures are related through a reflection across the North-East diagonal. 
 The proof of proposition \ref{rewriteTHM1} corresponds to an algebraic verification of this symmetry.

\subsection{Alternative factorization of the Tsuchiya Determinant }

It follows from the previous observations that the two factors $\mc{F}^{\pm}$ can be identified with the
partition function of the six-vertex model with reflecting ends. The latter has been computed, in terms of a
determinant, by Tsuchiya \cite{TsuchiyaPartitionFunctWithReflecEnd}. 
More precisely, let $\mc{C}^-$ be given by \eqref{definition matrice U moins}, then the aforementioned partition
function is defined by 
\beq
\mc{Z}_N\pa{ \{\xi_a\}_1^N  ; \{ \la_k \}_1^N ; \xi_- } \; \equiv \; \sbra{0} \mc{C}^{-}\!\pa{ \xi_1 }
\dots  \mc{C}^{-}\!\pa{\xi_N  }  \sket{\ov{0}}  \;. 
\enq
By using the relations provided in \cite{KozKitMailNicSlaTerElemntaryblocksopenXXZ}, it is readily seen that 
\beq
\sbra{\ov{0}} \mc{B}^{+}\!\pa{ \xi_1 } \dots  \mc{B}^{+}\!\pa{\xi_N  }  \sket{0}  = 
\pl{a=1}{N} \bigg\{  \f{ \s{2\xi_a + 2\eta} }{ \s{2\xi_a} } \bigg\}
\; \cdot \;  \mc{Z}_N\pa{ \{\xi_a\}_1^N  ; \{ \la_k \}_1^N ; \xi_+ } \;.
\enq
The Tsuchiya determinant representation for $\mc{Z}_N\pa{ \{\xi_a\}_1^N  ; \{ \la_k \}_1^N ; \xi_- }$ reads 
\beq
\mc{Z}_N\pa{ \{\xi_a\}_1^N  ; \{ \la_k \}_1^N ; \xi_- } =    
\f{ \pl{a=1}{N} \Big\{  \s{\xi_- + \la_a}  \s{2\xi_a} \Big\}\cdot  \pl{a,b=1}{N} \Big\{ \sd{\xi_a,\la_b} \cdot \sd{\xi_a+\eta,\la_b} \Big\}  }
{   \pl{a<b}{N} \Big\{ \sd{\la_b,\la_a}   \s{\xi_a-\xi_b} \s{\xi_a+\xi_b+\eta} \Big\} }  \cdot   
\ddet{N}{ \mc{N}} \;.
\label{ecriture Tsuchiya determinant}
\enq
$\mf{s}(\la,\mu)$ has been defined in \eqref{notation double sinus} whereas 
the entries of the matrix $\mc{N}$ are given by 
\beq
\mc{N}_{jk} = \f{  \s{\eta}  }{   \sd{\xi_j+\eta,\la_k}\sd{\xi_j,\la_k}  } 
%
%
%
\label{ecriture entree N et def double sinus}
\enq

The antisymmetry properties of the determinant along with its invariance under the transformations 
\beq
\{ \la_a \}_1^N \hookrightarrow \{\sg_a \la_a \}_1^N \qquad \e{and} \qquad 
\{ \xi_a \}_1^N \hookrightarrow \{ \eps_a \xi_a  -(\eps_a +1)\eta/2 \}_1^N \qquad \e{for}\;  \e{any} \qquad 
\eps_a, \sg_a \in \{ \pm \} 
\enq
ensures that the singularities present in the denominator of the right hand side effectively cancel out. 
Taking the homogeneous limit $\xi_a \tend -\tf{\be}{N}$ recasts \eqref{ecriture Tsuchiya determinant}
in terms of a Wronskian. Such a representation is however not adapted for our goals. Notwithstanding, 
by using the Cauchy determinant factorization techniques 
\cite{IzerginKitMailTerSpontaneousMagnetizationMassiveXXZ,KozKitMailSlaTerXXZsgZsgZAsymptotics}, 
one can provide a representation for $\mc{Z}_N$
that cancels out explicitly the apparent singularities at  $\xi_k=\xi_{\ell}$ and $\la_a=\la_b$. 

\begin{lemme}
The determinant of the matrix $\mc{N}$ can be decomposed into the below product of determinants 
\beq
\ddet{N}{ \mc{N} } = \pl{a=1}{N} \bigg\{ \f{ 1 }{ \s{2\la_a} }  \bigg\}  \cdot 
\det_{N}  \Big[ \f{1}{ \s{\xi_k-\la_j} \s{\xi_k+\la_j+\eta}} \Big] \cdot 
\det_{N} \big[\de_{jk} + \wt{U}_{jk} \big]
\label{ecriture fact det N}
\enq
Where the entries of the matrix $\wt{U}_{jk}$ read 
\bem
\wt{U}_{jk} = \f{  \s{2\la_k+\eta} }{ \s{\la_k+\la_j}\s{\la_j-\la_k+\eta} }  
\f{ \pl{a=1}{N} \s{\la_k-\la_a - \eta} }{ \pl{ \substack{a=1 \\ \not= k} }{N} \s{\la_k-\la_a}}
\cdot \pl{a=1}{N} \f{ \s{\la_a+\la_k} }{ \s{\la_a+\la_k+\eta} }   \\
\times \pl{a=1}{N}\f{ \s{\xi_a-\la_k}\s{\xi_a+\la_k+\eta} }{ \s{\xi_a+\la_k}\s{\xi_a-\la_k+\eta} } \;.
\label{definition matrice U tilde}
\end{multline}
\end{lemme}

\Proof 

Given the Cauchy matrix 
\beq
C_{kj} = \f{1}{ \s{\xi_k-\la_j} \s{\xi_k+\la_j+\eta}}  \;. 
\enq
Its inverse has entries
\beq
\big[ C^{-1} \big]_{jk} = \f{1}{ \s{\xi_k-\la_j} \s{\xi_k+\la_j+\eta}}  
\f{ \pl{a=1}{N} \Big\{ \s{\xi_a-\la_j}\s{\la_j+\xi_a+\eta} \s{\xi_k-\la_a}\s{\la_a+\xi_k+\eta}   \Big\} }
{  \pl{ \substack{a=1\\ \not= k} }{N}  \Big\{ \s{\xi_a-\xi_k} \s{\xi_k+\xi_a+\eta} \Big\}
 \pl{ \substack{a=1\\ \not= j} }{N} \Big\{  \s{\la_j-\la_a}\s{\la_j+\la_a+\eta}   \Big\} } \;.
\nonumber
\enq

It is readily seen that 
\beq
\sul{p=1}{N} \big[ C^{-1} \big]_{jp} \mc{N}_{pk}  = \f{ \pl{a=1}{N} \s{\xi_a-\la_j}\s{\la_j+\xi_a+\eta}   }
{  \pl{ \substack{a=1\\ \not= j} }{N}  \s{\la_j-\la_a}\s{\la_j+\la_a+\eta}  }  \cdot  \mc{S}_{jk}
\label{calcul produit matrices en terme Sjk}
\enq
with
\beq
\mc{S}_{jk} = \sul{p=1}{N}  \f{ \s{\eta} }{ \s{\xi_p-\la_j} \s{\xi_p+\la_j+\eta} \sd{\xi_p,\la_k}\sd{\xi_p+\eta,\la_k} }  
\cdot \f{ \pl{a=1}{N} \s{\xi_p-\la_a}\s{\la_a+\xi_p+\eta}  }
{  \pl{ \substack{a=1\\ \not= p} }{N}  \s{\xi_a-\xi_p} \s{\xi_p+\xi_a+\eta}  } \;.
%
\enq
Now, note that, on the one hand, the $i\pi$-periodicity of the below integrand leads to 
\beq
0= \Oint{ i \pi-\e{strip} }{} \hspace{-3mm} \f{ \dd \om }{ 4i\pi }  \cdot 
\f{ \s{\eta} \s{ 2\om + \eta } }{ \s{\om-\la_j} \s{\om+\la_j+\eta} \sd{\om,\la_k}\sd{\om+\eta,\la_k} }  
\pl{a=1}{N} \f{  \s{\om-\la_a}\s{\la_a+\om+\eta}  }
{   \s{\xi_a-\om} \s{\om+\xi_a+\eta}  } \;. 
\enq
On the other hand, the integral can be taken by computing the residues located in the $i\pi$-periodic strip. 
These two observations show that 
\bem
\mc{S}_{jk} = \de_{jk} \f{ 1 }{  \s{2\la_k} }  
\f{\prod_{a\not= k  }^{N} \s{\la_k-\la_a}    \s{\la_a+\la_k+\eta}  }
{ \prod_{a=1}^{N}  \s{\xi_a-\la_k} \s{\la_k+\xi_a+\eta}  }  \\
\; + \; 
\f{ 1 }{ \s{\la_k+\la_j} \s{\la_j-\la_k+\eta} \s{2\la_k} } 
\pl{a=1}{N} \f{  \s{\la_k + \la_a}\s{\la_k-\la_a-\eta}  }
{   \s{\xi_a+\la_k} \s{\xi_a-\la_k+\eta}  } \;. 
\label{formule alternative Sjk}
\end{multline}
Hence, after replacing $S_{jk}$ in \eqref{calcul produit matrices en terme Sjk} by \eqref{formule alternative Sjk}, 
pulling out the pre-factors and carrying out a similarity transformation, we obtain the desired representation. \qed

\begin{prop}
\label{Proposition representation limit homogene Tsuchiya}
Let $\{\la_a\}_1^N$ be a solution of the Bethe equations. Then, the homogeneous limit ($\xi_a\tend -\tf{\be}{N}$) of the 
partition function admits the below representation:
\bem
\mc{Z}_N\pa{ \{-\tf{\be}{N}\}_1^N ; \{ \la_a \}_1^N;\xi_-  } = \pl{a=1}{N}  \bigg\{ \f{ \s{-2\tf{\be}{N} } \s{\la_a+\xi_-} }{ \s{2\la_a} } \bigg\}
\cdot \pl{a<b}{N}  \f{\s{\la_a+\la_b+\eta}}{\s{\la_a+\la_b}}  \\
\times \pl{a=1}{N} \Big[ \s{\la_a-\tf{\be}{N}} \s{\eta -\la_a-\tf{\be}{N} }\Big]^{N}  \cdot 
\det_{N} \big[ \de_{jk} \; + \;  U_{jk} \big] \;.
\label{ecriture representation limit homogene fcton partition}
\end{multline}
where the matrix $U_{jk}$ reads 
\beq
U_{jk} = \f{ -  \ex{ - \f{ h }{T} }\s{2\la_k+\eta} }{ \s{\la_k+\la_j}\s{\la_j-\la_k+\eta} }  
\f{ \pl{a=1}{N} \s{\la_k-\la_a + \eta} }{ \pl{ \substack{a=1 \\ \not= k} }{N} \s{\la_k-\la_a}} \cdot 
\pl{a=1}{N} \f{ \s{\la_a + \la_k } }{ \s{\la_a + \la_k +\eta} } \;.
\label{definition matrice U}
\enq

\end{prop}

\Proof

Using that the parameters $\{\la_a\}_1^N$ satisfy the Bethe equations, it is readily seen that $\wt{U}_{jk}=U_{jk}$. 
The rest follows after straightforward algebra. \qed

\vspace{2mm}
As follows from proposition \ref{Proposition representation limit homogene Tsuchiya}, most of 
the apparent singularities of $\mc{Z}_N$ can be canceled out by means of the Cauchy determinant 
factorization. Yet, in 
\eqref{ecriture representation limit homogene fcton partition} there are still apparent singularities at $\la_k=-\la_j$. 
These are of course compensated by the zeroes of the determinant $\det_N\big[\de_{jk}+U_{jk}\big]$. 
However, their presence is problematic in respect to taking the infinite Trotter 
number limit. We now factor out these zeroes explicitly. 
\begin{lemme}
\label{Lemme factorisation determinant I+U}
The below factorisation holds
\beq
\ddet{N}{\de_{jk}+U_{jk}} = \pl{p=1}{N}[1+\wh{\mf{a}}(-\la_p)]^{\f{1}{2}} \cdot 
\paf{ 1+\wh{\mf{a}}\pa{0} }{ 1-\wh{\mf{a}}\pa{0} }^{\f{1}{4}} \cdot 
\ex{\mc{F}_N\pa{\{\la_a\}_1^N}} \; , 
 \label{lemme factorisation det ac Fn}
\enq
where $\wh{\mf{a}}$ has been defined in \eqref{definition counting function} and 
\bem
\mc{F}_N\pa{\{\la_a\}_1^N} = 
\sul{ k = 0  }{+\infty}   \Oint{ \msc{C}_1\supset\dots \supset \msc{C}_{2k+1} }{}
\hspace{-2mm} 
\sul{ n=k }{ +\infty } \f{ \Big[ \wh{f}\!\pa{\om_{2k+1}} \Big]^{n-k} }{2n+1}  \; \cdot  \; 
 \pl{p=1}{2k+1}\wh{U}(\om_p,\om_{p+1}) \; \f{ \dd^{2k+1} \om }{ \pa{2i\pi}^{2k+1} }  \\
\; -\; \sul{  k = 1   }{+\infty}   \Oint{ \msc{C}_1\supset\dots \supset \msc{C}_{2k} }{}
\sul{n=k}{+\infty} \f{ \Big[ \wh{f}\!\pa{\om_{2k}} \Big]^{n-k} }{2n} \; \cdot \;  \pl{p=1}{2k} \wh{U}(\om_p,\om_{p+1}) \; \f{ \dd^{2k} \om }{ \pa{2i\pi}^{2k} } \; . 
\label{ecriture F cal N trotter fini}
\end{multline}
Above we agree upon $\om_{n+1}\equiv \om_1$. Also, the integrands contain the function 
\beq
\wh{f}\pa{\om} = \ex{ -\f{2h}{T} } \pl{a=1}{N} \f{ \s{\la_a+\om-\eta} \s{\la_a-\om-\eta} }
				{ \s{\la_a+\om+\eta}\s{\la_a-\om+\eta}  }  
\enq
as well as the kernel 
\beq
\wh{U}(\om,\om^{\prime})  = \f{ - \ex{ -\f{h}{T} } \s{2\om^{\prime}+\eta} }
			{   \s{\om^{\prime} + \om }\s{\om-\om^{\prime} +\eta}}  
\pl{a=1}{N} \f{ \s{\la_a + \om^{\prime} } \s{\om^{\prime} - \la_a + \eta}  }
				{\s{\om^{\prime}-\la_a} \s{\la_a + \om^{\prime} +\eta} }    \;. 				
\enq
Finally, for any $p$, $\msc{C}_1 \supset \dots \supset \msc{C}_p$ are encased contours such that $\msc{C}_k$, 
for $k=1,\dots,p$ enlaces the roots $\la_1, \dots, \la_N$ but not the ones that are shifted by $\pm \eta$. 

\end{lemme}

We would like to stress that the series \eqref{ecriture F cal N trotter fini} might not be convergent. 
However, the function $\ex{ \mc{F}_N(\{\la_a\}_1^N) }$ is well-defined. More precisely, 
one can take 
\beq
\ex{\mc{F}_N\pa{\{\la_a\}_1^N}} = \pl{p=1}{N}[1+\wh{\mf{a}}(-\la_p)]^{-\f{1}{2}} \cdot 
\paf{ 1-\wh{\mf{a}}\pa{0} }{ 1+\wh{\mf{a}}\pa{0} }^{\f{1}{4}} \cdot  \ddet{N}{\de_{jk}+U_{jk}}
 \; , 
\enq
as its definition. It is readily seen by inspection of the zeroes and poles of the right hand side of this equation, 
that the left hand side does not have singularities at $\la_j = - \la_k$. More details about the precise
mechanism for this definition are given in the proof below.


\Proof

Let $| \kappa |$ be small enough, so that the logarithm of the determinant admits the series expansion 

\beq
\ln\ddet{N}{\de_{jk} + U_{jk} } = \sul{n=0}{+\infty} \f{ \kappa^{2n+1} }{2n+1} 
\sul{  \substack{ p_1,\dots, p_{2n+1} \\ =1} }{ N } U_{p_1p_2}\dots U_{p_{2n+1}p_1}
\; - \;  \sul{n=1}{+\infty} \f{ \kappa^{2n} }{2n} 
\sul{ \substack{ p_1,\dots, p_{2n}  \\ = 1}  }{ N } U_{p_1p_2}\dots U_{p_{2n}p_1}
\enq
We first derive a recurrence equation that allows one to compute the various traces. Namely,
given any $p_1, p_2, p_3 \in \intn{1}{N}$ and agreeing upon the shorthand notation 
$\ov{\la}_{ab}= \la_a+ \la_b$  and $\la_{ab}= \la_a -\la_b$
one gets
\beq
\sul{p_2=1}{N} \; U_{p_1p_2} U_{p_2p_3} \; =  -  \ex{ - \f{ 2 h }{T} }\s{2\la_{p_3}+\eta} 
\f{ \pl{a=1}{N} \s{\la_{p_3a} + \eta} }{ \pl{ \substack{a=1 \\ \not= p_3} }{N} \s{\la_{p_3a} }} \cdot 
\pl{a=1}{N} \f{ \s{ \ov{\la}_{a p_3} } }{ \s{ \ov{\la}_{a p_3} +\eta} }  \mc{S}_{p_1p_3}\;.
\enq
where,  
\bem
\mc{S}_{p_1p_3} = \sul{p_2=1}{N} 
\f{ - \s{2\la_{p_2}+\eta} }
			{ \s{\ov{\la}_{p_2p_3}}\s{\la_{p_2p_3}+\eta}   \s{\ov{\la}_{p_1p_2}}\s{\la_{p_1p_2}+\eta}}  
\f{ \pl{a=1}{N} \s{ \la_{p_2a} + \eta} }{ \pl{ \substack{a=1 \\ \not= p_2} }{N} \s{ \la_{p_2a} }} \cdot 
\pl{a=1}{N} \f{ \s{ \ov{\la}_{a p_2} } }{ \s{ \ov{\la}_{a p_2} +\eta} }  \\
=\Oint{\msc{C}_1}{} \f{ \dd \om }{ 2i\pi } \f{ - \s{2\om+\eta} }
			{ \s{\om+\la_{p_3}}\s{\om - \la_{p_3}+\eta}   \s{\om + \la_{p_1}}\s{\la_{p_1}-\om +\eta}}  
\pl{a=1}{N} \f{ \s{\la_a + \om } \s{\om -\la_a + \eta}  }{\s{\om-\la_a} \s{\la_a + \om +\eta} }  \\
 - \de_{p_1p_3} \f{  1 }{ \s{2\la_{p_1} +\eta} }  
 \f{  \pl{ a\not= p_1 }{N} \s{ \la_{p_1a} } }{ \pl{a=1}{N} \s{ \ov{\la}_{p_1a} } }
\pl{a=1}{N} \f{ \s{ \ov{\la}_{a p_1} - \eta}  }{ \s{ \la_{p_1 a}  -\eta} }   \;.
\end{multline}
Above, the contour $\msc{C}_1$ encircles all solutions $\{\la_a\}_1^N$, but not the ones that are sifted by $\pm \eta$, \textit{ie}
the sets $\{\la_a\pm \eta\}_1^N$. Thus, 
\bem
\sul{p_2=1}{N} \; U_{p_1p_2} U_{p_2p_3} \; = \wh{f}(\la_{p_1})  \de_{p_1p_3} \; + \; 
\Oint{\msc{C}_1}{} \f{ - \ex{ -\f{h}{T} } \wh{U}(\la_{p_1},\om)   \s{2\la_{p_3}+\eta}  }
			{ \s{\om+\la_{p_3}}\s{\om - \la_{p_3}+\eta} } \f{ \dd \om }{ 2i\pi }  \\
\times \f{ \pl{a=1}{N} \s{\la_{p_3a} + \eta} }{ \pl{ \substack{a=1 \\ \not= p_3} }{N} \s{ \la_{p_3a} }} \cdot 
\pl{a=1}{N} \f{ \s{ \ov{\la}_{a p_3} } }{ \s{ \ov{\la}_{a p_3} +\eta} } 
\end{multline}
When dealing with the contour integral-based term, one can compute the remaining sums over $\la_{p_1},\la_{p_3},\la_{p_4}\dots, \la_{p_n}$
by a similar contour integral provided that one successively uses an encased contour $\msc{C}_1\supset \dots \supset \msc{C}_n$. 
 This choice of contour ensures that the poles at $\om_p=-\om_{p+1}$  
do not contribute to the value of the integral. Ultimately, one gets 
\beq
\sul{ \substack{p_1,\dots,p_n \\ =1} }{ N } U_{p_1,p_2}\dots U_{p_n,p_1}= 
\hspace{-2mm} \Oint{\msc{C}_1\supset \dots \supset \msc{C}_n }{}  \hspace{-2mm} 
\pl{p=1}{n} \wh{U}(\om_p, \om_{p+1})  \cdot \f{ \dd^{n}\om }{ \pa{2i\pi}^n }
\; \; \; + \; \sul{ \substack{p_1,\dots, p_{n-2} \\ =1} }{ N } \wh{f}(\la_{p_1}) U_{p_1,p_2}\dots U_{p_{n-2},p_1} \;.
\enq
Where we agree upon $\om_{n+1}\equiv \om_1$. 

The induction can be solved and leads, in the even case, to
\beq
\sul{ \substack{p_1,\dots, p_{2n} \\ =1} }{ N } U_{p_1,p_2}\dots U_{p_{2n},p_1}= 
\sul{k=0}{n-1}\Oint{ \substack{ \msc{C}_1\supset \dots \\ \dots  \supset \msc{C}_{2(n-k)} } }{}  \hspace{-2mm} 
\Big[ \wh{f}\pa{\om_{2(n-k)}} \Big]^k  \pl{p=1}{2\pa{n-k}} \!\! \wh{U}(\om_p, \om_{p+1}) \cdot \!\!
\pl{p=1}{2(n-k)} \f{ \dd \om_p }{ 2i\pi }
\; \; \; + \; \sul{ \substack{p_1 =1} }{ N } \big[ \wh{f}(\la_{p_1}) \big]^n  \;.
\enq
Whereas, in the odd case, 
\beq
\sul{ \substack{p_1,\dots, p_{2n+1} \\ =1} }{ N } U_{p_1,p_2}\dots U_{p_{2n+1},p_1}= 
\sul{k=0}{n-1}\Oint{ \substack{ \msc{C}_1\supset \dots \\ \dots \supset \msc{C}_{2(n-k)+1} } }{}  \hspace{-4mm} 
\Big[ \wh{f}\pa{\om_{2(n-k)+1}} \Big]^{k} \pl{p=1}{2\pa{n-k}+1} \hspace{-2mm}\wh{U}(\om_p, \om_{p+1}) \,  \cdot  \!\!\!
\pl{p=1}{2(n-k)+1} \f{ \dd \om_p }{ 2i\pi }
\; \; \; + \; \sul{ p_1 =1 }{ N } \big[ \wh{f}(\la_{p_1}) \big]^n U_{p_1p_1} \;.
\enq
Yet, 
\beq
\sul{ p_1 =1 }{ N } \Big[ \wh{f}(\la_{p_1})\Big]^n  U_{p_1p_1} = \Oint{\msc{C}}{}  \Big[ \wh{f}\pa{\om} \Big]^n 
\wh{U}\pa{\om,\om} \f{\dd \om}{2i\pi } 
\; + \; \f{ \ex{-\f{h}{T}} }{2}  \Big[\wh{f}\pa{0}\Big]^n \pl{a=1}{N} \f{ \s{\la_a-\eta} }{ \s{\eta+\la_a} } \;. 
\enq
It then remains to observe that 
\beq
\wh{f}(0) = \big[ \, \wh{\mf{a}}(0)\,  \big]^2 \hspace{1cm} \e{and}  \hspace{1cm}  \wh{f} ( \la_{p_1} ) = - \wh{\mf{a}}(-\la_{p_{1}}) \;. 
\enq
what leads to 
\beq
\sul{ \substack{p_1,\dots, p_{2n+1} \\ =1} }{ N } U_{p_1,p_2}\dots U_{p_{2n+1},p_1}= 
\sul{k=0}{n}\Oint{ \substack{ \msc{C}_1\supset \dots \\ \dots \supset \msc{C}_{2(n-k)+1} } }{}  \hspace{-2mm} 
\Big[ \wh{f}\pa{\om_{2(n-k)+1}} \Big]^k  \pl{p=1}{2\pa{n-k}+1} \hspace{-2mm} \wh{U}(\om_p, \om_{p+1}) \cdot \f{ \dd^{2\pa{n-k}+1}\om }{ \pa{2i\pi}^{2\pa{n-k}+1} }
\; \; \; + \f{1}{2} \cdot \big[ \wh{\mf{a}}\pa{0} \big]^{2n+1} \;.
\enq
Finally after inserting the two formulae in the trace-like expansion for the determinant, weighting by 
appropriate powers of $\kappa$ and using 
\beq
\ln(1+x)-\ln(1-x) =  \sul{n=1}{+\infty}  \f{ 1+\pa{-1}^{n-1} }{n} x^n = 2 \sul{n=0}{+\infty} \f{x^{2n+1}}{2n+1} \;, 
\enq
we are led to the representation 
\beq
\ddet{N}{\de_{jk}+ \kappa U_{jk}}  \; = \;  \pl{p=1}{N}[1 + \kappa \wh{\mf{a}}(-\la_p)]^{\f{1}{2}} \cdot 
\paf{ 1 + \kappa \wh{\mf{a}}\pa{0} }{ 1 - \kappa  \wh{\mf{a}}\pa{0} }^{\f{1}{4}} \cdot 
\ex{\mc{F}_N^{(\kappa)} \pa{\{\la_a\}_1^N}} \; , 
\enq
where 
\bem
\mc{F}_N^{(\kappa)} \pa{\{\la_a\}_1^N} = 
\sul{ k = 0  }{+\infty}  \kappa^{2k+1} \Oint{ \msc{C}_1\supset\dots \supset \msc{C}_{2k+1} }{}
\hspace{-2mm} 
\sul{ n=k }{ +\infty } \f{ \Big[ \kappa^2 \wh{f}\!\pa{\om_{2k+1}} \Big]^{n-k} }{2n+1}  \; \cdot  \; 
 \pl{p=1}{2k+1}\wh{U}(\om_p,\om_{p+1}) \; \f{ \dd^{2k+1} \om }{ \pa{2i\pi}^{2k+1} }  \\
\; -\; \sul{  k = 1   }{+\infty}  \kappa^{2k}  \Oint{ \msc{C}_1\supset\dots \supset \msc{C}_{2k} }{}
\sul{n=k}{+\infty} \f{ \Big[ \kappa^2 \wh{f}\!\pa{\om_{2k}} \Big]^{n-k} }{2n} \; \cdot \;  \pl{p=1}{2k} \wh{U}(\om_p,\om_{p+1}) \; \f{ \dd^{2k} \om }{ \pa{2i\pi}^{2k} } \; . 
\label{ecriture F cal N kappa trotter fini}
\end{multline}

Since the kernel $\wh{U}$ and the function $\wh{f}$ are uniformly bounded on the sequence of integration contours, 
it is readily seen that the above series convegres for $|\kappa|$ small enough.   One thus has an equality between 
holomorphic functions of $\kappa$, in some neighborhood of $\kappa=0$
\beq
\ex{\mc{F}_N^{(\kappa)} \pa{\{\la_a\}_1^N}}  \; = \;  \pl{p=1}{N}[1 + \kappa \wh{\mf{a}}(-\la_p)]^{-\f{1}{2}} \cdot 
\paf{ 1 - \kappa \wh{\mf{a}}\pa{0} }{ 1 + \kappa  \wh{\mf{a}}\pa{0} }^{\f{1}{4}} \cdot 
\ddet{N}{\de_{jk}+ \kappa U_{jk}}   \; . 
\enq
The function in the \textit{rhs} is an analytic function on the closed unit disk. This thus ensures that the 
function $\kappa \mapsto \ex{\mc{F}_N^{(\kappa)} \pa{\{\la_a\}_1^N}}$ is analytic on the closed unit disk. 
It can thus be continued analytically from a neighborhood of $\kappa=0$ -where it is definitely licit to define
it through the formula \eqref{ecriture F cal N kappa trotter fini}- up to $\kappa=1$. If the series 
\eqref{ecriture F cal N kappa trotter fini}  is convergent  at $\kappa=1$, then we get the representation
\eqref{lemme factorisation det ac Fn} with $\mc{F}_N \pa{ \{\la_a\}_1^N }$ given by \eqref{ecriture F cal N trotter fini}.
Else, the representation \eqref{lemme factorisation det ac Fn} still holds but the definition of 
$\exp\big\{ \mc{F}_N \pa{\{\la_a\}_1^N} \big\}$ is to be understood in the sense of analytic continuation
from an open neighborhood of $\kappa=0$ up to $\kappa=1$ of the function  
$ \kappa \mapsto \exp \Big\{ \mc{F}_N^{(\kappa)} \pa{\{\la_a\}_1^N} \Big\}$ 
with $\mc{F}_N^{(\kappa)} \pa{\{\la_a\}_1^N}$ defined, for $| \kappa|$ small enough, by the  series
\eqref{ecriture F cal N kappa trotter fini}. 
\qed

\subsection{The representation at finite Trotter number}

By putting  together all of the previously obtained formulae, one arrives to the below representation for the 
finite Trotter number approximant of the surface free energy:
\bem
\ex{- \f{ f^{\pa{N}}_{\e{surf}} }{ T } }  =   \ex{\f{Nh}{T}}
\pl{ a=1 }{ N } \f{ \s{\xi_- + \la_a }  \s{\xi_+ + \la_a} }{ \s{\xi_-} \s{\xi_+} }  \cdot \pl{a=1}{N} \f{ \s{\eta} }{ \s{2\la_a +\eta} } \cdot 
\pl{a=1}{N} \f{ \s{-2\tf{\be}{N}}  }{  \s{2\la_a}    }  \\
\times   \pl{a,b=1}{N} \f{ \s{\la_a + \la_b +\eta} }{ \s{\la_a-\la_b +\eta} } \; \cdot \; 
\Bigg\{ \f{ \prod_{a \not=b }^{N} \s{\la_a-\la_b} }
{  \pl{a=1}{N} \tf{ \wh{\mf{a}}^{\, \prime}(\la_a) }{ \wh{\mf{a}}(\la_a) } }  \Bigg\}
\cdot \Bigg\{ \f{ \prod_{a=1}^{N} \big[ 1+\wh{\mf{a}}(-\la_a)\big] }{ \prod_{a,b}^{N}\s{\la_a+\la_b} }  \Bigg\} 
\paf{1+ \wh{\mf{a}}(0) }{ 1-\wh{\mf{a}}(0) }^{\f{1}{2}} \\
\times \paf{  \s{2\eta - 2\tf{\be}{N} } }{ \s{2\eta} }^{N}  \cdot \f{  \ex{2\mc{F}_N\pa{\{\la_a\}_1^N}}  }{  \ddet{\msc{C}}{I+\ov{K} } } \;.
\label{equation representation energie libre finite Trotter}
\end{multline}

\section{Taking the infinite Trotter number limit}
\label{Section Taking infinite Trotter number}

\subsection{Rewriting of the double products}

The representation \eqref{equation representation energie libre finite Trotter}
constitutes a good starting point for taking the infinite Trotter number limit. For this, as it is customary in the QTM approach, 
one should represent all simple and double products over the Bethe roots for the largest eigenvalue
of the QTM in terms of contour integrals involving the function $\wh{\mf{a}}$ defined in \eqref{definition counting function}. 
Once such a representation is obtained, 
the infinite Trotter number limit can be easily taken. The purpose of the proposition below is to provide such a contour 
integral representation. 
\begin{prop}
The below products admit the alternative representations
\bem
\pl{ \substack{ a,b \\ a\not= b}}{ N} \s{\la_a-\la_b}   = 
\pl{b=1}{N} \f{  \wh{\mf{a}}^{ \, \prime} (\la_b) }{ \wh{\mf{a}}\,(\la_b) }  \, \cdot \,
\paa{ \f{ \s{\eta} \s{-2\tf{\be}{N}} }{ \s{ \eta -2\tf{\be}{N} } } }^{N^2} \cdot 
\pl{a=1}{N} \paa{  \f{\s{\eta-\la_a - \tf{\be}{N} } }{ \s{\eta+\la_a + \tf{\be}{N} }  } }^N  \; \cdot \; 
\bigg( \f{ \ex{-\f{h}{T}} }{  \big( 1+\ex{-\f{h}{T}} \big)^2  } \bigg)^N \\
\exp\Bigg\{ -2N \Oint{ \msc{C} }{} \hspace{-1mm} \f{\dd \om }{ 2i\pi }  \ln[1+\wh{\mf{a}}\pa{\om}] \coth\pa{\om +\tf{\be}{N}} 
\; - \;  \Oint{\msc{C}}{} \hspace{-1mm} \f{\dd \om}{2i\pi} \Oint{\msc{C}^{\prime}\subset \msc{C}}{} \hspace{-2mm} 
\f{\dd \om^{\prime}}{2i\pi} 
\coth^{\prime}(\om^{\prime} - \om)   \ln[1+\wh{\mf{a}}\pa{\om}] \ln[1+\wh{\mf{a}}(\om^{\prime})] \Bigg\} \;.
\label{first factorization formula}
\end{multline}
Also 
\bem
\pl{  a,b }{ N} \s{\la_a+\la_b}   = 
\pac{ \s{-2\tf{\be}{N}}  }^{N^2} \cdot \pl{b=1}{N} \f{ 1+ \wh{\mf{a}}\pa{-\la_b}  }{ \big(  1+\ex{-\f{h}{T}}   \big)^2} 
\exp\Bigg\{ -2N \Oint{ \msc{C} }{} \hspace{-1mm} \f{\dd \om }{ 2i\pi }  \ln[1+\wh{\mf{a}}\pa{\om}] \coth\pa{\om - \tf{\be}{N}} \Bigg\} 
\\
\; \times  \; 
 \exp\Bigg\{ \Oint{\msc{C}}{} \hspace{-1mm} \f{\dd \om}{2i\pi} \Oint{\msc{C}^{\prime}\subset \msc{C}}{} \hspace{-2mm} 
\f{\dd \om^{\prime}}{2i\pi} 
\coth^{\prime}(\om+\om^{\prime})   \ln[1+\wh{\mf{a}}\pa{\om}]\ln[1+\wh{\mf{a}}(\om^{\prime})]  \Bigg\} \;.
\label{equation representation produit double avec plus}
\end{multline}
\beq
\pl{a=1}{N} \f{\s{2\la_a}}{ \s{-2\tf{\be}{N}} } =   \f{  1+\wh{\mf{a}}\pa{0} }{ 1+\ex{-\f{h}{T}}  } \cdot 
\exp\bigg\{  -2 \Oint{\msc{C}}{} \coth\pa{2\om}  \ln \big[ 1+\wh{\mf{a}}\pa{\om} \big] \f{\dd \om}{2i\pi} \bigg\}  \;. 
\enq
Finally, 
\bem
\pl{a,b=1}{N} \f{ \s{\la_a+\la_b+\eta} }{ \s{\la_a-\la_b+\eta} } =
\paa{ \f{ \s{\eta-2\tf{\be}{N}}}{\s{\eta}} }^{N^{2}}  
\pl{a=1}{N} \paa{  \f{ \s{\eta+\la_a + \tf{\be}{N} }  }{\s{\eta-\la_a - \tf{\be}{N} } } }^N \\
\times \exp\Bigg\{ -2N \Oint{ \msc{C} }{} \hspace{-1mm} \f{\dd \om }{ 2i\pi }  \ln\pa{1+\wh{\mf{a}}\pa{\om}} 
\pac{  \coth\pa{\om -\tf{\be}{N} +\eta} \, - \,   \coth\pa{\om +\tf{\be}{N} +\eta}   }  \Bigg\} \\
\times \exp\Bigg\{   \Oint{ \msc{C}  \times  \msc{C} }{} \hspace{-1mm} \f{\dd \om}{2i\pi} \f{\dd \om^{\prime}}{2i\pi} 
  \ln\pa{1+\wh{\mf{a}}\pa{\om}}\ln\pa{1+\wh{\mf{a}}(\om^{\prime})} 
 \Dp{\om} \Dp{\om^{\prime}} \ln\paf{ \s{\om+\om^{\prime}+\eta} }{ \s{\om-\om^{\prime}+\eta} }    \Bigg\} \;.
\label{formule representation produit double regulier}
\end{multline}
The contour $\msc{C}$ is the one occurring in the non-linear integral equation \eqref{ecriture NLIE fonction a hat regime general}. 
It is such 
that the poles of the integrand at $\om=\pm \tf{\be}{N}-\eta$, resp. $\om^{\prime}+\om +\eta=0$, are all located outside of 
$\msc{C}$, or $\msc{C}\times \msc{C}$, depending on the integral of interest. 
\end{prop}

We do stress that the encased contour $\msc{C}^{\prime}\supset\msc{C}$ corresponds to a small deformation 
of $\msc{C}$ such that the poles in $\om^{\prime} = \pm \om$ are all always outside of $\msc{C}^{\prime}$.

\Proof 

The representation for the fourth product is the easiest to obtain. Namely,
\bem
\sul{a,b=1}{N} \ln \pac{  \f{ \s{\la_a+\la_b +\eta} }{   \s{\la_a-\la_b +\eta} } } =
\sul{a=1}{N} \Oint{ \msc{C} }{} 
\ln \paf{ \s{\om+\la_a +\eta} }{   \s{\om-\la_a +\eta} }  \; \cdot \f{ \wh{\mf{a}}^{\, \prime}(\om) }{ 1 + \wh{\mf{a}}\pa{\om} }
\f{\dd \om }{ 2i\pi } 
\; + \; 
N \sul{a=1}{N} \ln \paf{ \s{-\tf{\be}{N}+\la_a +\eta} }{   \s{-\tf{\be}{N}-\la_a +\eta} } \\
=  \Oint{ \msc{C} }{} 
\ln \paf{ \s{\om+\om^{\prime} +\eta} }{   \s{\om-\om^{\prime} +\eta} }  \; \cdot \f{ \wh{\mf{a}}^{\prime}(\om) }{ 1 + \wh{\mf{a}}(\om) }
\f{ \wh{\mf{a}}^{\,  \prime}(\om^{\prime}) }{ 1 + \wh{\mf{a}}(\om^{\prime} ) }  \f{\dd \om \dd \om^{\prime} }{ \pa{2i\pi}^2 } 
\; + \; 
 N \Oint{ \msc{C} }{} 
\ln \paf{ \s{\om - \tf{\be}{N} + \eta} }{   \s{\om +  \tf{\be}{N}  +\eta} }  
\; \cdot \f{ \wh{\mf{a}}^{ \,  \prime}\!\pa{\om} }{ 1 + \wh{\mf{a}}\pa{\om} } \f{\dd \om }{ 2i\pi }  \\
\; + \; 
N \sul{a=1}{N} \ln \paf{ \s{+\tf{\be}{N}+\la_a +\eta} }{   \s{-\tf{\be}{N}-\la_a +\eta} }  
\; + \;  N^{2} \ln \paf{ \s{\eta-2\tf{\be}{N} } }{ \s{\eta} }  
\; + \; 
 N \Oint{ \msc{C} }{} 
\ln \paf{ \s{\om + \eta - \tf{\be}{N} } }{   \s{\om +  \tf{\be}{N}  +\eta} }  
\; \cdot \f{ \wh{\mf{a}}^{ \,  \prime}\!\pa{\om} }{ 1 + \wh{\mf{a}}\pa{\om} } \f{\dd \om }{ 2i\pi }  \;.
\end{multline}
Formula \eqref{formule representation produit double regulier} then follows after an integration by parts. 
Doing so is licit in as much as $\ln^{\prime}\big[1+\wh{\mf{a}}\pa{\om} \big]$ has a vanishing monodromy along $\msc{C}$.
Note that the contour $\msc{C}$ is chosen precisely so that the poles of the integrand in respect to $\om$ (resp. $\om^{\prime}$)
at $\om  \pm \om^{\prime} + \eta = 0 \; \e{mod}[i\pi]$ are all located outside of $\msc{C}$, this for any $\om^{\prime} \in \msc{C}$ 
(resp. $\om \in \msc{C}$).

Next, we consider 
\beq
\mc{S}_1\pa{u} = \sul{a,b=1}{N} \ln \pac{ \s{\la_a+\la_b+u} }\;.
\label{ecriture form depart S1}
\enq
It is then easy to see that, for $u$ small enough, 
\bem
\mc{S}_1^{\prime}\pa{u} = \sul{a=1}{N} \Oint{\msc{C}}{} \f{\dd \om}{2i\pi}  \coth\pa{\la_a+\om+u} \cdot
 \f{  \wh{\mf{a}}^{ \,  \prime}\pa{\om} }{ 1+ \wh{\mf{a}}\pa{\om} } \; + \; 
\sul{a=1}{N} N \coth\pa{\la_a - \tf{\be}{N} + u}   
\; - \; \sul{a=1}{N}   \f{  \wh{\mf{a}}^{ \,  \prime}\pa{-\la_a-u} }{ 1+\wh{\mf{a}}\pa{-\la_a-u} } \\
= \Oint{\msc{C}}{} \hspace{-1mm} \f{\dd \om}{2i\pi} \Oint{\msc{C}^{\prime}\subset \msc{C}}{} \hspace{-2mm} 
\f{\dd \om^{\prime}}{2i\pi}  \coth(\om+\om^{\prime}+u) \cdot
 \f{  \wh{\mf{a}}^{ \,  \prime}\pa{\om} }{ 1+\wh{\mf{a}}\pa{\om} }  
 \f{  \wh{\mf{a}}^{ \,  \prime}(\om^{\prime}) } { 1+\wh{\mf{a}}(\om^{\prime}) } \; + \; 
2N \Oint{\msc{C}}{} \f{\dd \om}{2i\pi}  \coth\pa{\om+u-\tf{\be}{N}} \cdot
 \f{  \wh{\mf{a}}^{ \,  \prime}\pa{\om} }{ 1+\wh{\mf{a}}\pa{\om} }  \\
 \; + \; N^2 \coth\pa{u-2\tf{\be}{N}} \; - \;  N  \f{  \wh{\mf{a}}^{ \,  \prime}\pa{\tf{\be}{N}-u} }{ 1+\wh{\mf{a}}\pa{\tf{\be}{N}-u} }  
\; - \; \sul{a=1}{N}   \f{  \wh{\mf{a}}^{ \,  \prime}\pa{-\la_a-u} }{ 1+\wh{\mf{a}}\pa{-\la_a-u} } \;.
\end{multline}
Above, the contour $\msc{C}^{\prime}$ is such that it does not encircle the points $-\om -u$, with $\om \in \msc{C}$. 
Taking the integral over $u$ and carrying out integrations by parts, we obtain 
\bem
\mc{S}_{1}(u) =   \Oint{\msc{C}}{} \hspace{-1mm} \f{\dd \om}{2i\pi} \Oint{\msc{C}^{\prime}\subset \msc{C}}{} \hspace{-2mm} 
\f{\dd \om^{\prime}}{2i\pi}  \coth^{\prime}(\om+\om^{\prime}+u) \cdot
\ln[ 1+\wh{\mf{a}}\pa{\om} ]  \ln[ 1+\wh{\mf{a}}(\om^{\prime}) ] \; - \; 
2N \Oint{\msc{C}}{} \f{\dd \om}{2i\pi}  \coth\pa{\om+u-\tf{\be}{N}} \cdot
\ln [ 1+\wh{\mf{a}}\pa{\om} ]  \\
\; + \; C_1  \; + \; N^2 \ln \big[ \sinh  \pa{u-2\tf{\be}{N}} \big]
\; + \;  N  \ln [ 1+\wh{\mf{a}}\pa{\tf{\be}{N}-u} ]  
\; + \; \sul{a=1}{N}   \ln[ 1+\wh{\mf{a}}\pa{-\la_a-u} ] \;.
\label{ecriture S1 rep alternative ac cste}
\end{multline}
There $C_1$ is an integration constant that ought to be fixed. Taking $u\tend +\infty $ in the original representation 
\eqref{ecriture form depart S1}, we get that $\mc{S}_1(u) = N^2 \big[u - \ln(2) \big] \, +  \,  2N\sum_{a=1}^{N} \la_a \; +\; \e{o}(1)$. 
The same limit can be taken on the level of the representation \eqref{ecriture S1 rep alternative ac cste}. 
Indeed, one can always send $u \tend +\infty$ by deforming it (and the contours $\msc{C}, \msc{C}^{\prime}$ if necessary) 
from $u=0$ along a path that keeps the properties of the contours entering in the double integral unaltered. 
Then, it remains to use that, for any fixed $z$, $\wh{\mf{a}}(z-u) = \ex{-\f{h}{T}}(1+\e{o}(1))$, whereas
\beq
- \Oint{ \msc{C} }{} \ln[1+\wh{\mf{a}}(\om)] \cdot \f{ \dd \om }{  2i\pi } 
\; = \;  \Oint{ \msc{C} }{} \om \f{ \wh{\mf{a}}^{\, \prime}(\om) }{1+\wh{\mf{a}}(\om)} \cdot \f{ \dd \om }{  2i\pi }
\; = \; \be \; + \; \sul{a=1}{N}\la_a   \;.  
\enq
These, in the $u\tend +\infty$ limit, lead to 
\beq
\mc{S}_1(u) \; = \; N^2 \big( u - \ln 2 \big) \; + \; 2N \sul{a=1}{N} \la_a \; + \; 
2N \ln[1+\ex{-\f{h}{T}}] \; +\; C_1 \; + \e{o}(1) \;. 
\enq
Thus, in order that the alternative representation \eqref{ecriture S1 rep alternative ac cste} produces
the correct large $u$ asymptotics, one has to set $C_1 = - 2N \ln[1+\ex{-\f{h}{T}}]$. 
Further, setting $u=0$ in \eqref{ecriture S1 rep alternative ac cste} and  using that $\wh{\mf{a}}\pa{\tf{\be}{N}}=0$ 
leads to equation \eqref{equation representation produit double avec plus}. 

We now establish the factorization formula \eqref{first factorization formula}. For this, we introduce
\beq
\mc{S}_2\pa{u} = \sul{a,b=1}{N} \ln \pac{\s{\la_a-\la_b+u}} \qquad \e{so} \; \e{that} \qquad 
\pl{a\not= b}{N} \s{\la_a-\la_b} = \lim_{u\tend 0} \bigg\{ \f{ \ex{\mc{S}_2\pa{u}} }{\sinh^N(u) }  \bigg\} \;. 
\label{definition S2}
\enq
Then, 
\bem
\mc{S}_2^{\prime}\pa{u} = \sul{a=1}{N} \Oint{\msc{C}}{} \f{\dd \om}{2i\pi}  \coth\pa{\la_a-\om+u} \cdot
 \f{  \wh{\mf{a}}^{\, \prime}\pa{\om} }{ 1+\wh{\mf{a}}\pa{\om} } \; + \; 
\sul{a=1}{N} N \coth\pa{\la_a + \tf{\be}{N} + u}   
\; + \; \sul{a=1}{N}   \f{  \wh{\mf{a}}^{ \, \prime}\pa{\la_a+u} }{ 1+\wh{\mf{a}}\pa{\la_a+u} } \\
= \Oint{\msc{C}}{} \hspace{-1mm} \f{\dd \om}{2i\pi} \Oint{\msc{C}^{\prime}\subset \msc{C}}{} \hspace{-2mm} 
\f{\dd \om^{\prime}}{2i\pi}  \coth(\om^{\prime}-\om+u) \cdot
 \f{  \wh{\mf{a}}^{\, \prime}\pa{\om} }{ 1+\wh{\mf{a}}\pa{\om} }  \f{  \wh{\mf{a}}^{\, \prime}(\om^{\prime}) }{ 1+ \wh{\mf{a}}(\om^{\prime}) } 
\; + \; \sul{a=1}{N}   \f{  \wh{\mf{a}}^{\, \prime}\pa{\la_a+u} }{ 1+\wh{\mf{a}}\pa{\la_a+u} }  \; + \; N^2 \coth\pa{u} \\
\; + \; N \Oint{\msc{C}}{} \f{\dd \om}{2i\pi}  \Big[ \coth\pa{\om+u+\tf{\be}{N}} - \coth\pa{\om+\tf{\be}{N}-u}  \Big]\cdot
 \f{  \wh{\mf{a}}^{\, \prime}\pa{\om} }{ 1+\wh{\mf{a}}\pa{\om} }  
  \; - \;  N  \f{  \wh{\mf{a}}^{\, \prime}\pa{-\tf{\be}{N}-u} }{ 1+\wh{\mf{a}}\pa{-\tf{\be}{N}-u} }  
 \;.
\end{multline}
Here, the contour $\msc{C}^{\prime}$ is such that $\om-u$ lies outside of $\msc{C}^{\prime}$ for any $\om \in \msc{C}$. 
Thus, taking the anti-derivative in respect to $u$ and then carrying out integrations by parts, one gets 
\bem
\mc{S}_2\pa{u} = 
- \Oint{\msc{C}}{} \hspace{-1mm} \f{\dd \om}{2i\pi} \Oint{\msc{C}^{\prime}\subset \msc{C}}{} \hspace{-2mm} 
\f{\dd \om^{\prime}}{2i\pi}  \coth^{\prime}(\om^{\prime}-\om+u) \cdot \ln \big[ 1+\wh{\mf{a}}(\om^{\prime}) \big]
 \ln \big[ 1+\wh{\mf{a}}(\om) \big]
\; + \; \sul{a=1}{N}    \ln \big[ 1+\wh{\mf{a}}\pa{\la_a+u} \big]   \; + \; N^2 \ln [\s{u}] \\
+ \; C_2  \; + \;  N  \ln \big[ 1+\wh{\mf{a}}\pa{-\tf{\be}{N}-u} \big] 
\; - \; N \Oint{\msc{C}}{}  \Big[ \coth\pa{\om+u+\tf{\be}{N}} + \coth\pa{\om+\tf{\be}{N}-u}  \Big]\cdot
\ln \big[ 1 + \wh{\mf{a}}\pa{\om} \big] \f{\dd \om}{2i\pi}   \;.
\end{multline}
The integration constant $C_2$ can be fixed by carrying out much the same reasoning as for $\mc{S}_1(u)$. 
One gets that $C_2 = -2N \ln[1+\ex{-\f{h}{T}} ]$. It then remains to take the exponent and
compute the $u\tend 0$ limit as in \eqref{definition S2}. This can be done by observing that 
\beq
\lim_{u \tend 0} \bigg\{ \pl{a=1}{N} \f{ 1+\wh{\mf{a}}\pa{\la_a + u } }{  \s{u}} \bigg\} =
 \pl{a=1}{N} \wh{\mf{a}}^{\, \prime}\!(\la_a)   \; ,
\enq
as well as 
\bem
\sinh^N\pa{u} \pac{ 1 + \wh{\mf{a}}\pa{-u -\tf{\be}{N}}} \underset{u\tend 0}{ \displaystyle{\sim} }
\ex{-\f{h}{T}}\, \pl{a=1}{N} \f{ \s{-\la_a+\eta-\tf{\be}{N}} }{ \s{-\la_a-\eta-\tf{\be}{N}} } 
\cdot \bigg\{ \f{\s{-\eta}\s{u}}{\s{\eta-2\tf{\be}{N}}\s{-u} } \bigg\}^N \; \cdot \;  \big[ \s{-2\tf{\be}{N}} \big]^N \\
\underset{u\tend 0}{ \displaystyle{\tend} } 
\pa{-1}^N \ex{-\f{h}{T}} \pl{a=1}{N} \f{ \s{\eta-\la_a-\tf{\be}{N}} }{ \s{\eta+ \la_a+\tf{\be}{N}} } 
\cdot \bigg\{ \f{\s{\eta}}{\s{\eta-2\tf{\be}{N}} } \bigg\}^N  \cdot  \big[ \s{-2\tf{\be}{N}} \big]^N  \;.
\end{multline}

Finally, we compute the last product. Setting $\mc{S}_3\pa{u} = \sul{a=1}{N} \ln \pac{ \s{2\la_a+u} } $ we get that :
\beq
\mc{S}^{\prime}_3\!\pa{u} = \Oint{\msc{C}}{} \coth\pa{2\om + u} 
 \f{  \wh{\mf{a}}^{\, \prime}\pa{\om} }{ 1+\wh{\mf{a}}\pa{\om} } \f{\dd \om}{2i\pi}
\; -\; \f{1}{2} \cdot \f{  \wh{\mf{a}}^{\prime}\pa{-\tf{u}{2}} }{ 1+\wh{\mf{a}}\pa{-\tf{u}{2}} } + N \coth \pa{-2\tf{\be}{N} + u } \; . 
\enq
Hence,  after an integration in respect to $u$, 
\beq
\mc{S}_3\!\pa{u} = -2 \Oint{\msc{C}}{} \coth\pa{2\om + u}  \ln \big[ 1+\wh{\mf{a}}\pa{\om} \big] \f{\dd \om}{2i\pi}
\;+ \; \ln \big[ 1+\wh{\mf{a}}\pa{-\tf{u}{2}} \big]  + N \ln \sinh \pac{ \pa{-2\tf{\be}{N} + u } } \; + \; C_3\;. 
\enq
The integration constant is fixed by comparing the $u\tend +\infty$ asymptotics of the two representations for $\mc{S}_3(u)$. 
One gets, on the one hand, 
\beq
\mc{S}_3(u)= 2 \sul{a=1}{N} \la_a  \; + \; N(u-\ln 2) \; + \; \e{o}(1) \;, 
\enq
whereas, on the other hand 
\beq
 \mc{S}_3(u)= 2 \Big( \sum_{a=1}^{N} \la_a \; + \; \be \Big)  \;+\; \ln(1+\ex{-\f{h}{T}}) 
  \; + \; N(u-\ln 2-2 \tf{\be}{N}) \; + \; C_3 \; + \; \e{o}(1) \; . 
\enq
This implies that $C_3 = - \ln(1+\ex{-\f{h}{T}})$. \qed

\subsection{A smooth representation}

Inserting the previous formulae into \eqref{equation representation energie libre finite Trotter} leads to 
\beq
\ex{- \f{ f_{\e{surf}}^{\pa{N}} }{ T } } =
\paf{ \s{\eta} \s{2\eta-2 \tf{\be}{N}}   }
{ \s{\eta-2\tf{\be}{N}} \s{2\eta}    }^{N}
\f{ \exp\Big\{- \tf{  \big[  \wh{\mc{B}}(\xi_+)  \; +\;  \wh{\mc{B}}(\xi_-) \big] }{T} \Big\} }
 {\sqrt{ 1-[\wh{\mf{a}}(0)]^2}   }    \cdot 
 \f{   1\; + \; \ex{-\tf{h}{T}}   }{ \ddet{\msc{C}}{I+\ov{K}}}  
  \exp\Big\{ 2 \mc{F}_{N}\pa{ \{\la_a\}_1^N }  \; + \; \wh{ \mc{I}} \, \Big\} \;. 
\enq

Above, we have set
\beq
\wh{\mc{B}}(\xi) \; = \;   -  N \ln \Big(  \f{ \s{\xi - \tf{\be}{N}} }{ \s{\xi} } \Big) 
- T \de_{\xi} \ln \pa{ 1+\wh{\mf{a}}(-\xi) } \; + \;
T \Oint{ \msc{C} }{}  \ln\pa{1+\wh{\mf{a}}\pa{\om}} \coth\pa{\om+\xi}  \f{\dd \om}{2i\pi}  
\enq
where we agree that 
\beq
\de_{\xi}  = \left\{ \ba{cc} 
		1 & \e{if}\;   -\xi  \; \e{is} \; \e{located} \; \e{inside} \; \e{of} \; \e{the} \; \e{contour} \; \; \msc{C} \\
		     0  & \e{otherwise} \ea \right.  \;. 
\label{definition fonction indicatrice  delta xi}
\enq
Furthermore, we have introduced 
\bem
\wh{\mc{I}} \; = \; 
 2 \Oint{\msc{C}}{} \hspace{-1mm} \f{\dd \om}{2i\pi} 
  \ln\pa{1+\wh{\mf{a}}\pa{\om}} \Big[\coth\pa{2\om} + \coth\pa{2\om +\eta} \Big]   
 \\
 -2N \Oint{ \msc{C} }{} \hspace{-1mm} \f{\dd \om }{ 2i\pi }  \ln\pa{1+\wh{\mf{a}}\pa{\om}} 
\pac{ \coth\pa{\om +\tf{\be}{N} } \, - \,   \coth\pa{\om -\tf{\be}{N} } \; + \; \coth\pa{\om -\tf{\be}{N} +\eta} \, - \,   \coth\pa{\om +\tf{\be}{N} +\eta}   }  \\
\; + \;    \Oint{\msc{C}}{} \hspace{-1mm} \f{\dd \om}{2i\pi}   
\hspace{-1mm} \Oint{\msc{C}^{\prime} \subset \msc{C}}{} \hspace{-1mm}  \f{\dd \om^{\prime}}{2i\pi} 
  \ln\pa{1+\wh{\mf{a}}\pa{\om}}\ln\pa{1+\wh{\mf{a}}(\om^{\prime})} \bigg[ 
- \, \coth^{\prime}(\om-\om^{\prime})
 \, +\, \Dp{\om} \Dp{\om^{\prime}} \ln\Big[\f{ \s{\om+\om^{\prime}+\eta} }{ \s{\om-\om^{\prime}+\eta} }  \Big]  
  \, - \,  \coth^{\prime}(\om+\om^{\prime})  
\bigg]   
 \;, 
\end{multline}
and we insist that the contour $\msc{C}$ arizing in the second line is such that the points $ i \tf{\pi}{2} $, 
$-\tf{\eta}{2}$ mod$[i\pi]$ are outside of $\msc{C}$. 

Finally, the function $\mc{F}_N\pa{ \{\la_a\}_1^N}$ is given as in \eqref{ecriture F cal N trotter fini}
with the sole difference that now, the kernel $\wh{U}$, is given by
\bem
\wh{U}(\om,\om^{\prime}) = \f{ - \ex{-\f{h}{T}} \s{2\om^{\prime}+\eta} }{ \s{\om+\om^{\prime}} \s{\om-\om^{\prime}-\eta}}
\cdot \bigg[ \f{ \s{\om^{\prime}- \tf{\be }{N}} \s{\om^{\prime}+ \tf{\be }{N}+\eta} }
	{ \s{\om^{\prime}+ \tf{\be }{N}} \s{\om^{\prime} - \tf{\be }{N}+\eta} }  \bigg]^N \\ 
\exp\Bigg\{  - \Oint{ \msc{C}_U }{} \hspace{-1mm} \f{\dd \tau }{ 2i\pi }  \ln\big[ 1+\wh{\mf{a}}\pa{\tau} \big] 
\Big[ \coth(\tau +\om^{\prime} ) \, + \, \coth( \om^{\prime} -\tau )
 \, -  \, \coth(\tau +\om^{\prime} +\eta) \, - \, \coth(\om^{\prime} -\tau +\eta)  \Big] \Bigg\}	\;, 
\end{multline}
whereas the function $\wh{f}\pa{\om}$ is recast in the form
\bem
\wh{f}\pa{\om} = \ex{ -\f{2h}{T} } \bigg[ \f{ \s{\om- \tf{\be }{N}-\eta} \s{\om+ \tf{\be }{N}+\eta} }
	{ \s{\om+ \tf{\be }{N} - \eta } \s{\om - \tf{\be }{N}+\eta} }  \bigg]^N \\ 
\exp\Bigg\{  - \Oint{ \msc{C}_U }{} \hspace{-1mm} \f{\dd \tau }{ 2i\pi }  \ln\big[ 1+\wh{\mf{a}}\pa{\tau} \big] 
\Big[ \coth(\tau +\om -\eta) \, + \, \coth(\om -\tau -\eta)
 \, -  \, \coth(\tau +\om +\eta) \, - \, \coth(\om -\tau +\eta)  \Big]  \Bigg\}	\;. 
\end{multline}
The contour $\msc{C}_U$ is such that given any $\om^{\prime}, \om \in \msc{C}_p$,
where $\msc{C}_p$ refers to any of the encasted contours introduced in 
lemma \ref{Lemme factorisation determinant I+U}, the points
\beq
 \pm \om^{\prime},\quad  \pm (\om^{\prime}+\eta) \quad \pm (\om^{\prime}-\eta)
\enq
 are not surrounded by $\msc{C}_U$. The latter loop encircles however the points $\{\la_a\}_1^N$ as well as the origin.

\subsection{Representation for the surface free energy}

In order to take the infinite Trotter number limit, one should first send $N\tend +\infty$ on the level 
of the non-linear integral equation 
for the function $\wh{\mf{a}}$. There $N$ only appears in the driving term which has a well defined $N\tend +\infty$
limit. It thus appears highly plausible that $\wh{\mf{a}} \tend \mf{a}$ where $\mf{a}$ is the solution to 
\beq
\ln \mf{a}(\om)  \; =   \; 
-\f{h}{T} \; + \; \f{2 J \s{\eta}}{T} \big\{ \coth(\om+\eta) \; - \; \coth(\om) \big\} 
\; + \; \Oint{ \msc{C} }{}   \theta^{\prime}(\om -\mu) \cdot \ln \big[ 1+\mf{a}(\mu) \big] \cdot \f{ \dd \mu }{ 2\pi } \;. 
\label{NLIE-Karol}
\enq

Once the question of the limit of the function $\wh{\mf{a}}$ is settled, 
it is not a problem to send $N\tend +\infty$ in the above formulae. One gets 
\bem
f_{\e{surf}}   \; =  \;  \mc{B}(\xi_+) \; + \; \mc{B}(\xi_-) 
 - 2\be T \big[ \coth (\eta) \; - \; \coth (2\eta) \big]
\; -\; 2 T \cdot \mc{F}  \;  -  \;  T  \cdot \mc{I}  \\
+ \; \f{T}{2}\ln [ 1 - \mf{a}_{\e{reg}}^2(0) ]  \; + \;  T \ln\Big( \det_{\msc{C}}[I+\ov{K}]  \Big)  
\; - \; T  \ln\big[ 1+\ex{-\f{h}{T}} \big] 
\label{ecriture surf free energy at infinite trotter}
\end{multline}
In this representation, we agree upon
\beq
\mc{B}(\xi) = T \be \coth(\xi)  \; - \;  T \de_{\xi}  \cdot \ln \big[ 1+ \mf{a}(-\xi)\big]   
\; + \; T \Oint{ \msc{C}}{} \ln \big[ 1+ \mf{a}(\om)\big]   \coth(\om + \xi ) \cdot \f{ \dd \om }{ 2i\pi } \;. 
\label{definition fonction B}
\enq
with $\de_{\xi}$ as defined in \eqref{definition fonction indicatrice  delta xi}. 
$\mf{a}_{\e{reg}}(0)$ corresponds to the $N\tend +\infty$ limit of $\wh{\mf{a}}(0)$. We do stress that the 
Trotter limit and the $\om \tend 0$ limits do not commute for $\wh{\mf{a}}(\om)$. In  fact, 
$\mf{a}_{\e{reg}}(0)$ corresponds to the regular part of the function $\mf{a}(\om)$ at $\om=0$, \textit{viz}.
\beq
\ln \mf{a}_{\e{reg}}(0)  \; =   \; \lim_{\om\tend 0} \Big[ \ln \mf{a}(\om)  + \f{2 J }{T} \s{\eta} \coth(\om) \Big] \; = \; 
-\f{h}{T} \; + \; \f{2 J }{T} \cosh(\eta)
\; + \; \Oint{ \msc{C} }{}   \theta^{\prime}(\mu) \cdot \ln \big[ 1+\mf{a}(\mu) \big] \cdot \f{ \dd \mu }{ 2\pi } \;. 
\enq

Furthermore, one has 
\bem
\mc{I} \; = \;   \Oint{\msc{C}   }{} 
  \ln\pa{1+\mf{a}\pa{\om}} \Big[2 \coth\pa{2\om} + 2 \coth\pa{2\om +\eta} 
\, - 4 \be \coth^{\prime}(\om ) \, + 4 \be  \coth^{\prime}(\om  +\eta)    \Big]   \f{\dd \om}{2i\pi}    \\
\;+\;     \Oint{\msc{C}}{} \hspace{-1mm} \f{\dd \om}{2i\pi} 
\hspace{-1mm} \Oint{\msc{C}^{\prime} \subset \msc{C}}{} \hspace{-1mm}  \f{\dd \om^{\prime}}{2i\pi} 
  \ln\pa{1+\mf{a}\pa{\om}}\ln\pa{1+\mf{a}(\om^{\prime})} \bigg[ 
 \Dp{\om} \Dp{\om^{\prime}} \ln\Big[\f{ \s{\om+\om^{\prime}+\eta} }{ \s{\om-\om^{\prime}+\eta} }  \Big]  
 \, -  \, \coth^{\prime}(\om-\om^{\prime}) \, - \,  \coth^{\prime}(\om+\om^{\prime})  \bigg]    
\end{multline}

Also, the kernel $\ov{K}$ is defined as in \eqref{definition noyau K bar cas discret} with the sole difference
that the function $\wh{\mf{a}}$ ought to be replaced by the function $\mf{a}$ defined by \eqref{NLIE-Karol}. 
Finally, $\mc{F}$ is given by 
\bem
\mc{F} = 
\sul{ k = 0   }{+\infty} \Oint{ \msc{C}_1\supset\dots \supset \msc{C}_{2k+1} }{}
\sul{n=k}{+\infty} \f{ \Big[ f\pa{\om_{2k+1}} \Big]^{n-k} }{2n+1} \; \cdot \; 
\pl{p=1}{2k+1}U(\om_p,\om_{p+1}) \; \f{ \dd^{2k+1} \om }{ \pa{2i\pi}^{2k+1} }  \\
\; -\; \sul{ k = 1   }{+\infty}   \Oint{ \msc{C}_1\supset\dots \supset \msc{C}_{2k} }{}
\sul{n=k}{+\infty} \f{ \Big[ f\pa{\om_{2k}} \Big]^{n-k} }{2n} \; \cdot \; 
\pl{p=1}{2k}U(\om_p,\om_{p+1}) \; \f{ \dd^{2k} \om }{ \pa{2i\pi}^{2k} } \; ,
\label{ecriture series limite thermo fction F caligraphique}
\end{multline}
The kernel $U$ defining the function $\mc{F}$ is given by 
\bem
U(\om,\om^{\prime}) = \f{ - \ex{-\f{h}{T}} \s{2\om^{\prime}+\eta} }{ \s{\om+\om^{\prime}} \s{\om-\om^{\prime}-\eta}}
\cdot \exp\big\{  -2\be \big[ \coth(\om^{\prime}) - \coth(\om^{\prime}+\eta) \big]  \big\} \\ 
\times \exp\Bigg\{  - \Oint{ \msc{C}_U }{} \hspace{-1mm} \f{\dd \tau }{ 2i\pi }  \ln\big[1+\mf{a}\pa{\tau}\big]
\cdot \Big[ \coth(\tau +\om^{\prime} ) \, + \, \coth( \om^{\prime} -\tau )
 \, -  \, \coth(\tau +\om^{\prime} +\eta) \, - \, \coth(\om^{\prime} -\tau +\eta) \Big] \Bigg\}	\; , 
\label{ecriture noyau U apres Trotter}
\end{multline}
and the function $f\pa{\om}$ reads 
\bem
f\pa{\om} = \ex{ -\f{2h}{T} } \exp \Big\{ -2\be\,  [\coth(\om - \eta) - \coth(\om+\eta) ]  \Big\}  \\
\exp\Bigg\{  - \Oint{ \msc{C}_U }{} \hspace{-1mm} \f{\dd \tau }{ 2i\pi }  \ln\big[1+\mf{a}\pa{\tau}\big] 
\cdot \Big[ \coth(\tau +\om^{\prime} -\eta) \, + \, \coth(\om^{\prime} -\tau -\eta)
 \, -  \, \coth(\tau +\om^{\prime} +\eta) \, - \, \coth(\om^{\prime} -\tau +\eta) \Big] \Bigg\}	\;. 
\label{ecriture fct f apres Trotter}
\end{multline}
The encased contours $\msc{C}_p$ are as defined in lemma \ref{Lemme factorisation determinant I+U} whereas 
the contour $\msc{C}_U$ is such that given any $\om^{\prime}, \om \in \msc{C}_p$, $p\in \mathbb{N}$, 
the points $\pm \om^{\prime}, \pm (\om^{\prime}+\eta)
\pm (\om^{\prime}-\eta)$ are not encircled by $\msc{C}_U$. The latter however encircles 
the region where the numbers $\la_1,\dots, \la_N$ condensate and, in particular, the origin.







\section{The boundary magnetization}
\label{Section Boundary magnetization}
\subsection{An integral representation at finite temperature}

It follows from general considerations that the boundary magnetization can be obtained from $\xi_-$ partial derivatives of the 
partition function. More precisely, one has
\beq
\moy{\sg_1}_{T;M} \; = \; \f{ \sinh^2(\xi_-) }{ \be }  \cdot \f{ \Dp{}  }{ \Dp{} \xi_- } \big[  \ln Z_M  \big] \;. 
\enq
Hence, using that 
\beq
Z_M \simeq \exp \bigg\{  -\f{ M f_{\e{bk}} }{ T } \; - \; \f{ f_{\e{surf} } }{ T }   \; + \; \e{O}\Big( \f{1}{M} \Big)   \bigg\}  
\enq
and assuming that it is licit to exchange the $M\tend \infty$ limit with differentiation, we get that, in the thermodynamic limit
\beq
\moy{ \sg_1 }_{T} \; = - \; \f{ \sinh^2(\xi_-) }{ J \sinh(\eta) }  \cdot \f{ \Dp{}  }{ \Dp{} \xi_- } f_{\e{surf}} \;. 
\label{representation spin site 1 comme derivee partielle}
\enq
In fact, solely the function $\mc{B}(\xi_-)$ defined in \eqref{definition fonction B} gives a non-zero contribution
 to \eqref{representation spin site 1 comme derivee partielle}:
\beq
\moy{\sg_1}_{T} = 1  \; + \; \f{ T \sinh^2(\xi_-) }{ J \s{\eta} }  \cdot \f{ \de_{\xi_-} \mf{a}^{\prime}(-\xi_-) }{ 1+ \mf{a}(-\xi_-) }   
\; + \; T \f{\sinh^{2}(\xi_-)  }{J \sinh(\eta)  } 
\Oint{ \msc{C}}{}   \f{  \ln \big[ 1+ \mf{a}(\om)\big]  }{ \sinh^2(\om + \xi_- ) } 
		\cdot \f{ \dd \om }{ 2i\pi } \;. 
\label{representation boundary spin mean value}
\enq

\subsection{Numerics}

\label{sec:numerics}

Building on the previous integral representation, we present numerical
calculation issued plots for the boundary 
magnetization.
We do not attempt a full exploration of the parameter space;
our aim is to demonstrate that our results can be easily implemented
by suitable computer programs and that the known $T\to 0$ limit  is
reproduced numerically. Therefore we constrain our investigations to
 one of the possible domains in parameter space, namely the so-called massive regime ($\cosh(\eta)>1$, $\eta\in \R$) of the chain\footnote{The reason for this name
is that the model has a finite gap between the (possibly doubly degenerate) ground state and 
the first excited states at \textit{sufficiently small} external magnetic field. The excitations become however massless
for sufficiently strong values of $h$ }. 

We choose to express quantities 
in units of $J$ what amounts to setting $J=1$ in the numerical analysis.
Also, we found it convenient to recast the parametrization of the boundary magnetic fields in \eqref{ecriture Hamiltonien XXZ ac bord}
as $-h_b^\pm\,= \, 2\sinh\eta\coth\xi_\pm $. 

We remind that, in the massive regime, it is more practical to consider the rotated counting function $\mf{a}_m(\la) \equiv \mf{a}(i\la)$. 
The latter satisfies 
\begin{equation}
\label{NLIE3}
\begin{split}
\log \mf{a}_m(\lambda)=&-\frac{h}{T}
+\frac{2}{T} \frac{\sinh^2\eta}{\sin(\lambda)\sin(\lambda-i\eta)}
+  \Oint{\msc{C}}{} 
\frac{\sinh2\eta \log(1+\fa_m(\omega))}{\sin(\lambda-\omega+i\eta)\sin(\lambda-\omega-i\eta)}
\cdot \frac{ \dd \omega}{2\pi }   \; . 
\end{split}
\end{equation}
There, the integration contour $\msc{C}$ consists of two intervals:
\begin{equation*}
\msc{C} \; = \;   [-i\alpha-\pi/2, -i\alpha+\pi/2]\cup
   [i\alpha-\pi/2, i\alpha+\pi/2] \;,
\end{equation*}
since the right and left lateral contours cancel out in virtue of the $\pi$-periodicity of the integrands. Finally, 
the parameter $\alpha<\eta/2$ is a real number which has to be chosen large
enough so that all the roots parametrizing the dominant eigenvalue of the QTM lie inside
the contour. 
We have solved the integral equation \eqref{NLIE3} by the iteration method.


Within these new notations, one gets the representation for the boundary spin 
$  \big< \sigma_1^z \big> $ \eqref{representation boundary spin mean value} 
\begin{equation}
\begin{split}
\label{expect2c}
   \big< \sigma_1^z \big> =1-T\frac{\sinh^2\xi_-}{\sinh \eta}
 \Oint{\msc{C}}{} 
\frac{\log(1+\mf{a}_m(\omega))}{\sin^2(\omega-i\xi_-)} \cdot \frac{d\omega}{2\pi}
\; - \;T  \delta_{\xi_-} \cdot \frac{\sinh^2\xi_-}{\sinh\eta} \frac{i\mf{a}_m^{\prime} (i\xi_-)}{ 1 + \mf{a}_m(i\xi_-)}
\end{split}
\end{equation}
We remind that $\delta_{\xi_-}=1$ if $\xi_-$ lies inside the contour and $\de_{\xi_-}=0$ otherwise. 
%
Since it is not convenient to evaluate derivatives on the numerical level, we have rather used the expression below
for a numeric evaluation of the last term in \eqref{expect2c}
\begin{equation}
\label{NLIExiD2}
\begin{split}
\frac{i \fa^{\prime}_m(i\xi_-)}{\fa_m(i\xi_-)}=
8\frac{\sinh^2\eta}{T}
\frac{\sinh(-2\xi_-+\eta)}{[\cosh(-2\xi_-+\eta)-\cosh\eta]^2}
-  \Oint{\msc{C}}{} 
 \log(1+\fa_m(\omega))
\frac{4 \sinh 2\eta\sinh(2(\xi_-+i\omega))}{ [\cosh(2(\xi_-+i\omega))-\cosh(2\eta)]^2} \cdot \frac{d\omega}{2\pi } 
\end{split}
\end{equation}

\begin{figure}
  \centering
\psfrag{sigmaz}{$\vev{\sigma_1^z}$}
\psfrag{alpha}{$\alpha$}
\psfrag{Np20}{$N_p=20$}
\psfrag{Np100}{$N_p=100$}
\includegraphics{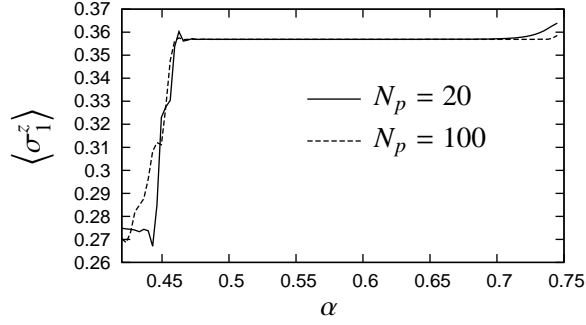}
\caption{In this figure we demonstrate 
the dependence of the numerical results on the parameter $\alpha$,
which describes the position of the integration contours. 
Here we consider one example with $\eta=1.5$, $T=1$ and $h=2$. The integral has been discretized 
into $N_p=20$ and $N_p=100$ points. 
 The theoretical requirement is that $\alpha<\eta/2$
 and $\alpha$ should be big enough so that the contour encircles all
 the Bethe roots of the Quantum Transfer Matrix.  In the present case
 the first singularities appear around $\alpha\approx 0.47$. Note 
 that numerical errors arise also around $\alpha=\eta/2=0.75$, however
 their magnitude decreases with growing $N_p$. 
The numerical value of the boundary magnetization is
$\big< \sigma_1^z \big>=0.356912$, and it changes less than
$10^{-6}$ in the middle regime, even with $N_p=20$.
}
\label{fig:alpha}
\end{figure}

We numerically evaluated the boundary magnetization as a function of the
magnetic fields for different values of the temperature and for
$\eta=1.5$. 
The choice $\alpha=0.9 \eta/2$ appeared
appropriate for all the values of $T$ and $h$ that we have
considered. In Fig.~\ref{fig:alpha} we consider one specific point in
the parameter space ($\eta=1.5$, $T=1$, $h=2$, $h_b^-=0$) and show 
that indeed there is an interval $\alpha^*<\alpha<\eta/2$ where the numerical result
for the boundary magnetization does not depend on $\alpha$. 

\begin{figure}
  \centering
\psfrag{-sigmaz}{$\vev{\sigma_1^z}$}
\psfrag{H}{$h$}
\psfrag{T0}{$T=0$}
\psfrag{T0.15}{$T=0.15$}
\psfrag{T0.5}{$T=0.5$}
\psfrag{T1}{$T=1$}
\psfrag{T2}{$T=2$}
\includegraphics{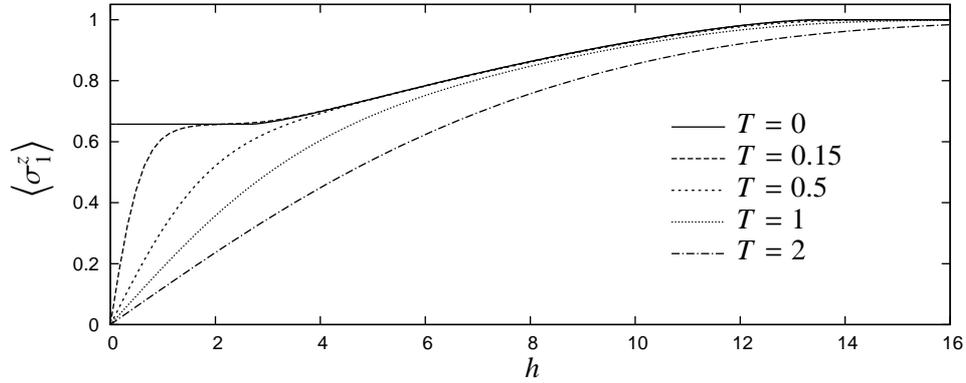}
\caption{The boundary magnetization as a function of the bulk field
  $h$, for different fixed temperatures ($\eta=1.5$ and
  $h_b^-=0$). The zero-temperature result is given by equation
  \eqref{ecriture boundary mag intermediate regime massive}. The
  properties of the ground state change at the two critical values
  $h_{cr}^1=2.6585$ and $h_{cr}^2=13.410$.
}
\label{fig:h}
\end{figure}

\begin{figure}
  \centering
\psfrag{-sigmaz}{$\vev{\sigma_1^z}$}
\psfrag{hbndry}{$h_b^-$}
\psfrag{T0}{$T=0$}
\psfrag{T0.25}{$T=0.25$}
\psfrag{T0.5}{$T=0.5$}
\psfrag{T1}{$T=1$}
\psfrag{T2}{$T=2$}
\includegraphics{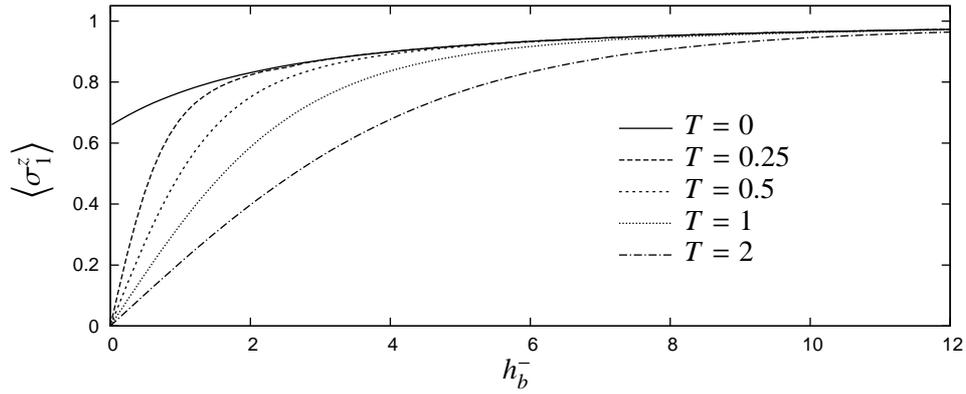}
\caption{The boundary magnetization as a function of the boundary field
  $h_b^-$, for different fixed temperatures ($\eta=1.5$ and $h=0$). 
 The zero-temperature result is given by equation \eqref{JMT0}.}
\label{fig:hb}
\end{figure}

Our plots are shown in Fig.~\ref{fig:h}, in what concerns $\big<
\sg_1^z \big>$ as a function of the bulk  
magnetic field at $h_b^-=0$, and Fig.~\ref{fig:hb}  in what concerns
$\big< \sg_1^z \big>$ as a function of the boundary 
magnetic field at $h=0$. 
We have also added the plots corresponding to the zero-temperature limit. The latter have been 
extracted by using the $T=0$ representation obtained in
\cite{JimboKedemKonnoMiwaXXZChainWithaBoundaryElemBlcks,
  KapustinSkorikStructureGSOpenXXZMassless,KozKitMailNicSlaTerElemntaryblocksopenXXZ}.

There arize two values of the external magnetic field in the description of the ground state of the Hamiltonian 
\eqref{ecriture Hamiltonien XXZ ac bord}. At $h_b^{+}=h_b^{-}=0$ and for 
$h>h_{cr}^{(2)}=4(1+\cosh\eta)$ the ground state
is completely polarized and $\big< \sigma_1^z \big> = 1$. Then for $h_{cr}^{(1)} \leq h \leq h_{cr}^{(2)}$, with 
\beq
h_{\e{cr}}^{(1)} \; = \; 4 \sinh{\eta} \sul{ n \in \mathbb{Z} }{} \f{(-1)^n}{ \cosh(n\eta)} \;, 
\enq
the model is massless. Then at $h_b^{+}=h_b^{-}=0$, the boundary magnetization admits the integral representation:
\beq
\big< \sigma_1^z \big> \; = \; 1 \; -  \; \Int{-q}{q} \cot(\nu-i\tf{\eta}{2}) g(\nu) \dd \nu
\label{ecriture boundary mag intermediate regime massive}
\enq
There the function $h$ solves the linear integral equation 
\beq
g(\la)\;  - \; \Int{-q}{q} \f{ \s{2\eta} \cdot  g(\tau)  }{\sin(\la-\tau+i\eta) \sin(\la-\tau-i\eta)}  \cdot
\f{ \dd \tau }{ 2\pi } \; = \;  \f{ \Dp{} }{ \Dp{} s }  \cdot 
\Big\{ \f{ \s{\eta}}{\pi \sin(\la + s + i\tf{\eta}{2}) \sin(\la +s - i\tf{\eta}{2})} \Big\}_{\mid s=0} \;. 
\label{ecriture eqn int density massive}
\enq

Note that the endpoint of integration $q \in \intff{0}{\tf{\pi}{2}}$ 
is defined as the solution to the equation $\eps_0(q)=0$, where the dressed
energy $\eps_0$ is given by the solution to the integral equation:
\beq
\eps_0(\la)\;  - \; \Int{-q}{q} \f{ \s{2\eta}}{\sin(\la-\tau+i\eta) \sin(\la-\tau-i\eta)} \eps_0(\tau) \cdot
\f{ \dd \tau }{ 2\pi } \; = \; h \; -\; \f{ 2J \sinh^2(\eta) }{ \sin(\la +i \tf{\eta}{2}) \sin(\la -i \tf{\eta}{2}) } \;. 
\label{ecriture eqn int dress energy massive}
\enq
One has $q=0$ for $h=h_{cr}^{(2)}$ and $q=\tf{\pi}{2}$ for $h=h_{cr}^{(1)}$. At $h=h_{cr}^{(1)}$,
the model starts to become massive and, for any $0<h<h_{cr}^{(1)}$, 
one should set $q=\tf{\pi}{2}$ in \eqref{ecriture eqn int density massive} and \eqref{ecriture eqn int dress energy massive}. 
In such a case, the integral equation \eqref{ecriture eqn int density massive} becomes explicitly solvable 
\textit{via} Fourier transformation.  The boundary magnetization is then given by 
\eqref{ecriture boundary mag intermediate regime massive} with $q=\tf{\pi}{2}$. 
In this case, one can compute it in a closed form, 
this for any value of the boundary magnetic field $h_b^-$ \cite{JimboKedemKonnoMiwaXXZChainWithaBoundaryElemBlcks}:
\begin{equation}
\label{JMT0}
\big<  \sigma_1^z \big> \; = \;  1+2(1-r)^2\sum_{\ell=1}^\infty 
\frac{(-1)^{\ell} \ex{2\ell \eta} }{(1-\ex{2\ell \eta}r)^2}
\qquad \e{where} \qquad \frac{h_b^-}{2} \; = \; \frac{1+r}{1-r}  \cdot \sinh\eta   \;. 
\end{equation}

\begin{figure}
  \centering
\psfrag{sigmaz}{$\vev{\sigma_1^z}$}
\psfrag{hb}{$h_b^-$}
\psfrag{h0}{$h=0$}
\psfrag{h8}{$h=8$}
\psfrag{h4}{$h=4$}
\psfrag{h-8}{$h=-8$}
\psfrag{h-4}{$h=-4$}
\includegraphics{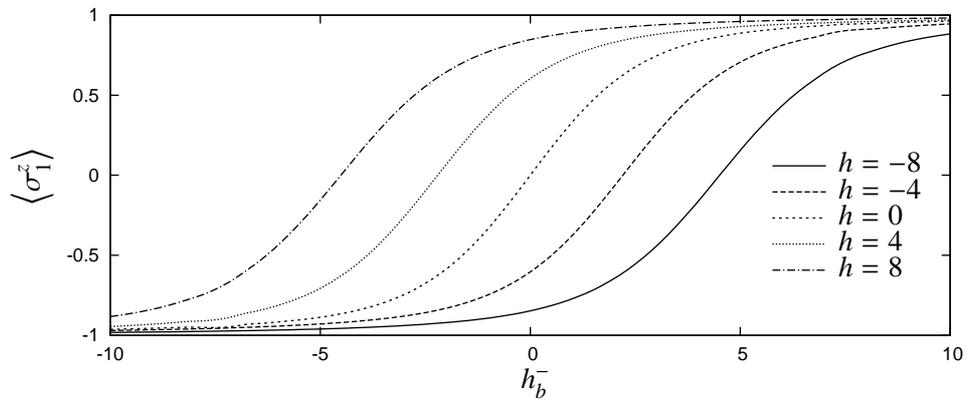}
\caption{
The boundary magnetization as a function of the boundary field
$h_b^-$, for different bulk fields ($\eta=1.5$ and $T=1$). 
}
\label{fig:both}
\end{figure}

In Fig.~\ref{fig:both} we plot results for the general situation with
both the bulk and the boundary magnetic field present (here we fix the
temperature to $T=1$).

\begin{figure}
  \centering
\psfrag{sigmaz}{$\vev{\sigma_1^z}$}
\psfrag{eta}{$\eta$}
\psfrag{h0.2}{$h_b^-=0.2$}
\psfrag{h1}{$h_b^-=1$}
\psfrag{h2}{$h_b^-=2$}
\psfrag{h4}{$h_b^-=4$}
\subfigure[Boundary magnetization in the ground state ($T=0$)]{    \includegraphics{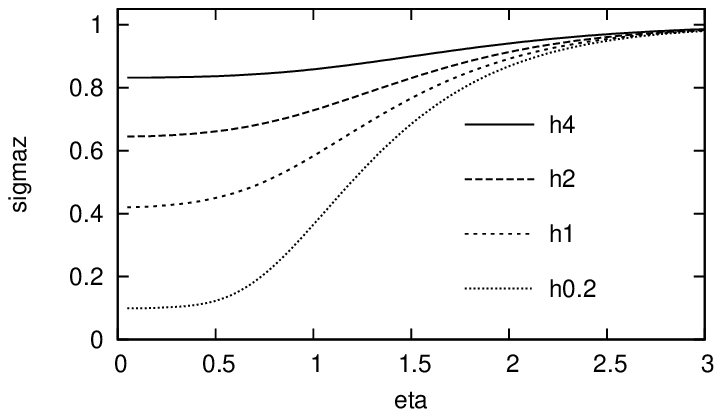}\label{subb1}}
\hspace{1cm}
\subfigure[Finite temperature results ($T=0.5$)]{    \includegraphics{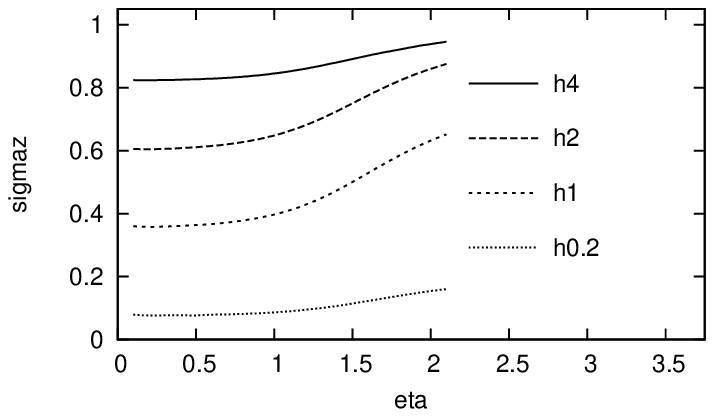}}
\caption{The boundary magnetization as a function of the anisotropy
  parameter $\eta$, for different boundary magnetic fields. The bulk
  field is set to $h=0$. The zero-temperature result depicted in Fig. \ref{subb1} is calculated using  equation \eqref{JMT0}. }
\label{vs-eta}
\end{figure}

In Fig.~\ref{vs-eta} we also plotted the boundary magnetization as a
function of the anisotropy parameter $\eta$, in the case of the ground
state ($T=0$) and also in a finite temperature situation ($T=0.5$). In
the latter case our data is limited to smaller values of $\eta$,
because our simple iteration algorithm loses its good convergence properties at higher
values of the ratio $\cosh(\eta)/T$.

\subsection{Low-$T$ expansion of the boundary dependent part}

\begin{prop}
In the massless regime of the chain $\eta=-i\zeta$ with $\zeta \in \intoo{0}{\pi}$, the boundary field dependent integral 
given by \eqref{definition fonction B} admits the low-$T$ asymptotic behavior
\bem
\mc{B}(\xi) = T \be \coth(\xi)  \; + \; 
 \Int{\mc{C}_{\xi} }{  } \coth(\la + \xi + i\tf{\zeta}{2}) \veps(\la ) \cdot \f{\dd\la}{2i\pi}   \\
\; + \;i   \f{\pi T^2 }{12  \veps_0^{\prime}(q) } \cdot \f{ \s{2 \xi + i\zeta}   }{\s{q + \xi + i\tf{\zeta}{2}}\s{q-\xi-i\tf{\zeta}{2}} }
\; + \; \e{O}(T^4) \;. 
\label{ecriture low T behavior B de xi}
\end{multline}
where 
\beq
\ba{cc} 
\mc{C}_{\xi} = \intff{-q}{q} \cup \Ga(-i\tf{\zeta}{2} - \xi)  & \e{if} \;\; 0 < -\Im(\xi)< \tf{\zeta}{2} \\ 
\mc{C}_{\xi} = \intff{-q}{q}   & \e{otherwise} \ea \;. 
\enq
Furthermore, we denote by $\Ga(z)$ a small counterclockwise loop around $z$.

\end{prop}

\Proof

\begin{figure}[h]

\begin{pspicture}(15,5)

\psline[linestyle=dashed, dash=3pt 2pt]{->}(1,3)(15,3)
\rput(14.5,2.5){$\R$}

\psline{-}(1,5)(15,5)
\psline[linewidth=2pt]{->}(6,5)(5.8,5)
\rput(14.5,4.5){$\msc{C}_{\ua}= -\R +i\a$}

\psline{-}(1,1)(15,1)
\psline[linewidth=2pt]{->}(5.8,1)(6,1)
\rput(14.5,1.5){$\msc{C}_{\da}=\R-i\a$}

\end{pspicture}

\caption{Contour $\msc{C}$ for the non-linear integral equation for $\mf{a}$ in the massless regime.\label{Contour NLIE massless regime} }
\end{figure}
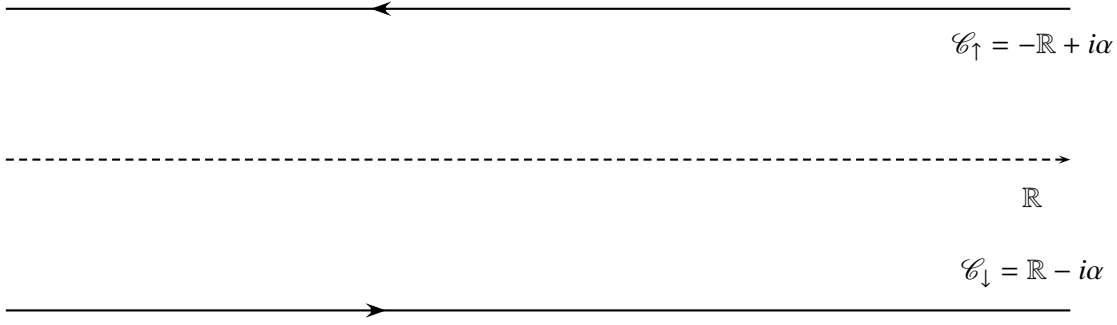

In order to obtain the low-$T$ expansion of the boundary field dependent part of the surface free energy, 
we first need to obtain the first few terms of the low-$T$ asymptotic expansion of the function $\mf{a}$. 
In the massless regime, within the parametrization $\eta=-i\zeta$ with $0<\zeta<\pi$, the  
non-linear integral equation satisfied by this function takes the form
\beq
\ln \mf{a}(\om)  \; =   - \f{ e_0(\om -i\tf{\zeta}{2}) }{T }
\; + \; \Oint{ \msc{C} }{}  \f{ \dd \mu }{ 2\pi } \, K(\om -\mu) \cdot \ln \big[ 1+\mf{a}(\mu) \big] \;. 
\label{equation fction a pour regime massif}
\enq
Here we agree upon
\beq
e_0(\la) =  h \; - \;  \f{2 J \sin^2(\zeta)}{\s{\la + i\tf{\zeta}{2} }  \s{\la - i\tf{\zeta}{2} }  } 
\quad \e{and} \; \qquad 
K(\mu) =  \f{ \sin(2\zeta) }{ \s{\mu + i\zeta }  \s{\mu - i\zeta }  } \;. 
\enq
The integration contour $\msc{C}$ can be chosen as depicted in Fig.~\ref{Contour NLIE massless regime}.  
There $\a<\e{min}( \tf{\zeta}{2}, \tf{(\pi-\zeta)}{2} )$, and is large enough so that the contour $\msc{C}$ encircles all the roots
describing the dominant eigenvalue. 

We assume that we are in the anti-ferromagnetic regime of the chain meaning that the magnetic field is not sufficiently strong 
enough so as to polarize the chain to the ferromagnetic state, namely
\beq
0 < h <h_c  \quad \e{with} \quad 
h_c = \f{2J\sin^2(\zeta) }{ \sin^2(\tf{\zeta}{2}) } = 8J \cos^2(\tf{\zeta}{2}) \;. 
\enq

In such a situation, the function $e_0(\la)$ admits two (symmetric) zeroes $\pm q_0$ on $\R$. Moreover it satisfies 
\beq
{ e_0 }_{\mid \intoo{-q_0}{q_0} } <0 \qquad \e{and} \qquad 
{ e_0}_{\mid \R\setminus \intff{ -q_0 }{ q_0 }  } >0 \;. 
\enq
We would like to analyse the behavior of the solution $\mf{a}(\om)$ at low temperatures. 
There, it appears that the dominant contribution at $T\tend 0$ (up to $\e{O}(T^{\infty})$
corrections) will stem from the line $\R + i\tf{\zeta}{2}$. It is apparent from the previous 
discussion that the contour $\msc{C}_{\ua}$ can be readily deformed towards this line for $\zeta \in  \intof{0}{\tf{\pi}{2}}$.

Indeed, suppose that $0<\zeta<\tf{\pi}{2}$. One expects 
that, at low-temperatures, $\mf{a}(\om)=\e{O}(T^{\infty})$ whenever $\om$ belongs to the region 
encircled by the curve $\msc{C}$ that moreover lies strictly in the lower-half plane. 
The function of $\om$ defined by the integral 
\beq
\Int{\msc{C}}{} \theta^{\prime}(\om-\mu) \ln \big[ 1+\mf{a}(\mu)\big] \f{ \dd \mu }{ 2\pi }
\enq
can be analytically continued from $\om$ lying inside of $\msc{C}$ up to $\om \in \R + i\tf{\zeta}{2}$. One can even deform 
the integration contour $\msc{C}_{\ua}$ up to $\R + i\tf{\zeta}{2}$. 
Then, agreeing upon 
\beq
\veps (\la) \equiv - T \ln \big[ \mf{a}(\la +i \tf{\zeta}{2}) \big] \qquad \e{for} \quad \la \in \R \;, 
\enq
we are lead to the below integral equation 
\beq
\veps(\la)  \; =   e_0(\la)  
\; + \;  T \Int{ \R }{}  \f{ \dd \mu }{ 2\pi } \, K(\la -\mu) \cdot \ln \big[ 1+\ex{-\f{\veps(\mu)}{T}} \big] 
\; - \;  T \Int{ \R }{}  \f{ \dd \mu }{ 2\pi } \, K(\la -\mu+i \tf{\zeta}{2} + i \a) \cdot 
\ln \big[ 1+ \mf{a}(\mu-i \a ) \big]  \;. 
\label{ecriture NLIE veps un peu transforme}
\enq
One should, in fact, choose $\a$ in such a way that $\mf{a}(\mu-i\a)= \e{O}(T^{\infty})$ uniformly in $\mu \in \R$, 
this in the $\big( L^1\cap L^{\infty} \big) \big( \R \big)$ sense. 
This condition should be checked 
\textit{a posteriori} once that $\mf{a}$ has been computed.  If its holds, this means that the low-$T$ analysis of the non-linear integral equation 
reduces to the one of an equation quite close to the Yang-Yang equation
\beq
\veps(\la)  \; =   e_0(\la)  
\; + \;  T \Int{ \R }{} \, K(\om -\mu) \cdot \ln \big[ 1+\ex{-\f{\veps(\mu)}{T}} \big]  \cdot  \f{ \dd \mu }{ 2\pi } 
\; + \; \e{O}(T^{\infty}) \;. 
%
%
\label{ecriture NLIE veps a ordre T infty pres}
\enq
%
%
%
%
%
%
%
%
%
%
%

Although it is unclear to us how such a deformation procedure would work for $\zeta \in  \intoo{\tf{\pi}{2}}{\pi}$, we shall
nonetheless work under the hypothesis that this can be done. In other words, we shall assume that 
equation \eqref{ecriture NLIE veps a ordre T infty pres} 
holds for the whole regime $0 < \zeta < \pi$ and that the power-law part of the $T\tend 0$ asymptotics contribution stemming from 
integrations of analytic functions versus $\ln\big[ 1+\mf{a}(\la)\big]$ along $\msc{C}$ always issues from a deformation of the contour
$\msc{C}_{\ua}$ to $\R+i\tf{\zeta}{2}$.

The equation \eqref{ecriture NLIE veps a ordre T infty pres} resembles to the Yang-Yang equation
\cite{Yang-YangNLSEThermodynamics}. Techniques for extracting the low-$T$ asymptotic behavior of its solution are well known
\cite{EsslerFrahmGohmanKlumperKorepinOneDimensionalHubbardModel,KlumperNLIEfromQTMDescrThermoXYZOneUnknownFcton}. 
A rigorous approach to the extraction of the low-$T$ asymptotic expansion of the solution to the Yang-Yang equation 
has been given in \cite{KozProofexistenceAEYangYangEquation}. 
The setting of this last paper can be directly applied here when $\tf{\pi}{2} \leq \zeta < \pi$,  as then
$K(\la)<0$ for $ \la \in \R$. For $0<\zeta<\tf{\pi}{2}$, the change in the sign of Lieb's kernel $K$
breaks certain properties used in \cite{KozProofexistenceAEYangYangEquation}. However, one can still compute
the asymptotic expansion on a formal level, by assuming the a priori existence thereof. 
Independently of the value of $\zeta$ we shall give a formal derivation of the asymptotic expansion. 
Hence, we assume that, just as $e_0$, $\veps$ admits two roots $\wh{q}^{\,(\pm)}$ on $\R$. Likewise, we shall take for granted that 
\beq
\hspace{1cm} { \veps }_{\mid \intoo{ \qm  }{ \qp } } <0 \hspace{1cm} \e{whereas} \hspace{1cm} 
{ \veps }_{\mid \R\setminus \intff{\qm }{\qp  }} >0 \;. 
\enq
It follows from \eqref{ecriture NLIE veps un peu transforme}, by slightly shifting $\a$
if necessary, that $\veps(\la)$ is holomorphic in some open neighborhood of the real axis. 
The size of this neighborhood does not depend on $T$. 
Thus, according to the results of lemma \ref{Lemme estimation integrales avec une fonction u}, one has 
\beq
T \Int{ \R }{}   K(\om -\mu) \cdot \ln \big[ 1+\ex{-\f{\veps(\mu)}{T}} \big]  \cdot \f{ \dd \mu }{ 2\pi } \; = \; 
- \Int{ \wh{q}^{(-)} }{ \wh{q}^{(+)} }  K(\om -\mu)   \veps(\mu) \f{ \dd \mu }{ 2\pi }  
\; + \;  \f{\pi^2 T^2 }{12 \pi } 
\Bigg\{ \f{  K(\la-\qp)  }{ \veps^{\prime}( \qp  ) }  \, -  \, 
\f{  K(\la-\qm)  }{ \veps^{\prime}( \qm ) }  \Bigg\} 
\; + \; \e{O}\big(T^{4}\big)\;. 
\label{equation DA ordre 1 terme integral NLIE}
\enq
In other words, the function $\veps$ solves the integral equation 
\beq
\veps(\la)  \; + \;  \Int{ \qm }{ \qp }  K(\om -\mu) \,   \veps(\mu) \f{ \dd \mu }{ 2\pi }    =   e_0(\la)  
\; + \;  \f{\pi^2 T^2 }{12 \pi } 
\bigg\{ \f{  K(\la-\qp)  }{ \veps^{\prime}( \qp  ) } - 
\f{  K(\la-\qm)  }{ \veps^{\prime}( \qm  ) }  \bigg\} 
\; + \; \e{O}\big(T^{4}\big) \;. 
\label{equation inegrale lineaire DA veps}
\enq
There, the $\e{O}\big(T^{4}\big)$ is in $(L^{1}\cap L^{\infty})(\R)$. We assume that $\veps$ admits the low-$T$ asymptotic expansion
\beq
\veps (\la )  \; = \;  \veps_0(\la) \; + \; T\veps_1(\la) \; + \;  T^2  \veps_2(\la) + \e{O}(T^3) 
\label{ecriture DA low T veps}
\enq
and that $\wh{q}^{\, (\pm)} \tend \pm q$ in the $T\tend 0$ limit. It is easy to see that, under such assumptions,
$\veps_0$ solves the linear integral equation
\beq
\veps_0\pa{\la} \; + \;  \Int{-q}{q} K(\la-\mu)\, \veps_0\pa{\mu} \f{\dd \mu}{2\pi}     = e_0(\la) 
\qquad \e{with} \; q \; \e{fixed} \; \e{by} \; \e{the} \; \e{condition} \;\;  \veps_{0}(\pm q) =0\;. 
\label{ecriture equation dressed energy}
\enq
In other words, as one could have expected, $\veps_0$ is to be identified with the dressed energy of the 
particle/hole type excitations
above the ground state of an open $XXZ$ spin-$\tf{1}{2}$ chain at finite magnetic field $h$. 
Note that, when $\tf{\pi}{2} < \zeta < \pi$, one can use the techniques developed in \cite{KozProofexistenceAEYangYangEquation}
so as to prove the unique solvability of \eqref{ecriture equation dressed energy} for $\veps_0$ and $q$. 
Equations \eqref{ecriture DA low T veps} and \eqref{equation inegrale lineaire DA veps} allow one 
to fix the dependence of the endpoints $\wh{q}^{\,(\pm)}$ on $T$. Namely,
\beq
0= \veps(\wh{q}^{\,(\pm)}) = \veps_0(\pm q) \; + \;  (\wh{q}^{(\pm)} \mp q) \veps_0^{\prime}(\pm q) 
\; + \;  T \veps_1(\pm q) \; + \; 
\e{O}\big(T^2 + (\wh{q}^{\,(\pm)} \mp q )^2 \big) \;. 
\label{ecriture estimation distance q hat a q}
\enq
Hence $(\wh{q}^{\,(\pm)} \mp q)$ goes to zero at least as $T$ (this if $\veps_1(\pm q) \not=0$, otherwise it 
goes to zero even faster, \textit{ie} at least as $T^{2}$). 
Inserting the above expansion \eqref{ecriture estimation distance q hat a q} into the linear integral 
equation \eqref{equation inegrale lineaire DA veps} and keeping terms that are 
at most a $\e{O}(T^2)$, we get
\beq
\Int{ \qm }{ \qp }\hspace{-1mm}  K(\om -\mu)   \veps(\mu) \f{ \dd \mu }{ 2\pi } = 
\Int{ -q }{ q }  \hspace{-1mm} K(\om -\mu)   \veps(\mu) \f{ \dd \mu }{ 2\pi } 
\; + \; (\qp - q) K(\om -\qp )   \f{ \veps( \qp ) }{ 2\pi }
\; - \; (\qm + q) K(\om -\qm )   \f{ \veps( \qm ) }{ 2\pi }
\; +  \; 
\e{O}\big( (\wh{q}^{\,(\pm)} \mp q )^2 \big) \;. 
\label{ecriture DA contribution bords a integral}
\enq
The first two terms vanish since  $\veps(\wh{q}^{\,(\pm)})=0$. Hence, 
the corrections stemming from the displacement of the endpoints in respect to $\pm q$ produce at most $\e{O}(T^{2})$
corrections in \eqref{equation DA ordre 1 terme integral NLIE}. 
 As a consequence $\veps_1=0$. This means that, in fact,
the corrections stemming from the displacement of the endpoints are at least a $\e{O}(T^2)$. 
Hence, as follows from \eqref{ecriture DA contribution bords a integral}, they generate 
$\e{O}(T^{4})$ corrections in \eqref{equation DA ordre 1 terme integral NLIE}. 
After rearranging all of the expansions, we get that 
\beq
\veps(\la) = \veps_0(\la) + \f{ T^2 \pi }{ 12 \veps_0^{\prime}(q) }  \bigg\{ R(\la,q) \; + \; R(\la,-q)  \bigg\}
\; + \; \e{O}\big( T^4 \big) \;, 
\enq
where $R(\la,\mu)$ is the kernel of the resolvent operator to $I+\tf{K}{(2\pi)}$, \textit{ie}
$\big(I-\tf{R}{(2\pi)}\big) \cdot \big(I+\tf{K}{(2\pi)}\big) =I$.

In order to access to the low-$T$ asymptotic behavior of $\mc{B}(\xi)$ given in \eqref{definition fonction B}, 
just as in our analysis of the non-linear 
integral equation for the function $\mf{a}$, we first deform the upper part of the contour to 
$\R \; + \; i\tf{\zeta}{2}$ and, if necessary, slightly shift upwards the lower part of the contour. 
Upon applying lemma \ref{Lemme estimation integrales avec une fonction u}, this leads to 
\bem
\mc{B}(\xi) = T \be \coth(\xi)   \; - \;  T \de_{\xi}   \ln \big[ 1+ \ex{-\f{\veps(-i\tf{\zeta}{2}-\xi)}{T} }\big]    \; + \; 
 \Int{\qm}{\qp} \coth(\la + \xi + i\tf{\zeta}{2}) \veps(\la ) \f{\dd\la}{2i\pi}   \\
\; - \; \f{\pi^2 T^2 }{2i\pi 6} \bigg\{ \f{  \coth(\qp+\xi + i\tf{\zeta}{2})  }{ \veps^{\prime}(\qp) } \;  - \;  
\f{  \coth(\qm + \xi + i\tf{\zeta}{2})  }{ \veps^{\prime}(\qm) }   \bigg\} \; + \; \e{O}(T^4) \;. 
\end{multline}
One can simplify the logarithm using that 
\beq
\ba{cc} \Re[ \veps(-i\tf{\zeta}{2}-\xi) ]<0  & \e{for} \;\; 0 < -\Im(\xi)< \tf{\zeta}{2} \vspace{2mm} \\

\Re[ \veps(-i\tf{\zeta}{2}-\xi) ]>0 & \e{otherwise} \ea \; .
\enq
 Also, the endpoints  $\wh{q}^{(\pm)}$ can be replaced by $\pm q$
since the former deviated from the latter by $\e{O}(T^2)$ corrections 
and one also has that $\veps(\wh{q}^{(\pm)})=0$. This leads to 
\bem
\mc{B}(\xi) = T \be \coth(\xi)  \; + \;  T \bs{1}_{A}(\xi) \f{\veps(-i\tf{\zeta}{2}-\xi)}{T}    \; + \; 
 \Int{ - q }{ q } \coth(\la + \xi + i\tf{\zeta}{2}) \; \veps(\la ) \f{\dd\la}{2i\pi}   \\
\; + \;i   \f{\pi T^2 }{12  \veps_0^{\prime}(q) } \Big\{ \coth(q + \xi + i\tf{\zeta}{2})    \;  + \;  
\coth(q - \xi - i\tf{\zeta}{2})    \Big\} \; + \; \e{O}(T^4) \;. 
\label{ecriture fct B a ordre 4 en low T}
\end{multline}
Where $\bs{1}_{A}$ represents the indicator function of the set 
$A = \{ z \in \Cx \; : \; 0\leq - \Im(z) < \tf{\zeta}{2} \}$. It is clear on the level of 
the expansion \eqref{ecriture fct B a ordre 4 en low T} that the second term can be 
recast as a contour integral leading to \eqref{ecriture low T behavior B de xi}. \qed

%
%
%
%
%
%
%
%
%
%
%
%

\vspace{3mm}
We can now apply the previous proposition so as to obtain the low-temperature expansion for the boundary magnetization. 
Indeed, inserting the low-$T$ asymptotic expansion of $\mc{B}$ in the reconstruction formula 
\eqref{representation spin site 1 comme derivee partielle} for $\big< \sg_1^{z} \big>$ one gets  
\bem
\big< \sg_1^z \big> \;  =  \;  1 + \f{ \sinh^2(\xi_-) }{J i  \sin(\zeta) } \Int{ \mc{C}_{\xi_{-}} }{}
 \f{ \veps_0(\la) }{ \sinh^{2}(\la+\xi_- + i\tf{\zeta}{2})} \cdot \f{\dd \la}{2i\pi}  \\
 + T^2 \f{ \pi \sinh^2(\xi_-) }{ 12 J \sin(\zeta ) \veps_0^{\prime}(q) }   \cdot  \f{ \Dp{} }{ \Dp{} \xi_- } 
  \f{  \s{2\xi_- +i \zeta} }{ \s{q \, + \, \xi_- \, + \, i\tf{\zeta}{2}} \s{q \,-\, \xi_- \, - \,i\tf{\zeta}{2}} } 
\; + \; \e{O}(T^4)   \; .
\label{ecriture formule bound mag}
\end{multline}

One can check that the $T=0$ limit in this above expansion does reproduce the 
expression obtained in \cite{KozKitMailNicSlaTerElemntaryblocksopenXXZ} at finite magnetic field $h$. Indeed, let 
us recast the first line of \eqref{ecriture formule bound mag} 
in the language of that paper. One can first carry out an integration by parts. 
The boundary terms coming from an integration along $\intff{-q}{q}$ vanish since $\veps_0(\pm q)=0$ whereas 
the contour integral $\Ga(-i\tf{\zeta}{2}-\xi_-)$, should it be present,  has no boundaries. Thus 
\beq
\Int{ \mc{C}_{\xi_{-}} }{}
 \f{  \sinh(\xi_-)  \veps_0(\la) }{ \sinh^{2}(\la+\xi_- + i\tf{\zeta}{2})} \cdot \dd \la  \; = \; 
-  \Int{ \mc{C}_{\xi_{-}} }{} \big[ \Dp{\la}\veps_0(\la)  \big] \sinh(\xi_-) \cdot  \coth(\la+\xi_- + i\tf{\zeta}{2}) \cdot \dd \la 
\enq
%
%
One then has that 
\beq
 \sinh(\xi_-) \cosh(\la+\xi_- + i\tf{\zeta}{2}) =  \cosh(\xi_-)  \sinh(\la+\xi_- + i\tf{\zeta}{2})  - \sinh(\la + i\tf{\zeta}{2}) \;. 
\label{equation decomposition produit sinh cosh}
\enq
Clearly, the first term in the \textit{r.h.s.}  of \eqref{equation decomposition produit sinh cosh} 
will only have a vanishing contribution to the integral so that 
\beq
\big< \sg_1^z \big>_{\mid T=0}  \; = \;  1 - \f{ \sinh(\xi_-) }{J 2\pi \sin(\zeta) } \Int{ \mc{C}_{\xi_{-}} }{}
 \f{ \veps_0^{\prime}(\la) \sinh(\la+i\tf{\zeta}{2}) }{ \sinh(\la+\xi_- + i\tf{\zeta}{2})} \cdot \dd \la \;. 
\enq
It is easy to check that $\veps_0^{\prime}(\la)$ solves the linear integral equation
\beq
\veps_0^{\prime}(\la) \;  + \;  \Int{-q}{q} K(\la-\mu) \veps_0^{\prime}(\mu) \f{ \dd \mu }{2\pi} \; = \; 
-2\pi J \sin(\zeta)  \f{\Dp{} }{ \Dp{} \la } \Big(  \f{ \sin(\zeta) }{ \pi \sinh(\la+i\tf{\zeta}{2}) \sinh(\la-i\tf{\zeta}{2})  } \Big) \;. 
\enq
The so-called inhomogeneous density of Bethe roots in the ground state $\rho(\la;\xi)$ which was used
in \cite{KozKitMailNicSlaTerElemntaryblocksopenXXZ}, satisfies the linear integral equation 
\beq
\rho(\la;\xi) \;  + \;  \Int{-q}{q} K(\la-\mu) \rho(\la;\xi) \f{ \dd \mu }{2\pi} \; = \; 
 \f{ \sin(\zeta) }{ \pi \sinh(\la- \xi ) \sinh\cosh(\eta)(\la-\xi-i\zeta)  }  \;. 
\enq
Thence 
\beq
\Dp{\xi}\rho(\la;\xi) \mid_{\xi = -i\tf{\zeta}{2}} \;  =  \; 
\f{\veps_0^{\prime}(\la) }{2\pi J \sin(\zeta) } \;, 
\enq
and, upon substitution, we get 
\beq
\moy{\sg_1}_{T=0} =  1 - \Int{ \mc{C}_{\xi_{-}} }{}
 \f{ \sinh(\xi_-) \sinh(\la+i\tf{\zeta}{2}) }{ \sinh(\la+\xi_- + i\tf{\zeta}{2})} 
 { \f{ \Dp{} }{ \Dp{} \xi }\rho(\la;\xi) }_{\mid_{\xi = -i\tf{\zeta}{2}}} \hspace{-3mm} \dd \la \;. 
\enq
which is precisely the representation found in \cite{KozKitMailNicSlaTerElemntaryblocksopenXXZ}.


\section*{Conclusion}

In this paper we have studied the so-called boundary magnetization of an XXZ spin-$\tf{1}{2}$ chain subject to 
diagonal boundary fields on each of its boundaries. Starting from a representation for its finite Trotter number 
approximant obtained by G\"{o}hmann, Bortz and Frahm \cite{BortzFrahmGohmannSurfaceFreeEnergy} we have recast it as
a product of partition functions of the six-vertex model with reflecting ends. Using the determinant representations of the
latter partition functions obtained by Tsuchiya \cite{TsuchiyaPartitionFunctWithReflecEnd} along with the Cauchy determinant
factorization trick \cite{IzerginKitMailTerSpontaneousMagnetizationMassiveXXZ,KozKitMailSlaTerXXZsgZsgZAsymptotics} 
we have been able to recast the resulting expression into a form that allowed us to take the infinite Trotter number limit explicitly. 
We then applied our result to the computation of the boundary magnetization at finite temperature. 
In such a way, we were able to check that the zero temperature limit of our result does reproduce the known integral 
representations for $\big< \sg_1^z \big>$ and also to draw some curves for the latter quantity in the massive regime. 

A natural continuation of our study would be to address the question of building an effective approach to the computation of the 
correlation functions in models subject to diagonal boundary conditions at finite temperature. 
Such a work would provide one with the boundary analog of the representations obtained in 
\cite{GohmannKlumperSeelFinieTemperatureCorrelationFunctionsXXZ} and an extension of the representations 
for the correlation functions at $T=0K$ obtained in 
\cite{KozKitMailNicSlaTerElemntaryblocksopenXXZ,KozKitMAilNicSlaTerResummationsOpenXXZ}.

It would also be interesting to confront the predictions issuing from the 
thermodynamic Bethe Ansatz with our results. The full boundary free energy has
not yet been obtained in the latter approach. More precisely, certain boundary magnetic field 
independent terms (which are expected to arize see 
\cite{sajat-g,woynarovich,woynarovich-uj}) have not yet been
considered in the case of the spin chain. It is in fact not clear 
whether the thermodynamic Bethe Ansatz in its present setting is capable of providing the full boundary free energy.
A less ambitious problem would be to consider the boundary
magnetization in the framework of thermodynamic Bethe Ansatz. Indeed, as opposed to the full free
energy, the boundary magnetization is typically given by simple
integrals and not by a series of multiple integrals. Therefore, the
comparison to our present results is probably a manageable task. We
leave these problems for further research.

Lastly, it would be desirable to extract the low-T behavior of the surface free energy 
directly out of its representation \eqref{ecriture surf free energy at infinite trotter}. 
For the simple and double integrals, this can be done rather straightforwardly 
following the standart method of low-T asymptotic analysis. Yet, the series of multiple integrals
characterizing the function $\mc{F}$ \eqref{ecriture series limite thermo fction F caligraphique}
poses serious problems for extracting the low-T asymptotic analysis. We expect that 
all terms of the series will contribute to the leading orders, and thus a more
refined method of low-T  asymptotic analysis, possibly in the spirit of \cite{KozKitMailSlaTerXXZsgZsgZAsymptotics}, 
should be implemented.

\section*{Acknowledgements}

We are grateful to F. G\"{o}hmann and A. Kl\"umper for drawing our
attention to the present problem and for further discussions.
We are also indebted to C. Matsui and F. Toninelli for useful discussions. 

K. K. K. is supported by CNRS. 
K. K. K. would like to thank J.-S. Caux for an invitation to the Institute for Theoretical Physics 
at the University of Amsterdam which allowed us to initiate this work. 
K. K. K. would like to thank the department of Mathematical sciences of IUPUI for hospitality during 
the time part of this work was done. Also, when the first part of this research was carried 
K. K. K. was supported by the EU Marie-Curie Excellence Grant MEXT-CT-2006-042695 and would like 
to thank the DESY theory group for hospitality during that time. 

B. P. was supported by the VENI Grant 016.119.023 of the NWO.



\appendix

\section{Low-temperature behavior of integrals}

In this appendix, we briefly recall a lemma  that allows
one to carry out the asymptotic analysis of integrals involving the function $\ln[1+\mf{a}(\om)]$
integrated versus some regular function. 
This lemma can be proven along the lines given in \cite{KozProofexistenceAEYangYangEquation}. 

\begin{lemme}
\label{Lemme estimation integrales avec une fonction u}
Assume that

\begin{itemize}

\item  $u$ is holomorphic in a neighborhood U of $\R$,

\item  $u$ has two simple zeroes $ \wh{q}^{\,(\pm)}$ in $U$ and $\Re\big[u(\la)\big] \limit{\Re(\la)}{+\infty} +\infty $ with $\la \in U$;

\item $\ln\Big[ 1 + \ex{-\f{u(\la)}{T}} \Big] $ is holomorphic on $U$.

\item for any $k$, \;  $\ex{-\f{u(\la) }{T}} = T^k \e{O}\pa{ \la^{-\infty} }$  uniformly in $\Re\pa{\la} \tend \pm \infty$ in $U$.
\end{itemize}

For any function $f$ holomorphic on $U$ with a polynomial growth along $\Re\pa{\la} \tend \pm \infty $ in $U$, $f=\e{O}\big( \la^k \big)$,
$k\in \mathbb{N}$ one has
\beq
 \Int{\R}{}  f(\la) \ln\Big[ 1 + \ex{-\f{u\pa{\la}}{T}} \Big] \dd \la = 
 - \f{1}{T} \Int{ \qm }{ \qp  }\! \! f(\la) u(\la) \cdot  \dd \la
\;  + \; \f{\pi^2 T }{6}
 \bigg\{ \f{f}{u^{\prime}} \big( \qp \big) - \f{f}{u^{\prime}} \big( \qm \big)  \bigg\} \; + \;  \e{O}(T^3) \;.
\label{ecriture DA integrale avec u lisse}
\enq
\end{lemme}

We stress that the $\e{O}$ is uniform in respect to any auxiliary parameter over which $f$ depends. 
If the $l.h.s.$ in \eqref{ecriture DA integrale avec u lisse}
is integrable in respect to this parameter so is the $r.h.s$. In such a case, the $\e{O}(T^3)$ remains unaffected by such an integration.


\begin{thebibliography}{10}

\bibitem{AffleckCFTPreForLargeSizeCorrPartitionFctonAndLowTBehavior}
I.~Affleck, \emph{{"Universal term in the free energy at a critical point and
  the conformal anomaly."}}, Phys. Rev. Lett. \textbf{56} (1986), 746--748.

\bibitem{Affleck-XXZ-boundary-QFT}
I.~Affleck, S.S. Eggert, S.~Fujimoto, N.~Laflorencie, and J.~Sirker,
  \emph{{"Thermodynamics of impurities in the anisotropic Heisenberg spin-1/2
  chain."}}, J. Stat. Mech. (2008), P02015.

\bibitem{AlcarazBatchelorBaxterQuispelCBAopenXXZ}
F.C. Alcaraz, N.M. Batchelor, R.J. Baxter, and G.~R.W. Quispel, \emph{{"Surface
  exponents of the quantum XXZ, Ashkin-Teller and Potts models."}}, J. Phys. A:
  Math. Gen. \textbf{20} (1987), 6397--6409.

\bibitem{BatchelorKlumperFirstIntoNLIEForFiniteSizeCorrectionSpin1XXZAlternati%
veToRootDensityMethod}
M.~T. Batchelor and A.~Kl\"{u}mper, \emph{{"An analytic treatment of
  finite-size corrections in the spin-1 antiferromagnetic XXZ chain."}}, J.
  Phys. A: Math; Gen. \textbf{\bf{23}} (1990), L--189--195.

\bibitem{BetheSolutionToXXX}
H.~Bethe, \emph{{"On the theory of metals: Eigenvalues and Eigenfunctions of a
  linear chain of atoms."}}, Zeitschrift f$\ddot{u}$r Physik \textbf{\bf 71}
  (1931), 205--226.

\bibitem{BortzFrahmGohmannSurfaceFreeEnergy}
F.~G\"{o}hmann, M.~Bortz and H.~Frahm, \emph{{"Surface free energy for systems
  with integrable boundary conditions."}}, J. Phys. A: Math. Gen. \textbf{\bf
  38} (2005), 10879--10892.

\bibitem{Bortz:JPhysAMathGen38:2005}
M.~Bortz and J.~Sirker, \emph{{"Boundary susceptibility in the open
  XXZ-chain."}}, J. Phys. A: Math. Gen. 38 (2005), 5957.

\bibitem{Sirker:JStatMech0601P01007:2006}
\bysame, \emph{{"The open XXZ-chain: bosonisation, Bethe Ansatz and logarithmic
  corrections."}}, J.Stat.Mech. (2006), P01007.

\bibitem{CardyConformalExponents}
J.L. Cardy, \emph{{"Conformal invariance and universality in finite-size
  scaling."}}, J. Phys. A: Math. Gen. \textbf{\bf 17} (1984), L385--387.

\bibitem{CherednikReflectionEquationFactorisabilityOfScattering}
V.I. Cheredink, \emph{{"Factorizing particles on a half-line and root
  systems."}}, Theor. Math. Phys. \textbf{\bf{61}} (1984), 977.

\bibitem{TsvelikOpenXXZandKondoimpurityeffect}
P.~de~Sa and A.~M. Tsvelik, \emph{{"Anisotropic spin-1/2 Heisenberg chain with
  open boundary conditions."}}, Phys. Rev. B \textbf{52} (1995), 3067--3070.


\bibitem{Dorey:2004xk}
P.~Dorey, D.~Fioravanti, C.~Rim, and R.~Tateo, \emph{{"Integrable quantum
  field theory with boundaries: the exact g-function."}}, Nucl. Phys.
  \textbf{B696} (2004), 445--467.


\bibitem{DorlasLewisPuleRigorousProofYangYangThermoEqnNLSE}
T.C. Dorlas, J.T. Lewis, and J.V. Pul\'{e}, \emph{{"The Yang-Yang thermodynamic
  formalism and large deviations."}}, Comm. Math. Phys. \textbf{\bf{124}, 3}
  (1989), 365--402.
  
    
\bibitem{Fujimoto:PhysRevLett92:2004}
S.~Eggert and S.~Fujimoto, \emph{Boundary susceptibility in the spin-1/2
  chain: Curie like behavior without magnetic impurities}, Phys. Rev. Lett. 92
  (2004), 037206.


\bibitem{EsslerFrahmGohmanKlumperKorepinOneDimensionalHubbardModel}
F.~H.L. Essler, H.~Frahm, F.~G\"{o}hmann, A.~Kl\"{u}mper, and V.~E. Korepin,
  \emph{{"The one-dimensional Hubbard model."}}, Cambridge University Press,
  2005.

\bibitem{Frahm-Zvyagin}
H.~Frahm and A.~A. Zvyagin, \emph{{"The open spin chain with impurity: an exact
  solution."}}, J. Phys.: cond. mat. \textbf{9} (1997), no.~45, 9939.

\bibitem{FurusakiHikihara}
A.~Furusaki and T.~Hikihara, \emph{{"Boundary contributions to specific heat
  and susceptibility in the spin-1/2 XXZ chain."}}, Phys. Rev. B \textbf{69}
  (2004), no.~9, 094429.

\bibitem{GaudinTBAXXZMassiveInfiniteSetNLIE}
M.~Gaudin, \emph{{"Thermodynamics of a Heisenberg-Ising ring for $\De \geq
  1$."}}, Phys. Rev. Lett \textbf{\bf 26} (1971), 1301--1304.

\bibitem{GohmannPrivateCommJustOfQTMMethod}
F.~G\"{o}hmann, \emph{{Private communication.}}

\bibitem{GohmannKlumperSeelFinieTemperatureCorrelationFunctionsXXZ}
F.~G\"{o}hmann, A.~Kl\"{u}mper, and A.~Seel, \emph{{"Integral representations
  for correlation functions of the XXZ chain at finite temperature."}}, J.
  Phys. A: Math. Gen. \textbf{\bf 37} (2004), 7625--7652.

\bibitem{HaldaneLuttingerLiquidCaracterofBASolvableModels}
F.D.M. Haldane, \emph{{"Demonstration of the ``Luttinger liquid`` character of
  Bethe-Ansatz soluble models of 1-D quantum fluids"}}, Phys. Lett. A
  \textbf{\bf 81} (1981), 153--155.

\bibitem{IzerginPartitionfunction6vertexDomainWall}
A.G. Izergin, \emph{{"Statistical sum of the six-vertex model in finite
  volume."}}, Dokl. Akad. Nauk SSSR \textbf{\bf{297}} (1987), 331--333.

\bibitem{IzerginKitMailTerSpontaneousMagnetizationMassiveXXZ}
A.G. Izergin, N.~Kitanine, J.M. Maillet, and V.~Terras, \emph{{"Spontaneous
  magnetization of the XXZ Heisenberg spin 1/2 chain."}}, Nucl. Phys. B
  \textbf{\bf{554}} (1999), 679--696.

\bibitem{JimboKedemKonnoMiwaXXZChainWithaBoundaryElemBlcks}
M.~Jimbo, R.~Kedem, T.~Kojima, H.~Konno, and T.~Miwa, \emph{{"XXZ chain with a
  boundary."}}, Nucl. Phys. B \textbf{\bf 441} (1995), 437--470.

\bibitem{KapustinSkorikStructureGSOpenXXZMassless}
A.~Kapustin and S.~Skorik, \emph{{"Surface excitations and surface energy of
  the antiferromagnetic $XXZ$ chain by the Bethe Ansatz approach."}}, J. Phys.
  A: Math. Gen \textbf{\bf 29} (1996), 1629.

\bibitem{KozKitMailNicSlaTerElemntaryblocksopenXXZ}
N.~Kitanine, K.K. Kozlowski, J.-M. Maillet, G.~Niccoli, N.A. Slavnov, and
  V.~Terras, \emph{{"Correlation functions of the open XXZ chain I."}}, J.
  Stat. Mech.: Th. and Exp. (2007), P10009.


\bibitem{KozKitMAilNicSlaTerResummationsOpenXXZ}
N.~Kitanine, K.K. Kozlowski, J.-M. Maillet, G.~Niccoli, N.A. Slavnov, and
  V.~Terras, \emph{{"Correlation functions of the open XXZ chain II."}}, J.
  Stat. Mech.: Th. and Exp. (2008), P07010.


\bibitem{KozKitMailSlaTerXXZsgZsgZAsymptotics}
N.~Kitanine, K.K. Kozlowski, J.-M. Maillet, N.A. Slavnov, and V.~Terras,
  \emph{{"Algebraic Bethe Ansatz approach to the asymptotics behavior of
  correlation functions."}}, J. Stat. Mech: Th. and Exp. \textbf{04} (2009),
  P04003.

\bibitem{KlumperNLIEfromQTMDescrThermoXYZOneUnknownFcton}
A.~Kl\"{u}mper, \emph{{"Thermodynamics of the anisotropic spin-$1/2$ Heisenberg
  chain and related quantum chains."}}, Zeit. F\"{u}r Phys.: Cond. Mat.
  \textbf{\bf B 91} (1993), 507--519.
  

\bibitem{KlumperBatchelorPearceCentralChargesfor6And19VertexModelsNLIE}
A.~Kl\"{u}mper, M.~T. Batchelor, and P.~A. Pearce, \emph{{"Central charges for
  the 6- and 19-vertex models with twisted boundary conditions."}}, J. Phys. A:
  Math. Gen. \textbf{\bf 24} (1991), 3111--3133.

\bibitem{KlumperWehnerZittartzConformalSpectrumofXXZCritExp6Vertex}
A.~Kl\"{u}mper, T.~Wehner, and J.~Zittartz, \emph{{"Conformal spectrum of the
  6-vertex model."}}, J. Phys. A: Math. Gen. \textbf{\bf 26} (1993),
  2815--2827.

\bibitem{KomaIntroductionQTM6VertexForThermodynamicsOfXXX}
T.~Koma, \emph{{"Thermal Bethe-Ansatz method for the one-dimensional Heisenberg
  model."}}, Prog. Theor. Phys. \textbf{\bf 78} (1987),
  1213--1218.

\bibitem{KomaIntroductionQTM6VertexForThermodynamicsOfXXZ}
\bysame, \emph{{"Thermal Bethe-Ansatz method for the spin-$\tf{1}{2}$ XXZ
  Heisenberg model."}}, Prog. Theor. Phys. \textbf{\bf 81} (1989), 783--809.

\bibitem{KorepinNormBetheStates6-Vertex}
V.~E. Korepin, \emph{{"Calculation of norms of Bethe wave-functions."}}, Comm.
  Math.Phys. \textbf{\bf 86} (1982), 391.

\bibitem{KozProofexistenceAEYangYangEquation}
K.~K. Kozlowski, \emph{{"Low-$T$ asymptotic expansion of the solution to the
  Yang-Yang equation."}}, math-ph: 11126199.

\bibitem{KozMailletSlaLowTLimitNLSE}
K.~K. Kozlowski, J.-M. Maillet, and N.~A. Slavnov, \emph{{"Low-temperature
  limit of the long-distance asymptotics in the non-linear Schr\"{o}dinger
  model."}}, J.Stat.Mech. (2011), P03019.

\bibitem{LiebLinigerCBAForDeltaBoseGas}
E.H. Lieb and W.~Liniger, \emph{{"Exact analysis of an interacting Bose gas. I.
  The general solution and the ground state."}}, Phys. Rev. \textbf{\bf{130}}
  (1963), 1605--1616.

\bibitem{McCoySomeAsymptoticsForXYCorrelators}
B.~M. McCoy, \emph{{"Spin correlation functions in the XY model."}}, Phys. Rev.
  \textbf{\bf{173}} (1968), 531--541.




\bibitem{sajat-g}
B.~Pozsgay, \emph{{On O(1) contributions to the free energy in Bethe Ansatz
  systems: the exact g-function}}, JHEP \textbf{08} (2010), 090.

\bibitem{RuelleRigorousResultsForStatisticalMechanics}
D.~Ruelle, \emph{{"Statistical mechanics: rigorous results"}}, W.A. Benjamin,
  Inc., 1969.

\bibitem{SklyaninABAopenmodels}
E.~K. Sklyanin, \emph{{"Boundary conditions for integrable quantum systems."}},
  J. Phys. A: Math. Gen. \textbf{\bf 28} (1988), 2375--2389.

\bibitem{SuzukiCorrespondence(D+1)StatPhysDQuantumHamiltonians}
J.~Suzuki, \emph{{"Relationship between $d$-dimensional quantal spin systems
  and $(d+1)$-dimensional Ising systems."}}, Prog. Theor. Phys. \textbf{\bf 56}
  (1976), 1454--1469.

\bibitem{SuzukiArgumentsForInterchangeabilityTrotterAndVolumeLimitInPartFcton}
M.~Suzuki, \emph{{"Transfer-matrix methods and Monte Carlo simulation in
  quantum spin systems."}}, Phys. Rev. B \textbf{\bf 31} (1985), 2957--2965.

\bibitem{Takahashi-book}
M.~Takahashi, \emph{{"Thermodynamics of one-dimensional solvable models."}},
  Cambridge University Press, 1999.

\bibitem{TakahashiTBAforXXZFiniteTinfiniteNbrNLIE}
M.~Takahashi, \emph{{"One-dimensional Heisenberg model at finite
  temperature."}}, Prog. Theor. Phys. \textbf{\bf 42} (1971), 1289.

\bibitem{TakahashiThermoXXZInfiniteNbrRootsFromQTM}
\bysame, \emph{{"Correlation length and free energy of the S$=\tf{1}{2}$ XXZ
  chain in a magnetic field."}}, Phys. Rev. B \textbf{\bf 44} (1991),
  12382--12394.

\bibitem{TakahashiThermoXYZInfiniteNbrRootsFromQTM}
\bysame, \emph{{"Correlation length and free energy of the S$=\tf{1}{2}$ XYZ
  chain."}}, Phys. Rev. B \textbf{\bf 43} (1991), 5788--5797.

\bibitem{TracyVaidyaRedDensityMatrixSpaceAsymptImpBosonsT=0}
C.~A. Tracy and H.~Vaidya, \emph{{"One particle reduced density matrix of
  impenetrable bosons in one dimension at zero temperature."}}, J. Math. Phys.
  \textbf{\bf{20}}  (1979), 2291--2312.

\bibitem{TsuchiyaPartitionFunctWithReflecEnd}
O.~Tsuchiya, \emph{{"Determinant formula for the six-vertex model with
  reflecting end."}}, J. Phys. A: Math. Gen. \textbf{\bf 39} (1998),
  5946--5951.

\bibitem{woynarovich}
F.~Woynarovich, \emph{{"O(1) contribution of saddle-point fluctuations to the
  free energy of Bethe Ansatz systems."}}, Nucl. Phys. \textbf{B700} (2004),
  331.

\bibitem{woynarovich-uj}
\bysame, \emph{{"On the normalization of the partition function of Bethe Ansatz
  systems."}}, Nuclear Physics B \textbf{852} (2011), 269--286.

\bibitem{WuAsymptoticsSpinSpinIsingModel}
T.~T. Wu, \emph{{"Theory of Toeplitz determinants and the spin correlation
  functions of the two-dimensional Ising model I."}}, Phys. Rev.
  \textbf{\bf{149}} (1966), 380--401.

\bibitem{Yang-YangNLSEThermodynamics}
C.~N. Yang and C.~P. Yang, \emph{{"Thermodynamics of a one-dimensional system
  of bosons with repulsive delta-interactions."}}, J. Math. Phys. \textbf{\bf
  10} (1969), 1115--1122.

\bibitem{Zvyagin-Makarova}
A.~A. Zvyagin and A.~V. Makarova, \emph{{"Bethe-Ansatz study of the
  low-temperature thermodynamics of an open Heisenberg chain"}}, Phys. Rev. B
  \textbf{69} (2004), no.~21, 214430.

\end{thebibliography}
\end{document}